\documentclass[a4paper,11pt]{article}
\pdfoutput=1 

\usepackage{jheppub} 
\usepackage{bm} 

\usepackage[T1]{fontenc} 
\usepackage[pdftex]{color}

\title{\boldmath ${\overline{\rm MS}}$ renormalization of 
$S$-wave quarkonium wavefunctions at the origin}
\preprint{TUM-EFT 135/20}

\author{Hee~Sok~Chung}
\affiliation{Physik-Department, Technische Universit\"at M\"unchen,
James-Franck-Str. 1, 85748 Garching, Germany}
\affiliation{Excellence Cluster ORIGINS,
Boltzmannstrasse 2, D-85748 Garching, Germany}

\emailAdd{heesok.chung@tum.de}

\abstract{
We compute $S$-wave quarkonium wavefunctions at the origin in the
$\overline{\rm MS}$ scheme based on nonrelativistic effective field theories. 
We include the effects of nonperturbative long-distance behaviors  
of the potentials, 
while we determine the short-distance behaviors of the potentials 
in perturbative QCD. 
We obtain $\overline{\rm MS}$-renormalized quarkonium wavefunctions at the
origin that have the correct scale dependences that are expected from 
perturbative QCD, so that the scale dependences cancel in physical quantities. 
Based on the calculation of the wavefunctions at the origin, we make
model-independent predictions of decay constants and 
electromagnetic decay rates of $S$-wave charmonia and bottomonia, 
and compare them with measurements.
We find that the poor convergence of perturbative QCD corrections 
are substantially improved when we include corrections to the 
wavefunctions at the origin in the calculation of 
decay constants and decay rates. 
}

\begin{document} 
\maketitle
\flushbottom

\section{Introduction}
\label{sec:intro}

Heavy quarkonium production and decay processes are multiscale problems that
are sensitive to both short-distance and long-distance natures of QCD. 
As many of these processes have been measured experimentally, 
a great amount of effort has been made towards understanding them
theoretically~\cite{Brambilla:2010cs,Brambilla:2014jmp}. 
Much of the heavy quarkonium phenomenology is based on 
nonrelativistic effective field theories, which provide factorization 
formalisms that separate the perturbative short-distance physics from 
nonperturbative long-distance physics. 
In the nonrelativistic QCD (NRQCD) factorization
formalism~\cite{Caswell:1985ui, Bodwin:1994jh}, 
production or decay rates of a heavy quarkonium are given by sums of products 
of the perturbatively calculable short-distance coefficients (SDCs) 
and long-distance matrix elements (LDMEs). 
The LDMEs are nonperturbative quantities that correspond to the probability to 
find a heavy quark $Q$ and a heavy antiquark $\bar Q$ inside a quarkonium. 
The LDMEs have known scalings in $v$, the typical heavy-quark velocity 
inside the quarkonium, and the sum is organized in powers of $v$. 
While the SDCs can be computed in perturbative QCD, 
accurate determinations of the LDMEs, especially the ones that appear at the
lowest orders in $v$, are also important in making predictions of 
production or decay rates of heavy quarkonia. 

So far, many phenomenological studies on heavy quarkonium production and decay
have relied on model calculations of the LDMEs~\cite{Eichten:1978tg,
Buchmuller:1980su, Eichten:1995ch, Bodwin:2007fz, Chung:2010vz}. 
One major disadvantage of model calculations 
is that in general, 
they do not reproduce the correct ultraviolet (UV) behaviors of the LDMEs 
that are predicted in perturbative QCD. 
Perturbative QCD calculations show that 
the LDMEs contain UV divergences, which require
renormalization~\cite{Bodwin:1994jh}. 
That is, the LDMEs are renormalization scheme dependent. 
The scale associated with the renormalization of the LDMEs is often called
the NRQCD factorization scale. 
The SDCs also depend on the scheme in which the LDMEs are
renormalized, in the way that the scheme dependence cancels between the SDCs
and the LDMEs in the factorization formula. 
It has been found from perturbative QCD calculations that strong dependencies
on the factorization scale start to appear in the SDCs from two 
loops~\cite{Czarnecki:1997vz, Beneke:1997jm, Czarnecki:2001zc, Kniehl:2006qw}. 
Therefore, in order to make accurate predictions based on NRQCD, it is
critically important to determine the LDMEs that exhibit the correct 
scale dependence. Since perturbative QCD calculations are most
conveniently done in dimensional regularization (DR), the SDCs are usually 
computed in the ${\overline{\rm MS}}$ scheme. In order to be 
consistent with the ${\overline{\rm MS}}$ calculations of the SDCs, 
the LDMEs must also be determined in the ${\overline{\rm MS}}$ scheme.

While lattice QCD determinations of certain LDMEs exist~\cite{Bodwin:1993wf, 
Bodwin:1996tg, Bodwin:1996mf, Bodwin:2001mk, Bodwin:2004up, Bodwin:2005gg}, 
these calculations are usually done in quenched lattice QCD, 
and their results have large uncertainties. 
Moreover, the relations between the LDMEs in lattice and continuum are known 
only at one-loop level. Hence, existing lattice QCD determinations are not
accurate enough to reproduce the factorization-scale dependence that is
expected in perturbative QCD. 

It has been known that NRQCD LDMEs can be computed from quarkonium wavefunctions
at the origin~\cite{Bodwin:1994jh}. 
Rigorous formulations for quarkonium wavefunctions have been
developed in the potential NRQCD (pNRQCD) effective field theory
approach~\cite{Pineda:1997bj,Brambilla:1999xf,Brambilla:2004jw}. This
formalism provides a Schr\"odinger formulation, from which the quarkonium
wavefunctions can be computed. The potentials that appear in the
Schr\"odinger equation have field-theoretical definitions in terms of Wilson
loops, and they can be computed nonperturbatively in lattice
QCD~\cite{Brambilla:2000gk,Pineda:2000sz}. While this makes possible the
nonperturbative determination of quarkonium wavefunctions, there are still
challenges in computing the wavefunctions at the origin from
first principles. One major challenge is that the wavefunctions at the origin
involve divergences that require renormalization. These divergences are related
closely to the UV divergences that appear in the LDMEs. In order to
obtain the ${\overline{\rm MS}}$-renormalized LDMEs, the wavefunctions at the 
origin must also be renormalized in the same scheme. The problem is that this 
requires dimensionally regulated calculations, which are difficult to be done 
outside of perturbation theory. This is because calculations in DR are most
conveniently done in momentum space, while nonperturbative determinations
of the potentials are done in position space. 
For this reason, computations of quarkonium wavefunctions at the origin 
to two-loop accuracy have only been done within perturbative 
QCD~\cite{Hoang:1998xf,
Melnikov:1998pr,
Melnikov:1998ug,Penin:1998kx,
Yakovlev:1998ke,
Beneke:1999qg,
Nagano:1999nw,
Hoang:1999zc,
Penin:1998mx,
Hoang:2000yr,
Penin:2004ay}, 
where the nonperturbative, long-distance behavior of the potentials 
are ignored. 
However, many charmonium and bottomonium states are non Coulombic, so that 
their wavefunctions are sensitive to the nonperturbative long-distance behavior 
of the potentials. In such cases, the nonperturbative behavior of the 
potentials cannot be neglected. 

While a direct momentum-space calculation of the wavefunctions at the origin 
in DR with nonperturbative potentials may be difficult, 
position-space calculations are possible if we regulate the divergences in
position space. Then, we only need to convert the position-space 
regularization to the ${\overline{\rm MS}}$ scheme in order to obtain the 
${\overline{\rm MS}}$-renormalized wavefunctions at the origin.
The conversion from position-space regularization to the ${\overline{\rm MS}}$ 
scheme may be computed in perturbative QCD, 
because this depends only on the divergent short-distance behavior of the
potentials, which are determined completely by perturbative QCD.
This makes possible the nonperturbative calculations of 
${\overline{\rm MS}}$-renormalized wavefunctions at the origin
based on first principles.

In this paper, we compute the ${\overline{\rm MS}}$-renormalized quarkonium
wavefunctions at the origin for $S$-wave charmonium and bottomonium states. 
We compute the wavefunctions at the origin in two steps. First, we compute the
quarkonium wavefunctions in position space, by using potentials that have
nonperturbative long-distance behaviors that are determined in lattice QCD,
while the potentials are given by perturbative QCD at short distances. 
We use quantum-mechanical perturbation theory 
to first order in the expansion in powers of $1/m$
to compute the wavefunctions at the origin.
Then, we convert the position-space regularization to the 
${\overline{\rm MS}}$ scheme. 
Using the ${\overline{\rm MS}}$-renormalized quarkonium wavefunctions at 
the origin that we obtain, we determine the NRQCD LDMEs in the strongly coupled
pNRQCD formalism, which is valid for non Coulombic quarkonia. 
Based on the determinations of the LDMEs, we make model-independent 
predictions of 
decay constants and electromagnetic decay rates of 
$S$-wave charmonium and bottomonium states. 
In the NRQCD factorization formulas, we include loop corrections 
at leading order in $v$ to two-loop accuracy, 
as well as corrections of order $\alpha_s^0 v^2$, 
and, when available, we also include corrections of order $\alpha_s v^2$. 
We restrict the calculation of wavefunctions 
to $S$-wave states, because the two-loop
corrections to the SDCs are generally not available for the production or decay
rates of quarkonia with higher orbital angular momentum. 

This paper is organized as follows. In section~\ref{sec:LDMEs}, we
review the definitions of NRQCD LDMEs and the relations with 
wavefunctions at the origin. We outline the calculation of quarkonium
wavefunctions in position space in sec.~\ref{sec:wavefunctions}, 
which allow nonperturbative calculations of the wavefunctions at the origin 
with a position-space regulator. 
We compute the scheme conversion from position-space regularization to the 
${\overline{\rm MS}}$ scheme in sec.~\ref{sec:conversion}. 
In sec.~\ref{sec:results}, we compute the ${\overline{\rm MS}}$-renormalized 
wavefunctions at the origin, as well as electromagnetic decay rates and 
decay constants of $S$-wave charmonium and bottomonium states, 
which we compare with
experimental measurements and lattice QCD determinations. 
We conclude in sec.~\ref{sec:summary}.

\section{\boldmath NRQCD long-distance matrix elements} 
\label{sec:LDMEs}

In this section, we review the definitions of NRQCD LDMEs involving $S$-wave 
quarkonia and their relations to quarkonium wavefunctions that appear in
pNRQCD. 

In NRQCD factorization formulas for electromagnetic decay rates and
exclusive electromagnetic production rates of vector quarkonium
$V=J/\psi$ or $\Upsilon$, the following LDME appears at leading order in $v$:
\begin{equation}
\langle 0 | \chi^\dag \bm{\epsilon} \cdot \bm{\sigma} \psi | V \rangle, 
\end{equation}
where $\psi$ and $\chi$ are Pauli spinor fields that annihilate and create 
a heavy quark and a heavy antiquark, respectively, 
$|0 \rangle$ is the QCD vacuum, 
and $\bm{\epsilon}$ is the polarization vector of the quarkonium. 
We take nonrelativistic normalization for the state $|V\rangle$. 
At relative order $v^2$, the following LDME appears:
\begin{equation}
\langle 0 | \chi^\dag \bm{\epsilon} \cdot \bm{\sigma} 
(-\tfrac{i}{2} \overleftrightarrow{\bm{D}} )^2 
\psi | V \rangle, 
\end{equation}
where $\bm{D} = \bm{\nabla} -i g_s \bm{A}$, and 
$\chi^\dag \overleftrightarrow{\bm{D}} \psi 
 = \chi^\dag \bm{D} \psi - (\bm{D} \chi)^\dag \psi$. 
The leading-order LDME depends on the factorization scale $\Lambda$ 
from its renormalization.
This factorization scale dependence is given by the 
following evolution equation~\cite{Bodwin:1994jh, Czarnecki:1997vz, 
Beneke:1997jm, Kniehl:2006qw}
\begin{equation}
\label{eq:RG_vec}
\frac{d \log
\langle 0 | \chi^\dag \bm{\epsilon} \cdot \bm{\sigma} \psi | V \rangle
}{d \log \Lambda} 
= \alpha_s^2 C_F \left(\frac{C_F}{3} + \frac{C_A}{2} \right) 
- \frac{4 \alpha_s C_F}{3 \pi} 
\frac{\langle 0 | \chi^\dag \bm{\epsilon} \cdot \bm{\sigma} 
(-\tfrac{i}{2} \overleftrightarrow{\bm{D}} )^2 
\psi | V \rangle}{m^2 \langle 0 | \chi^\dag \bm{\epsilon} \cdot 
\bm{\sigma} \psi | V \rangle}
+ O(\alpha_s^3, \alpha_s^2 v^2),
\end{equation}
where $m$ is the heavy quark pole mass, 
$C_F = (N_c^2-1)/(2 N_c)$, $C_A = N_c$, and $N_c = 3$ is the number of colors. 
We reproduce the anomalous dimensions on the right-hand side of
eq.~(\ref{eq:RG_vec}) from NRQCD loop calculations 
in appendix~\ref{sec:anomalous_dimensions}. 

Analogously, in electromagnetic decay rates
and exclusive electromagnetic production rates of pseudoscalar quarkonium 
$P = \eta_c$ or $\eta_b$, the following LDME appears in factorization
formulas at leading order in $v$:
\begin{equation}
\langle 0 | \chi^\dag \psi | P \rangle, 
\end{equation}
and at relative order $v^2$, the LDME $\langle 0 | \chi^\dag 
(-\tfrac{i}{2} \overleftrightarrow{\bm{D}} )^2 \psi | P \rangle$ appears. 
The factorization scale dependence of the leading-order LDME is given 
by~\cite{Bodwin:1994jh,Czarnecki:2001zc,Kniehl:2006qw}
\begin{equation}
\label{eq:RG_ps}
\frac{d \log
\langle 0 | \chi^\dag \psi | P \rangle
}{d \log \Lambda} 
= \alpha_s^2 C_F \left(C_F + \frac{C_A}{2} \right) 
- 
\frac{4 \alpha_s C_F}{3 \pi} 
\frac{\langle 0 | \chi^\dag 
(-\tfrac{i}{2} \overleftrightarrow{\bm{D}} )^2 
\psi | P \rangle}{m^2 \langle 0 | \chi^\dag \psi | P \rangle}
+ O(\alpha_s^3, \alpha_s^2 v^2).
\end{equation}
We reproduce the anomalous dimensions on the right-hand side of
eq.~(\ref{eq:RG_ps}) from NRQCD loop calculations 
in appendix~\ref{sec:anomalous_dimensions}. 

In this paper, we aim to compute the LDMEs 
$\langle 0 | \chi^\dag \bm{\epsilon} \cdot \bm{\sigma} \psi | V \rangle$, 
$\langle 0 | \chi^\dag \bm{\epsilon} \cdot \bm{\sigma} 
(-\tfrac{i}{2} \overleftrightarrow{\bm{D}} )^2 
\psi | V \rangle$, 
$\langle 0 | \chi^\dag \psi | P \rangle$,
and $\langle 0 | \chi^\dag
(-\tfrac{i}{2} \overleftrightarrow{\bm{D}} )^2 \psi | P \rangle$ 
in pNRQCD. The pNRQCD effective field theory is obtained from NRQCD by
integrating out modes associated with energy scales larger than $mv^2$ 
(see ref.~\cite{Brambilla:2004jw} for a review). 
The pNRQCD formalism provides relations between decay LDMEs in NRQCD, 
which are given by expectation values of four-quark operators
on heavy quarkonium states, and quarkonium wavefunctions at the origin and
their derivatives. Since we are interested in computing the LDMEs for non
Coulombic quarkonia, for which the nonperturbative long-distance behavior of
the quarkonium wavefunctions are important, we work in the strongly coupled
regime, where we assume $mv \gtrsim \Lambda_{\rm QCD} \gg mv^2$.
The only degree of freedom of strongly coupled pNRQCD is the singlet field 
$S(\bm{x}_1, \bm{x}_2)$, which describe the heavy quark at position $\bm{x}_1$ 
and heavy antiquark at position $\bm{x}_2$ in a color-singlet state. 
The pNRQCD Lagrangian is given by 
\begin{equation}
{\cal L}_{\rm pNRQCD} = S^\dag [ i \partial_0 - h_S(\bm{x}_1,\bm{x}_2,
\bm{\nabla}_{\bm{x}_1}, \bm{\nabla}_{\bm{x}_2})] S, 
\end{equation}
where $h_S$ is the pNRQCD Hamiltonian, which is obtained by matching 
NRQCD and pNRQCD. In the case of strongly coupled pNRQCD, this matching is
done nonperturbatively~\cite{Brambilla:2000gk}. 
The Hamiltonian $h_S$ has the general form 
\begin{equation}
\label{eq:hamiltonian_singlet}
h_S = - \frac{\bm{\nabla}_{\bm{x}_1}^2}{2 m} 
 - \frac{\bm{\nabla}_{\bm{x}_2}^2}{2 m} 
+ V(\bm{r}, \bm{\nabla}), 
\end{equation}
where $\bm{r} = \bm{x}_1 - \bm{x}_2$ is the relative coordinate between the
quark and antiquark, and $\bm{\nabla} = \bm{\nabla}_{\bm{r}}$ is the derivative
with respect to $\bm{r}$. 
Here, $V(\bm{r}, \bm{\nabla})$ is the potential, which is the matching  
coefficient of pNRQCD. The potential is obtained as a formal expansion in
powers of $1/m$.\footnote{Since the matching calculations correspond to 
integrating out high energy degrees of freedom, the matching coefficients are 
independent on the low-energy dynamics of the effective field 
theory~\cite{Pineda:1998kj, Pineda:1998kn, Brambilla:1999qa, Brambilla:1999xf}. 
Hence, the potential in pNRQCD, when organized as an expansion in powers of
$1/m$, is obtained through matching independently on the specific power 
counting in pNRQCD~\cite{Brambilla:2000gk}.}
A heavy quarkonium state can be identified as an eigenstate of $h_S$. 
Due to translation symmetry in the potential, 
the wavefunction $\Psi_n(\bm{r})$ associated 
with the quarkonium state $n$ with binding energy $E_n$ can be defined as a 
function of the relative coordinate $\bm{r}$ through separation of variables. 
This wavefunction is an eigensolution of the Schr\"odinger equation
\begin{equation}
\label{eq:schroedinger_equation_general}
\left[ - \frac{\bm{\nabla}^2}{m} + V(\bm{r}, \bm{\nabla}) \right]
\Psi_n(\bm{r}) = E_n \Psi_n(\bm{r}), 
\end{equation}
where the potential $V(\bm{r}, \bm{\nabla})$ is the one that appears in $h_S$, 
and we take the wavefunction to be unit normalized 
($\int d^3r \, | \Psi_n(\bm{r}) |^2 =1$).

The NRQCD LDMEs can be computed in pNRQCD by matching the four-quark operators
in the NRQCD Lagrangian to the pNRQCD Hamiltonian $h_S$~\cite{Brambilla:2002nu}
(alternatively, the same result can be obtained by directly matching the 
NRQCD LDMEs to pNRQCD~\cite{Brambilla:2002nu, Brambilla:2020xod}). 
The pNRQCD expression for a decay LDME from a four-quark operator $\cal O$ 
on a heavy quarkonium state $H$ is given by
\begin{eqnarray}
\langle H | {\cal O} | H \rangle &= &
\int d^3r \; \int d^3r' \int d^3R 
\, \psi_H^* (\bm{r})
\nonumber\\ && \times 
[- V_{\cal O} (\bm{x}_1,\bm{x}_2;
\bm{\nabla}_{\bm{x}_1}, \bm{\nabla}_{\bm{x}_2}) 
\delta^{(3)}(\bm{x}_1 - \bm{x}_1') 
\delta^{(3)}(\bm{x}_1 - \bm{x}_1') ] 
\psi_H (\bm{r}), 
\end{eqnarray}
where $\psi_H(\bm{r})$ is the wavefunction associated with the state $H$, 
and $V_{\cal O} (\bm{x}_1,\bm{x}_2; \bm{\nabla}_{\bm{x}_1},
\bm{\nabla}_{\bm{x}_2})$ is the matching coefficient. 
This matching coefficient is a contact term, which is
proportional to the delta function $\delta^{(3)}(\bm{r})$ due to the fact that
$\cal O$ is a local operator. As a result, the LDME $\langle H | {\cal O} | H
\rangle$ is given in terms of the wavefunction at the origin 
$\Psi_H(\bm{0})$ and its derivatives. 
The contact term is obtained as a formal expansion in powers of $1/m$, and this 
can be considered as an expansion in powers of $v$ and $\Lambda_{\rm QCD}/m$,
which are the scales appearing in NRQCD divided by $m$. 
For electromagnetic decays or exclusive electromagnetic production processes,
the four-quark operators have the form ${\cal O} 
= \psi^\dag \kappa \chi |0\rangle \langle 0| \chi^\dag \kappa' \psi$, where 
$\kappa$ and $\kappa'$ are polynomials of Pauli matrices and covariant
derivatives. Hence, in this case, the LDMEs $\langle H | {\cal O} | H \rangle$
have the form of squares of quarkonium-to-vacuum matrix elements. 

The leading-order LDME for a vector quarkonium $V$ is given in strongly
coupled pNRQCD to relative order $v^2$ and $\Lambda_{\rm QCD}^2/m^2$ accuracy 
by~\cite{Brambilla:2002nu, Brambilla:2020xod} 
\begin{equation}
\label{eq:pNRQCD_vec}
\left| \langle 0 | \chi^\dag \bm{\epsilon} \cdot \bm{\sigma} \psi 
| V \rangle  \right|^2
= 2 N_c | \Psi_V(0)|^2 \left[ 
1- \frac{E_V}{m} \frac{2 {\cal E}_3}{9}
- \frac{2 {\cal E}_1 {\cal E}_3}{9 m^2} 
+ \frac{2 {\cal E}_3^{\rm (2,em)}}{3 m^2} 
+ \frac{c_F^2 {\cal B}_1}{3 m^2} + O(v^3) 
\right], 
\end{equation}
while the order-$v^2$ LDME is given at leading order in $v$ and 
$\Lambda_{\rm QCD}/m$ by 
\begin{equation}
\label{eq:pNRQCD_vec_v2_squared}
\frac{1}{2} 
\langle V | \psi^\dag \bm{\epsilon}^* \cdot \bm{\sigma} \chi | 0 \rangle 
\langle 0 | \chi^\dag \bm{\epsilon} \cdot \bm{\sigma}
(-\tfrac{i}{2} \overleftrightarrow{\bm{D}} )^2 \psi | V \rangle
+ {\rm c.c.} 
= 2 N_c m^2 | \Psi_V(0)|^2 \left[ \frac{E_V}{m} + O(v^3) \right].
\end{equation}
Here, $\Psi_V(r)$ is the quarkonium wavefunction for the $V$ state, 
which is a normalized eigenfunction of the Schr\"odinger equation
in eq.~(\ref{eq:schroedinger_equation_general}), and $E_V$ is the corresponding
eigenenergy, which scales like $m v^2$.
The constant $c_F$ is the short-distance coefficient
associated with the spin-dependent operator in the NRQCD
Lagrangian~\cite{Manohar:1997qy}, 
and c.c. stands for complex conjugated contribution of the preceding terms. 
The ${\cal E}_1$, ${\cal E}_3^{\rm (2,em)}$, and 
${\cal B}_1$ are nonperturbative gluonic correlators of mass dimension two,
which scale like $\Lambda_{\rm QCD}^2$. The quantity ${\cal E}_3$ is a
dimensionless gluonic correlator, which in general can be order one. 
The field-theoretical definitions of the gluonic correlators can be found in
refs.~\cite{Brambilla:2001xy, Brambilla:2002nu}.
If we compute the order-$v^2$ LDME $\langle 0 | \chi^\dag \bm{\epsilon} 
\cdot \bm{\sigma} (-\tfrac{i}{2} \overleftrightarrow{\bm{D}} )^2 \psi 
| V \rangle$ at leading order in $v$, as we do in 
eq.~(\ref{eq:pNRQCD_vec_v2_squared}), we can neglect the imaginary part 
which occurs from subleading orders in $v$.\footnote{
The quarkonium-to-vacuum LDMEs can develop imaginary parts when there
is a contribution from a cut diagram. Such contributions can arise at lowest
orders in $v$ from insertions of the dimension-5 operators 
in the NRQCD Lagrangian, which can induce transitions between quarkonium
states. In the standard power counting of NRQCD, such contributions are
suppressed by at least $v^2$~\cite{Bodwin:1994jh}. Hence, when we compute 
the quarkonium-to-vacuum LDMEs at leading orders in $v$, imaginary parts
arising from cut diagrams can be neglected.} 
Hence, at the current level of accuracy, we can write 
\begin{equation}
\label{eq:pNRQCD_vec_v2}
\langle 0 | \chi^\dag \bm{\epsilon} \cdot \bm{\sigma}
(-\tfrac{i}{2} \overleftrightarrow{\bm{D}} )^2 \psi | V \rangle
= \sqrt{2 N_c} m^2 | \Psi_V(0)| \left[ \frac{E_V}{m} + O(v^3) \right],
\end{equation}
which is valid at leading order in $v$ and $\Lambda_{\rm QCD}/m$. Here, we
utilize the freedom to choose the overall phase of the $|V \rangle$ state to
make $\langle 0 | \chi^\dag \bm{\epsilon} \cdot \bm{\sigma} \psi | V \rangle$
real and positive, so that at leading order in $v$ and $\Lambda_{\rm QCD}/m$, 
$\langle 0 | \chi^\dag \bm{\epsilon} \cdot \bm{\sigma} \psi | V \rangle
=\sqrt{2 N_c}|\Psi_V(0)|$. 
We can now compare the expressions in eqs.~(\ref{eq:pNRQCD_vec}) 
and (\ref{eq:pNRQCD_vec_v2}) with the evolution equation in
eq.~(\ref{eq:RG_vec}). 
Equation~(\ref{eq:pNRQCD_vec}) implies that the factorization scale
dependence in the leading-order LDME 
must come from ${\cal E}_3$ and $|\Psi_V(0)|$, because the gluonic
correlators of mass dimension two are scaleless power divergent in perturbative
QCD, and $E_V$ is finite. The order-$\alpha_s v^2$ contribution to the
anomalous dimension in eq.~(\ref{eq:RG_vec}) is consistent with the 
known scale dependence of ${\cal E}_3$~\cite{Brambilla:2001xy, 
Brambilla:2020xod}. 
As a result, the two-loop anomalous dimension
in the first term on the right-hand side of eq.~(\ref{eq:RG_vec}) must come
from the scale dependence of the wavefunction at the origin $|\Psi_V(0)|$. 

The leading-order LDME for a pseudoscalar quarkonium $P$ is given in 
strongly coupled pNRQCD to relative order
$v^2$ and $\Lambda_{\rm QCD}^2/m^2$ accuracy
by~\cite{Brambilla:2002nu, Brambilla:2020xod}
\begin{equation}
\label{eq:pNRQCD_ps}
\left| \langle 0 | \chi^\dag \psi | P \rangle  \right|^2
= 2 N_c | \Psi_P(0)|^2 \left[ 
1- \frac{E_P}{m} \frac{2 {\cal E}_3}{9}
- \frac{2 {\cal E}_1 {\cal E}_3}{9 m^2} 
+ \frac{2 {\cal E}_3^{\rm (2,em)}}{3 m^2} 
+ \frac{c_F^2 {\cal B}_1}{m^2} + O(v^3) 
\right],
\end{equation}
and the order-$v^2$ LDME is given at leading order in $v$ and
$\Lambda_{\rm QCD}/m$ by
\begin{equation}
\label{eq:pNRQCD_ps_v2_squared}
\frac{1}{2}
\langle P | \psi^\dag \chi | 0 \rangle
\langle 0 | \chi^\dag 
(-\tfrac{i}{2} \overleftrightarrow{\bm{D}} )^2 \psi | P \rangle
+ {\rm c.c.}
= 2 N_c m^2 | \Psi_P(0)|^2 \left[ \frac{E_P}{m} + O(v^3) \right],
\end{equation}
where $\Psi_P(r)$ is the quarkonium wavefunction for the $P$ state, 
and $E_P$ is the corresponding binding energy. 
Similarly to eq.~(\ref{eq:pNRQCD_vec_v2}), we can write
the order-$v^2$ LDME at leading order in $v$ as
\begin{equation}
\label{eq:pNRQCD_ps_v2}
\langle 0 | \chi^\dag 
(-\tfrac{i}{2} \overleftrightarrow{\bm{D}} )^2 \psi | P \rangle 
= \sqrt{2 N_c} m^2 | \Psi_P(0)| \left[ \frac{E_P}{m} + O(v^3) \right], 
\end{equation}
which is valid at leading order in $v$ and $\Lambda_{\rm QCD}/m$, 
with a suitable choice of the overall phase of the $|P\rangle$ state. 
Similarly to the case of vector quarkonia, 
by comparing the expressions in eqs.~(\ref{eq:pNRQCD_ps}) and 
(\ref{eq:pNRQCD_ps_v2}) with the evolution equation in eq.~(\ref{eq:RG_ps}), 
we see that the known scale
dependence of ${\cal E}_3$ coincides with the second term on the right-hand
side of eq.~(\ref{eq:RG_ps}), and so, the two-loop anomalous dimension in the
first term on the right-hand side of eq.~(\ref{eq:RG_ps}) must come from the
scale dependence of the wavefunction at the origin $|\Psi_P(0)|$. 

In order to obtain the wavefunctions at the origin $|\Psi_V(0)|$ and 
$|\Psi_P (0)|$ in the $\overline{\rm MS}$ scheme 
with the correct dependence on the scale, 
it is necessary to compute the quarkonium wavefunctions from the
Schr\"odinger equation with the potential that has the correct short-distance
behavior that is expected from perturbative QCD. For many 
charmonium and bottomonium states, it is also necessary to include the 
nonperturbative long-distance behavior of the potential that is not captured in
perturbative QCD calculations. 
As we have discussed previously, in order to include the
long-distance behavior of the potential, it is most convenient to work in
position space, where the divergences are regulated by a position-space
regulator. 
In the following sections, we discuss our strategy to compute 
the wavefunctions at the origin with a position-space regulator, and compute
the conversion from 
the position-space regularization to the $\overline{\rm MS}$ scheme.

\section{\boldmath $S$-wave quarkonium wavefunctions in position space}
\label{sec:wavefunctions} 

In this section, we compute $S$-wave quarkonium wavefunctions at the origin 
in position space by solving the Schr\"odinger equation given 
in eq.~(\ref{eq:schroedinger_equation_general}).
To do so, we need to obtain the potential $V(\bm{r},\bm{\nabla})$ to a
sufficient accuracy in the expansion in powers of $1/m$. 
At leading order in $1/m$, the potential $V(\bm{r},\bm{\nabla})$ 
is given by the static potential $V^{(0)} (r)$, 
which has a nonperturbative definition in terms of a Wilson 
loop~\cite{Wilson:1974sk, Susskind:1976pi, Brown:1979ya, Brambilla:1999xf}. 
For $r \ll 1/\Lambda_{\rm QCD}$, the static potential is completely determined 
by perturbative QCD, which gives $V^{(0)} (r) = - \alpha_s C_F/r$ 
at leading order in $\alpha_s$~\cite{Pineda:2003jv, Bazavov:2014soa}. 
As we will see later, if we keep only the static potential and neglect the
terms of higher powers in $1/m$, the potential diverges like $1/r$ at $r = 0$,
and as a result, the $S$-wave wavefunctions are finite at the origin. 
Hence, in order to reproduce the dependence on the renormalization scale 
in the wavefunctions at the origin 
that we expect from perturbative QCD, it is necessary to include terms of
higher orders in $1/m$ to the potential. 
The potential including the correction terms of order $1/m$ and $1/m^2$ 
can be written generically as 
\begin{equation}
\label{eq:potentials}
V(\bm{r},\bm{\nabla}) 
= V^{(0)} (r) + \frac{V^{(1)} (r)}{m} + \frac{1}{m^2} \bigg[ V_r^{(2)} (r) + 
\frac{1}{2} \{ V_{p^2}^{(2)} (r), -\bm{\nabla}^2 \}
+ V_{S^2}^{(2)} (r) \bm{S}^2 \bigg] 
- \frac{\bm{\nabla}^4}{4 m^3} + \cdots, 
\end{equation}
where we include only the contributions relevant for $S$-wave states up to
order $1/m^2$, and the ellipsis represent terms of higher orders in $1/m$. 
Here, $\bm{S}$ is the $Q \bar Q$ spin, 
$\bm{S}^2 = 2$ for the spin-triplet state, and $\bm{S}^2=0$ for the 
spin-singlet state. 
The effect of the $V^{(1)}(r)/m$ term to the wavefunction at
the origin arises from insertions of the spin-independent dimension-5 
operators in the NRQCD Lagrangian, and in the standard NRQCD power counting, 
such effects are suppressed by $v^2$~\cite{Bodwin:1994jh, Bodwin:2007fz}. 
Hence, in order to compute the wavefunctions at the origin to relative order 
$v^2$ accuracy, it is necessary to include the $V^{(1)}(r)/m$ term in the 
potential.\footnote{
In the more conservative power counting in
refs.~\cite{Brambilla:2000gk, Pineda:2000sz}, the $V^{(1)}(r)/m$ term can be of
the same order as the static potential, and in such case, both the static 
potential and the $V^{(1)}(r)/m$ term should be included at leading order.
This power counting is based on the assumption that $V^{(1)}(r)$ is
of order $(mv)^2$, which follows from dimensional analysis. 
However, it is possible that this power counting overestimates the effect of
the $V^{(1)}(r)/m$ term on the wavefunctions at the origin, because the
wavefunctions are sensitive only to the shape of the potential.  
It can be seen from lattice measurements of the $V^{(1)}(r)/m$ term 
that inclusion of the $V^{(1)}(r)/m$ term does 
not significantly change the slope of the potential at long 
distances~\cite{Koma:2007jq, Koma:2012bc}. 
Hence, in this paper, we adopt the standard NRQCD power counting in
refs.~\cite{Bodwin:1994jh, Bodwin:2007fz} and 
assume that the effect of the $V^{(1)}(r)/m$ term in
the potential to the wavefunctions at the origin is suppressed by $v^2$. } 
When computing the correction from the $V^{(1)}(r)/m$ term to the wavefunction,
it is necessary to also include terms of order $1/m^2$ to the potential, 
because unitary transformations can reshuffle the $1/m$ terms with the $1/m^2$ 
terms in the potential, and so, the $V^{(1)}(r)/m$ term can be determined
unambiguously only when the $1/m^2$ terms are included. 
Even though the last term in eq.~(\ref{eq:potentials}), which originates from
the relativistic correction to the kinetic energy, is suppressed by $1/m^3$, 
this term must be regarded as a $1/m$ contribution, because a power of 
$-\bm{\nabla}^2/m$ can be traded with a power of $E_n-V^{(0)}(r)$ 
by using the Schr\"odinger equation. 
In the same way, higher order corrections to the 
kinetic energy of the form $\bm{\nabla}^{2 n}/m^{2 n-1}$ for $n\geq 3$ 
are suppressed by at least $1/m^2$. If we assume that $E_n$ and 
$V^{(0)}(r)$ are of order $m v^2$, these higher order corrections are 
suppressed by higher powers of $v$ compared to the order $1/m$ and $1/m^2$
terms included in eq.~(\ref{eq:potentials}). 
Hence, we neglect the higher order corrections to the kinetic energy 
of the form $\bm{\nabla}^{2 n}/m^{2 n-1}$ for $n\geq 3$.

The form of the potential beyond the static one 
depends on the scheme in which the matching
between NRQCD and pNRQCD is done. Nonperturbative definitions of the potentials 
that appear in eq.~(\ref{eq:potentials}) are found from Wilson loop
matching~\cite{Brambilla:2000gk, Pineda:2000sz}. 
On the other hand, it is necessary to employ the results from on-shell
matching~\cite{Kniehl:2001ju, Kniehl:2002br} 
in order to obtain dimensionally regulated wavefunctions at the origin that is 
consistent with the SDCs, because dimensionally regulated
calculations of the SDCs are also done by matching on-shell amplitudes in
QCD and NRQCD. 
We note that the potentials are gauge invariant in both cases. 
The different forms of the potentials can be related by unitary 
transformations~\cite{Brambilla:2000gk, Brambilla:2002nu, Peset:2015vvi}. 

The order-$1/m$ and $1/m^2$ potentials in eq.~(\ref{eq:potentials}) are to be
considered perturbations in the quantum-mechanical perturbation theory (QMPT), 
where the wavefunctions and binding energies are first computed at leading 
order from the Schr\"odinger equation without including the corrections to the
potential that are suppressed by powers of $1/m$. 
Then, the corrections of higher orders in $1/m$ 
are included by using the Rayleigh-Schr\"odinger perturbation theory. 

If we ignore the terms suppressed by powers of $1/m$ and keep only the 
static potential in eq.~(\ref{eq:potentials}), then 
$S$-wave wavefunctions are finite at the origin $r=0$;
this is because the behavior of the wavefunctions near $r=0$ is determined by 
the short-distance behavior of the static potential, which diverges like $1/r$
at $r=0$. 
Therefore, the corrections to the wavefunctions at the origin are finite if the
corrections come from potentials that diverge at most like $1/r$. 
On the other hand, the $1/m$ and $1/m^2$ terms in eq.~(\ref{eq:potentials}) 
can produce divergences in the wavefunctions at the
origin if they diverge faster than the static potential at $r=0$. 
We list the short-distance behavior of the potentials 
from on-shell matching and Wilson loop matching in 
appendix~\ref{appendix:potentials}.
The divergent behavior of the
wavefunctions at $r=0$ can be inferred from nonrelativistic quantum mechanics. 
Since the $1/m$ potential $V^{(1)} (r)$ diverges like $1/r^2$ at $r=0$, 
the first order correction to the wavefunctions from the $1/m$ potential 
produces a logarithmic divergence that is proportional to $\alpha_s^2 \log r$ 
at $r=0$. 
As we will see later, the velocity-dependent potential $V_{p^2}^{(2)} (r)$ and 
the relativistic correction to the kinetic energy $-\bm{\nabla}^4/(4 m^3)$
also produce logarithmic divergences that are similar to the correction from
the $1/m$ potential. The $1/m^2$ potential includes the delta function 
$\delta^{(3)} (\bm{r})$, and this produces at first order in the QMPT 
a power divergence proportional to $1/(mr)$ to the wavefunctions at $r=0$. 

We compute the wavefunctions and the corrections of higher orders in $1/m$ in
the following way. 
We define the leading-order (LO) potential $V_{\rm LO} (r)$ from the static
potential $V^{(0)} (r)$ by subtracting the perturbative corrections of order
$\alpha_s^2$ and beyond, but keeping the long-distance nonperturbative
behavior. The specific form of the LO potential that we use will be given in
sec.~\ref{sec:results}. 
This makes $V_{\rm LO} (r)$ behave like $-\alpha_s C_F/r$ at short
distances, while it coincides with the static potential at long distances. 
The perturbative corrections of higher orders in $\alpha_s$ will be included as
perturbations in the QMPT. 
The LO wavefunction $\Psi_n^{\rm LO} (r)$ and the binding energy 
$E_n^{\rm LO}$ satisfy the LO Schr\"odinger equation
\begin{equation}
\label{eq:schroedinger_LO}
h_{\rm LO} (r,\bm{\nabla}) \Psi_n^{\rm LO} (\bm{r}) 
= E_n^{\rm LO} \Psi_n^{\rm LO} (\bm{r}), 
\end{equation}
where 
\begin{equation}
h_{\rm LO} (r,\bm{\nabla}) = - \frac{\bm{\nabla}^2}{m} + V_{\rm LO}(r). 
\end{equation}
To first order in the QMPT, the wavefunction $\Psi_n (r)$ for an $S$-wave
state $n$ is given in terms 
of the LO wavefunction $\Psi_n^{\rm LO} (r)$ by 
\begin{equation}
\label{eq:corr}
\Psi_n(r') = 
\Psi_n^{\rm LO} (r') + \delta \Psi_n (r')
= 
\Psi_n^{\rm LO} (r') 
- \int d^3r \, \hat{G}_n (\bm{r}',\bm{r}) \delta V(\bm{r},\bm{\nabla}) 
\Psi_n^{\rm LO} (r), 
\end{equation}
where $\delta V(\bm{r},\bm{\nabla}) = V(\bm{r},\bm{\nabla}) - V_{\rm LO} (r)$. 
$\hat{G}_n (\bm{r}',\bm{r})$ is the reduced Green's function for the 
eigenstate $n$, which is defined by 
\begin{equation}
\label{eq:redgreen_definition}
\hat{G}_n (\bm{r}',\bm{r}) 
= \sum_{k \neq n} 
\frac{\Psi_k^{\rm LO} (\bm{r}') \Psi_k^{\rm LO*} (\bm{r})}
{E_k^{\rm LO} - E_n^{\rm LO}},
\end{equation}
where the sum runs over all eigenstates of the LO Schr\"odinger equation 
except for the state $n$. Although the sum includes states with nonzero orbital 
angular momentum, only $S$-wave states contribute to the integral in
eq.~(\ref{eq:corr}) due to the rotational symmetry of 
$\delta V(\bm{r},\bm{\nabla})$. 
The reduced Green's function is related to the Green's function 
$G(\bm{r}',\bm{r};E)$ by 
\begin{equation}
\hat{G}_n (\bm{r}',\bm{r}) 
= \lim_{E \to E_n^{\rm LO}} 
\left[ G(\bm{r}',\bm{r};E) - 
\frac{\Psi_n^{\rm LO} (\bm{r}') \Psi_n^{\rm LO*} (\bm{r})}{E_n^{\rm LO} - E}
\right],
\end{equation}
while $G(\bm{r}',\bm{r};E)$ is defined for arbitrary complex $E$ by 
\begin{equation}
\label{eq:greenfunction1}
G(\bm{r}',\bm{r};E) 
= \sum_k 
\frac{\Psi_k^{\rm LO} (\bm{r}') \Psi_k^{\rm LO*} (\bm{r})}{E_k^{\rm LO} - E}.
\end{equation}
The Green's function satisfies the equation 
\begin{equation}
\label{eq:greenfunction2}
\big[ h_{\rm LO} (r,\bm{\nabla})-E \big] G(\bm{r}',\bm{r};E) 
=
\big[ h_{\rm LO} (r',\bm{\nabla}')-E \big] G(\bm{r}',\bm{r};E) 
= 
\delta^{(3)} (\bm{r}-\bm{r}'), 
\end{equation}
which implies 
\begin{equation}
\label{eq:redgreen_equation}
\left( - \frac{\bm{\nabla}^2}{m} + V_{\rm LO} (r) - E_n^{\rm LO} \right)
\hat{G}_n (\bm{r}',\bm{r}) 
= \delta^{(3)} (\bm{r}-\bm{r}') 
- \Psi_n^{\rm LO} (\bm{r}') \Psi_n^{\rm LO*} (\bm{r}). 
\end{equation}
The Green's function in position space can be computed by using the formal
definition in eq.~(\ref{eq:greenfunction1}), or by solving the differential
equation in eq.~(\ref{eq:greenfunction2}). 
Note that the reduced Green's function can be computed from 
$G(\bm{r}',\bm{r};E)$ by using 
\begin{equation}
\label{eq:redgreen_relation}
\hat{G}_n (\bm{r}',\bm{r}) 
= \lim_{\eta \to 0} 
\frac{1}{2} 
\left[ 
G(\bm{r}',\bm{r};E_n^{\rm LO}+\eta) + 
G(\bm{r}',\bm{r};E_n^{\rm LO}-\eta)
\right].
\end{equation}
We note that the reduced Green's function satisfies
\begin{equation}
\label{eq:redgreen_orthogonal}
\int d^3r \,\hat{G}_n (\bm{r}',\bm{r}) \Psi_n^{\rm LO} (\bm{r}) = 0, 
\end{equation}
which follows from the orthogonality of wavefunctions. 
The vanishing of eq.~(\ref{eq:redgreen_orthogonal}) also follows from the fact 
that adding or subtracting a constant to the potential 
$V(\bm{r},\bm{\nabla})$ have no effect on the wavefunctions $\Psi_n(\bm{r})$.

The corrections to the wavefunction $\delta \Psi_n (r')$ 
can be computed from eq.~(\ref{eq:corr}). 
The corrections from the velocity-dependent potential and the relativistic
correction to the kinetic energy contain $\bm{\nabla}^2$, which can be reduced
by using the Schr\"odinger equation and eq.~(\ref{eq:redgreen_equation}). 
The correction from the velocity-dependent potential reads 
\begin{eqnarray}
\label{eq:velpotential_corr}
&& \hspace{-5ex} - \int d^3 r \, \hat G_n (\bm{r}',\bm{r})
\frac{1}{2 m^2} \{ V_{p^2}^{(2)} (r), - \bm{\nabla}^2 \} 
\Psi_n^{\rm LO} (r) 
\nonumber \\
&=& 
\frac{1}{m} \int d^3 r \, \hat G_n (\bm{r}',\bm{r})
\left[ V_{p^2}^{(2)} (r) V_{\rm LO}(r) 
- E_n^{\rm LO} V_{p^2}^{(2)} (r) 
\right]
\Psi_n^{\rm LO} (r) 
\nonumber \\ && 
+ \frac{1}{2 m} \Psi_n^{\rm LO} (\bm{r}')  \int d^3 r \, V_{p^2}^{(2)} (r)
\left| \Psi_n^{\rm LO} (r) \right|^2 
- \frac{1}{2 m} V_{p^2}^{(2)} (r') \Psi_n^{\rm LO} (r').
\end{eqnarray}
It is clear that the first term in the last line is finite at $r'=0$, since 
$\Psi_n^{\rm LO} (r)$ is regular at $r=0$, and  
$V_{p^2}^{(2)} (r)$ diverges like $1/r$ at $r=0$. 
At $r'=0$, the last term in the last line of eq.~(\ref{eq:velpotential_corr}) 
requires knowledge of $V_{p^2}^{(2)} (0)$. 
While in dimensionally regulated perturbative QCD, the quantity 
$V_{p^2}^{(2)} (0)$, when computed as the Fourier transform of the
momentum-space expression, is scaleless power divergent, 
$V_{p^2}^{(2)} (0)$ may still not vanish nonperturbatively. 
In order to investigate the quantity $V_{p^2}^{(2)} (0)$ nonperturbatively, 
we use the 
nonperturbative expression for $V_{p^2}^{(2)} (r)$ in Wilson loop matching
given in ref.~\cite{Pineda:2000sz} in terms of a rectangular Wilson loop 
$W_{r \times T}$ with spatial size $r$ and time extension $T$, with insertions
of the chromoelectric field $\bm{E}^i = G^{i0}$, where $G^{\mu \nu}$ is the
gluon field-strength tensor. We show the explicit nonperturbative expression
for $V_{p^2}^{(2)} (r)$ in Wilson loop matching 
in appendix~\ref{appendix:potentials}.
By setting $r=0$ in the nonperturbative expression
for $V_{p^2}^{(2)} (r)$ in ref.~\cite{Pineda:2000sz}, we find 
\begin{equation}
\label{eq:velpot_wilsonloop_zero}
V_{p^2}^{(2)} (0) \big|^{\rm WL} =
2 i \hat{\bm{r}}^i \hat{\bm{r}}^j 
\frac{T_F}{N_c} 
\int_0^\infty dt\, t^2
\langle 0 | g_s \bm{E}^{i,a} (t,\bm{0}) 
\Phi_{ab}(t,0)
g_s \bm{E}^{j,b} (0,\bm{0}) | 0 \rangle, 
\end{equation}
where $\hat{\bm{r}} = \bm{r}/|\bm{r}|$, 
$T_F = 1/2$ and $\Phi_{ab}(t,0)$ is an adjoint Wilson line connecting the 
points $(0,\bm{0})$ and $(t,\bm{0})$. The right-hand side of
eq.~(\ref{eq:velpot_wilsonloop_zero}) is proportional to the 
gluonic correlator $i {\cal E}_2$ defined in refs.~\cite{Brambilla:2002nu,
Brambilla:2020xod}, which scales like $\Lambda_{\rm QCD}$. Hence, in Wilson
loop matching, $V_{p^2}^{(2)}(0)$ is a nonperturbative quantity that scales
like $\Lambda_{\rm QCD}$, and may be nonvanishing. 

Similarly, the correction from the $-\bm{\nabla}^4/(4 m^3)$ term is given by 
\begin{eqnarray}
\label{eq:kinetic_corr}
&& \hspace{-5ex} - \int d^3 r \, \hat G_n (\bm{r}',\bm{r})
\left( - \frac{\bm{\nabla}^4}{4 m^3} \right) 
\Psi_n^{\rm LO} (r) 
\nonumber \\
&=& 
\frac{1}{4 m} \int d^3 r \, \hat G_n (\bm{r}',\bm{r})
\left[ \left(V_{\rm LO}(r) \right)^2 
- 2 E_n^{\rm LO} V_{\rm LO}(r) \right]
\Psi_n^{\rm LO} (r) 
\nonumber \\ && 
+ \frac{1}{4 m} \Psi_n^{\rm LO}(r') \int d^3r \, V_{\rm LO} (r) 
\left|\Psi_n^{\rm LO} (r) \right|^2 
- \frac{1}{4 m} V_{\rm LO} (r') \Psi_n^{\rm LO} (r'). 
\end{eqnarray}
Again, it is clear that the first term in the last line is finite at $r'=0$. 
The LO potential at zero distance $V_{\rm LO} (0)$ vanishes in
dimensionally regulated perturbative QCD, because it is a scaleless power 
divergence. This quantity also vanishes nonperturbatively, which follows from 
the exact vanishing of the static potential at zero 
distance~\cite{Brambilla:2002nu}. This can be seen from the expression for the
static potential in terms of a Wilson loop, which reads~\cite{Wilson:1974sk, 
Susskind:1976pi, Brown:1979ya, Brambilla:1999xf} 
\begin{equation}
\label{eq:staticpot_wilsonloop}
V^{(0)} (r) = \lim_{T \to \infty} \frac{i}{T} \log \langle W_{r \times T}
\rangle,
\end{equation}
where $\langle \cdots \rangle$ stand for the average of the Yang-Mills action. 
If we set $r=0$, the right-hand side vanishes because 
$\langle W_{r \times T} \rangle|_{r =0} =1$, and therefore, $V^{(0)} (0) = 0$. 
The same conclusion can be obtained from the fact that the static potential is
purely perturbative at short distances~\cite{Pineda:2003jv, Bazavov:2014soa},
and so, the Fourier transform of the momentum-space expression vanishes in
dimensional regularization
because it is scaleless power divergent. Since the LO potential differs from
the static potential only by the loop corrections in perturbative QCD, the LO
potential also vanishes at $r=0$, because the loop corrections are also
scaleless power divergent at $r=0$. 

By using the expressions in eqs.~(\ref{eq:velpotential_corr}) and
(\ref{eq:kinetic_corr}), we can write the first order correction to the 
wavefunction as 
\begin{eqnarray}
\label{eq:corr_simplified} 
\delta \Psi_n (r') 
&=& 
- \int d^3r \, \hat G_n (\bm{r}',\bm{r}) 
\delta {\cal V} (r) \Psi_n^{\rm LO} (r) 
\nonumber \\ &&  
- \int d^3r \, \hat G_n (\bm{r}',\bm{r})
\left\{ \delta V_C(r) 
+ \frac{E_n^{\rm LO}}{m} 
\left[ V_{p^2}^{(2)} (r) + \frac{1}{2} V_{\rm LO}(r)
\right] 
\right\}
\Psi_n^{\rm LO} (r) 
\nonumber \\ && 
+ \frac{1}{2 m} \Psi_n^{\rm LO} (r')  \int d^3 r \, 
\left[ V_{p^2}^{(2)} (r) + \frac{1}{2} V_{\rm LO} (r) \right] 
\left|\Psi_n^{\rm LO} (r) \right|^2 
\nonumber \\ && 
- \frac{1}{2 m} V_{p^2}^{(2)} (r') \Psi_n^{\rm LO} (r')
- \frac{1}{4 m} V_{\rm LO} (r') \Psi_n^{\rm LO} (r'), 
\end{eqnarray}
where $\delta V_C(r) = V^{(0)} (r) - V_{\rm LO} (r)$, and 
\begin{equation}
\label{eq:deltav_position}
\delta {\cal V}(r) = 
\frac{V^{(1)}(r)}{m} - \frac{ V_{p^2}^{(2)} (r) V_{\rm LO}(r) }{m} 
- \frac{ \left(V_{\rm LO}(r) \right)^2 }{4m} 
+ \frac{V^{(2)}_{r} (r)}{m^2} + \frac{V^{(2)}_{S^2} (r) \bm{S}^2}{m^2}. 
\end{equation}
We note that for an $S$-wave state $n$, each term in 
eq.~(\ref{eq:corr_simplified}) is a function of $r'=|\bm{r}'|$, and does not
depend on the angles of $\bm{r}'$. 
Because of the explicit rotational symmetry of $\delta V(\bm{r},\bm{\nabla})$
and $\delta {\cal V} (r)$, eq.~(\ref{eq:corr_simplified}) is unchanged if we
only include $S$-wave states in the definition of the reduced Green's
function in eq.~(\ref{eq:redgreen_definition}). 

In eq.~(\ref{eq:corr_simplified}), 
the divergences at $r'=0$ are contained in the first integral. 
The second integral in eq.~(\ref{eq:corr_simplified}) is finite at 
$r'=0$, because 
the terms in the curly brackets diverge at most like $1/r$ at $r=0$. 
The UV divergence in the first integral can be cut off by setting $r'=r_0$ 
with $r_0 > 0$, instead of setting $r'=0$. 
This defines the finite-$r$ regularization~\cite{Kiyo:2010jm}, 
which is the position-space regularization that we use in this paper. 
We note that a similar version of position-space regularization has been 
used in refs.~\cite{Melnikov:1998pr, Yakovlev:1998ke, Nagano:1999nw, 
Penin:1998kx}. 
We define the correction to the $S$-wave wavefunctions at the origin 
in the finite-$r$ regularization by 
\begin{eqnarray}
\label{eq:corr_finiter}
\delta \Psi_n(0) \big|_{r_0}
&=& 
- \int d^3r \, \hat G_n (\bm{r}',\bm{r}) \delta {\cal V} (r) 
\Psi_n^{\rm LO} (r) \Big|_{|\bm{r}'| = r_0}
\nonumber \\ &&  
- \int d^3r \, \hat G_n (\bm{0},\bm{r}) 
\left\{ \delta V_C(r)
+ \frac{E_n^{\rm LO}}{m} 
\left[ V_{p^2}^{(2)} (r) + \frac{1}{2} V_{\rm LO}(r)
\right] 
\right\}
\Psi_n^{\rm LO} (r) 
\nonumber \\ && 
+ \frac{
\Psi_n^{\rm LO} (0) 
}{2 m} 
\int d^3 r \, 
\left[ V_{p^2}^{(2)} (r) + \frac{1}{2} V_{\rm LO} (r) \right] 
\left|\Psi_n^{\rm LO} (r) \right|^2 
- \frac{V_{p^2}^{(2)} (0)}{2 m} \Psi_n^{\rm LO} (0), 
\quad \quad
\end{eqnarray}
where the subscript $r_0$ implies that the divergences are regulated by a
finite distance $r_0$ between the $Q$ and $\bar Q$. 
Here, $\Psi_n(0)|_{r_0} = \Psi_n^{\rm LO} (0) + \delta \Psi_n(0) |_{r_0}$, 
and we used $V_{\rm LO} (0) = 0$ following the exact vanishing of the static
potential at $r=0$~\cite{Brambilla:2002nu}. 

In order to obtain the wavefunctions at the origin in the $\overline{\rm MS}$
scheme, we need to compute the scheme conversion from finite-$r$
regularization to DR. This is given by the 
difference between the two different schemes in the divergent integral 
$\int d^3r \, \hat G_n (\bm{0},\bm{r}) \delta {\cal V} (r) 
\Psi_n^{\rm LO} (r)$. We define the scheme conversion coefficient
$\delta Z$ through the relation 
\begin{equation}
\label{eq:conversion_relation}
\Psi_n (0) |_{\overline{\rm MS}}
= \Psi_n(0) |_{r_0} - \delta Z \times \Psi_n^{\rm LO} (0), 
\end{equation}
so that 
\begin{eqnarray}
\label{eq:conversion_definition0}
\delta Z
&=& 
\frac{1}{\Psi_n^{\rm LO} (0)} 
\bigg[ 
\int d^3r \, \hat G_n (\bm{0},\bm{r}) \delta {\cal V} (r) 
\Psi_n^{\rm LO} (r)
\Big|_{\overline{\rm MS}}
\nonumber \\ && \hspace{15ex}
- 
\int d^3r \, \hat G_n (\bm{r}',\bm{r}) \delta {\cal V} (r) 
\Psi_n^{\rm LO} (r) \Big|_{|\bm{r}'| = r_0} \bigg], 
\end{eqnarray}
where the divergent integral in the $\overline{\rm MS}$ scheme is first 
computed in momentum space in $d=4-2 \epsilon$ spacetime dimensions,
and then the UV poles are subtracted 
according to the prescription for renormalization in the $\overline{\rm MS}$
scheme. The prescription that we use for the $\overline{\rm MS}$ scheme 
is defined by associating a factor of 
$( \Lambda^2 \frac{e^{\gamma_{\rm E}}}{4 \pi} )^\epsilon$ with each loop
integration, and subtracting the $1/\epsilon$ poles after evaluating the loop
integrals in DR and expanding in powers of $\epsilon$. 
Here, $\gamma_{\rm E}$ is the Euler-Mascheroni constant. 
Then, 
$\Lambda$ is the renormalization scale in the $\overline{\rm MS}$ scheme. 
We note that in the calculation of the scheme conversion,
we employ the potentials determined from on-shell
matching in order to ensure the consistency with the SDCs computed in DR. 
Since $\Psi_n^{\rm LO} (r)$ is regular at $r=0$,
we can replace $\Psi_n^{\rm LO} (r)$ in the integrand by 
$\Psi_n^{\rm LO} (0)$ without affecting the right-hand side of 
eq.~(\ref{eq:conversion_definition0}), because this affects only the finite
parts of the divergent integrals, which cancel in $\delta Z$. 
Therefore, we can write $\delta Z$ as 
\begin{equation}
\label{eq:conversion_definition}
\delta Z
= \int d^3r \, \hat G_n (\bm{0},\bm{r}) \delta {\cal V} (r) 
\Big|_{\overline{\rm MS}}
- \int d^3r \, \hat G_n (\bm{r}',\bm{r}) \delta {\cal V} (r) 
\Big|_{|\bm{r}'| = r_0}. 
\end{equation}
Since in $\delta Z$ we are only interested in the divergences and the finite
contributions in the limit $r_0 \to 0$, we neglect any contributions to the 
right-hand side of eq.~(\ref{eq:conversion_definition}) that vanish as $r_0 \to
0$, such as positive powers of $r_0$.  We compute $\delta Z$ in the next
section. 

As we have argued based on the divergent behavior of the wavefunctions at
the origin from nonrelativistic quantum mechanics, 
the corrections to the wavefunctions at the origin 
that are divergent at $r_0 = 0$ in the finite-$r$ regularization 
can contain contributions that are not suppressed by powers of
$1/m$, even though the corrections come from $1/m$ and $1/m^2$ potentials.
For example, the logarithmically divergent correction from the $1/m$ potential
is proportional to $\alpha_s^2 \log r_0$, which is not suppressed by any power
of $1/m$. 
This behavior will be confirmed in the calculation of $\delta Z$ in the next
section.
The appearance of such large corrections is in accordance
with the fact that, unless $r \gg m^{-1}$, 
the $1/m$ and $1/m^2$ potentials can overpower the static potential at short
distances, so that the expansion in powers of $1/m$ is no longer valid. 
These corrections can potentially jeopardize the nonrelativistic 
power counting, unless they are subtracted through renormalization. 
Since $\delta Z$ reproduces the divergent small $r_0$ dependence of the 
wavefunctions at the origin in the finite-$r$ regularization, the 
divergences in $\Psi_n(0)|_{r_0}$ at small $r_0$ are subtracted completely 
by the scheme conversion $\delta Z \times \Psi_n^{\rm LO} (0)$, 
and hence, the nonrelativistic power counting is restored 
in the $\overline{\rm MS}$-renormalized wavefunctions at the origin. 

In the calculation of the wavefunctions at the origin, we have assumed that the
effect of the $1/m$ potential to the wavefunctions are suppressed by $v^2$,
based on the standard NRQCD power counting in refs.~\cite{Bodwin:1994jh, 
Bodwin:2007fz}. We have found that the first order correction to the 
wavefunctions at the origin involves a correction that scales like
$\Lambda_{\rm QCD}/m$, which comes from the velocity-dependent potential at
order $1/m^2$. Since corrections of similar form may arise at second order in 
the Rayleigh-Schr\"odinger perturbation theory, which can scale like
$\Lambda_{\rm QCD}^2/m^2$, we assume that the wavefunctions at the origin that
we compute in this section are accurate up to corrections of relative 
order $v^3$ and $\Lambda_{\rm QCD}^2/m^2$.

\section{\boldmath $S$-wave quarkonium 
wavefunctions at the origin in the ${\overline{\rm MS}}$ scheme }
\label{sec:conversion}

In this section, we compute the scheme conversion coefficient $\delta Z$ 
defined in eq.~(\ref{eq:conversion_definition}), which converts 
finite-$r$ regularization to the ${\overline{\rm MS}}$ scheme. 
We note that, 
since the reduced Green's function can be written as a linear combination of 
the Green's function $G(\bm{r}',\bm{r};E)$ for different $E$ by using 
eq.~(\ref{eq:redgreen_relation}), it is sufficient to compute 
\begin{equation}
\label{eq:conversion_alternate}
\delta Z_E = 
\int d^3r \, G (\bm{0},\bm{r};E) \delta {\cal V} (r) 
\Big|_{\overline{\rm MS}}
- \int d^3r \, G (\bm{r}',\bm{r};E) \delta {\cal V} (r)
\Big|_{|\bm{r}'| = r_0}. 
\end{equation}
Analogously to the definition of $\delta Z$ in 
eq.~(\ref{eq:conversion_definition}), we neglect in
eq.~(\ref{eq:conversion_alternate}) any contributions that vanish as $r_0 \to
0$, such as positive powers of $r_0$, because we are only interested in the
divergences and finite contributions that appear in the limit $r_0 \to 0$. 
We will later show that $\delta Z_E$ is independent of $E$, and therefore,
coincides with $\delta Z$ for all $S$-wave states $n$. 
We first compute eq.~(\ref{eq:conversion_alternate}) in perturbative QCD, 
and then show that the nonperturbative long-distance behaviors of the
potentials do not affect the result. 

\subsection{Green's function in dimensional regularization} 

In order to compute the divergent integral in
eq.~(\ref{eq:conversion_alternate}) in the $\overline{\rm MS}$ scheme, we work
in momentum space in $d = 4-2 \epsilon$ spacetime dimensions. 
For this purpose, we need an expression for the $d$-dimensional Green's 
function in momentum space $\tilde G(\bm{p}',\bm{p};E)$, which is related to
the position-space counterpart by 
\begin{equation}
G(\bm{r}',\bm{r};E) = 
\int_{\bm{p}} 
\int_{\bm{p}'} 
e^{i \bm{p}' \cdot \bm{r}'} 
e^{-i \bm{p} \cdot \bm{r}} 
\tilde G(\bm{p}',\bm{p};E),
\end{equation}
where we use the shorthand
\begin{equation}
\int_{\bm{p}} \equiv 
\int \frac{d^{d-1} p}{(2 \pi)^{d-1}}.
\end{equation}
Then, the divergent integral 
$\int d^3r \, G (\bm{0},\bm{r};E) \delta {\cal V} (r)$ in DR can be written as 
\begin{equation}
\label{eq:div_DR}
\int d^3r \, G (\bm{0},\bm{r};E) \delta {\cal V} (r) \Big|_{\rm DR} 
= \int_{\bm{p}} \int_{\bm{p}'}
\tilde G(\bm{p}',\bm{p};E) 
\delta \tilde {\cal V}(\bm{p}), 
\end{equation}
where $\delta \tilde {\cal V} (\bm{p})$ is the momentum-space counterpart of 
$\delta {\cal V}(r)$ in $d$ dimensions. 
Explicit expressions for $\delta \tilde {\cal V} (\bm{p})$ in DR 
will be given in the next section. 
The finite-$r$ regularized integral can also be expressed in terms of the
momentum-space Green's function as 
\begin{equation}
\label{eq:div_finiter}
\int d^3r \, G (\bm{r}',\bm{r};E) \delta {\cal V} (r) \Big|_{|\bm{r}'| = r_0}
= \int_{\bm{p}} \int_{\bm{p}'}
e^{i \bm{p}' \cdot \hat{\bm{n}} r_0} 
\tilde G(\bm{p}',\bm{p};E) 
\delta \tilde {\cal V}(\bm{p}), 
\end{equation}
where $\hat{\bm{n}}$ is an arbitrary unit vector. 
Invariance of $\delta \tilde {\cal V}(\bm{p})$ under rotations ensures that the
right-hand side of eq.~(\ref{eq:div_finiter}) is independent of $\hat{\bm{n}}$.
The integrals in the finite-$r$ regularization can be computed at $d=4$ 
because the UV divergence is regulated by $r_0 > 0$. 

We note that the UV divergences in eqs.~(\ref{eq:div_DR}) 
and (\ref{eq:div_finiter}) come from the 
behavior of $\tilde G(\bm{p}',\bm{p};E)$ and $\delta \tilde {\cal V}(\bm{p})$ 
at large $\bm{p}$ and $\bm{p}'$. 
We compute the Green's function in momentum space 
in order to determine its large-momentum behavior. 
The momentum-space Green's function satisfies the Lippmann-Schwinger equation 
\begin{equation}
\label{eq:lippmann-schwinger}
\left( \frac{\bm{p'}^2}{m} - E \right) \tilde G(\bm{p}',\bm{p};E) 
+ \int_{\bm{k}} \tilde V_{\rm LO} (\bm{k}) 
\tilde G(\bm{p}'-\bm{k},\bm{p};E) = (2 \pi)^{d-1} 
\delta^{(d-1)}(\bm{p}-\bm{p}'), 
\end{equation}
where $\tilde V_{\rm LO} (\bm{k})$ is the LO potential in momentum space. 
A formal solution of eq.~(\ref{eq:lippmann-schwinger}) can be found
iteratively, which reads 
\begin{eqnarray}
\label{eq:green_iter}%
\tilde G(\bm{p}',\bm{p};E) 
&=& 
- \frac{(2 \pi)^{d-1} \delta^{(d-1)} (\bm{p}-\bm{p}')}
{E - \bm{p}^2/m} 
- \frac{1}{E - \bm{p}'^2/m} 
\tilde V_{\rm LO} (\bm{p}'-\bm{p})
\frac{1}{E - \bm{p}^2/m} 
\nonumber \\
&&  
- \frac{1}{E - \bm{p}'^2/m} 
T (\bm{p}',\bm{p},E) 
\frac{1}{E - \bm{p}^2/m}, 
\end{eqnarray}
where the first term comes from the free propagation of the $Q \bar Q$, 
and the second term corresponds to a single exchange of the LO potential between
the $Q$ and the $\bar Q$. The quantity $T$ encodes two or more exchanges of the
LO potential to all orders:
\begin{equation}
T (\bm{p}',\bm{p},E) = 
\sum_{n=1}^\infty 
\int_{\bm{k}_1} 
\int_{\bm{k}_2} 
\cdots
\int_{\bm{k}_n} 
\tilde V_{\rm LO} (\bm{k}_1)
\prod_{i=1}^n \frac{
\tilde V_{\rm LO} ( \bm{k}_{i+1}-\bm{k}_i) 
}{\left[E - \frac{(\bm{p}'+\bm{k}_i)^2}{2 m} \right]}, 
\end{equation}
where $\bm{k}_{n+1} = \bm{p}-\bm{p}'$ for each $n$. 
The formal solution in eq.~(\ref{eq:green_iter}) is organized so that 
the large $\bm{p}$ and $\bm{p}'$ behavior in each term becomes less divergent 
as the number of exchanges of the LO potential increases~\cite{Beneke:2013jia}. 
This greatly simplifies the calculation of $\delta Z$, since the divergent
contributions in eqs.~(\ref{eq:div_DR}) and (\ref{eq:div_finiter})
come only from the first few 
terms in eq.~(\ref{eq:green_iter}), and the non-divergent contributions 
coming from higher numbers of exchanges of the LO potential cancel 
in eq.~(\ref{eq:conversion_alternate}). 
Hence, for the purpose of computing $\delta Z$, it suffices to consider only
the first few terms in eq.~(\ref{eq:green_iter}).

\subsection{Potentials in dimensional regularization}

A necessary ingredient in computing $\delta Z$ is the potentials in momentum
space in $d$ spacetime dimensions. In order to obtain the correct 
$d$-dimensional 
expressions, it is necessary to compute the potentials in the on-shell matching
scheme in momentum space, which is done in perturbative QCD. 
The momentum-space potential $\tilde V (\bm{p}', \bm{p})$
appears in the $d$-dimensional momentum-space 
Schr\"odinger equation in the form 
\begin{equation}
\left( \frac{\bm{p}'^2}{m} - E_n \right) \tilde \Psi_n (\bm{p}')
+ \int_{\bm{p}} \tilde V (\bm{p}',\bm{p}) \tilde \Psi_n (\bm{p}) 
= 0, 
\end{equation}
where $\tilde \Psi_n(\bm{p})$ is the momentum-space wavefunction.
In $d=4$ dimensions, the momentum-space potential 
is related to the position-space counterpart 
$V(\bm{r},\bm{\nabla})$ by 
\begin{equation}
\label{potential_FT}
\tilde V (\bm{p}', \bm{p}) 
= \int d^3r \, e^{i \bm{p}' \cdot \bm{r}}
V(\bm{r},\bm{\nabla}) e^{-i \bm{p} \cdot \bm{r}}.
\end{equation}
The $d$-dimensional expression 
for $\tilde V (\bm{p}', \bm{p})$ to two-loop accuracy has been 
obtained in refs.~\cite{Beneke:1999qg, Beneke:2013jia}, which we display here:
\begin{eqnarray}
\label{eq:potentials_mom}
\tilde V (\bm{p}', \bm{p}) &=& 
- \frac{4 \pi \alpha_s C_F}{\bm{q}^2} 
\bigg[ 1 + \delta \tilde V_C(\bm{q}^2) - 
\frac{\alpha_s}{4 \pi} 
\frac{\pi^2 |\bm{q}|}{m} 
\left( \frac{\Lambda^2}{\bm{q}^2} \right)^{\epsilon} 
\left( \frac{C_F}{2} (1-2 \epsilon) - C_A (1-\epsilon) \right)
c_\epsilon 
\nonumber \\ && \hspace{12ex} 
- \frac{ s_\epsilon +2 }{4} 
\frac{\bm{q}^2}{m^2} 
+ \frac{\bm{p}'^2+\bm{p}^2}{2 m^2} 
\bigg]
- (2 \pi)^{d-1} \delta^{(d-1)} (\bm{q}) \frac{\bm{p}^4}{4 m^3}, 
\end{eqnarray}
where $\bm{q} = \bm{p}'-\bm{p}$, and the scale $\Lambda$ 
comes from associating a factor of 
$\left( \Lambda^2 \frac{e^{\gamma_{\rm E}}}{4 \pi} \right)^\epsilon$ 
with each loop integration. 
The expression in eq.~(\ref{eq:potentials_mom}) corresponds to the
calculation of the potential in weakly coupled pNRQCD, where $\alpha_s \sim v$ 
(see ref.~\cite{Pineda:2011dg} for a review). 
The constant $c_\epsilon$ is defined by 
\begin{equation}
c_\epsilon = \frac{e^{\gamma_{\rm E} \epsilon} \Gamma(\tfrac{1}{2}-\epsilon)^2
\Gamma(\tfrac{1}{2}+\epsilon)}{\pi^{3/2} \Gamma(1-2 \epsilon)}
= 1+2 \epsilon \log 2 + O(\epsilon^2),
\end{equation}
where in the last equality, we expanded in powers of $\epsilon$ up to order
$\epsilon$. The constant $s_\epsilon$ depends on $\bm{S}$; for spin
triplet ($\bm{S}^2=2$), 
\begin{equation}
\label{eq:dr_spintriplet}
s_{\epsilon} \big|_{\textrm{spin triplet}}
= \frac{10-7 d+d^2}{1-d} 
= \frac{2}{3} + \frac{10}{9} \epsilon+O(\epsilon^2), 
\end{equation}
and for spin singlet ($\bm{S}^2=0$), 
\begin{equation}
\label{eq:dr_spinsinglet}
s_{\epsilon} \big|_{\textrm{spin singlet}}
= \frac{50-15d+d^2}{1-d}
= - 2 -6 \epsilon + O(\epsilon^2). 
\end{equation}
The constant $s_\epsilon$ for the spin singlet case can be obtained 
by using the results in ref.~\cite{Beneke:2013jia} 
to compute the $d$-dimensional spin projection according to the treatment of
Pauli matrices in DR in ref.~\cite{Braaten:1996rp}. 
Equation~(\ref{eq:dr_spinsinglet}) agrees through order $\epsilon$ with
refs.~\cite{Czarnecki:2001zc,Kiyo:2010jm}. 
To order-$\epsilon$ accuracy, $s_\epsilon +2$ can be written in terms of
$\bm{S}^2$ as 
\begin{equation}
s_{\epsilon} +2 
= \frac{4}{3} \bigg[ \bm{S}^2 - \epsilon \left(\frac{9}{2}-\frac{8}{3}
\bm{S}^2 \right) \bigg] + O(\epsilon^2). 
\end{equation}
Equation~(\ref{eq:potentials_mom}) implies that the LO potential in momentum
space is given by 
\begin{equation}
\tilde V_{\rm LO} (\bm{q}) = - \frac{4 \pi \alpha_s C_F}{\bm{q}^2},
\end{equation}
which is valid in $d=4-2 \epsilon$ dimensions. 
The term $\delta \tilde V_C(\bm{q}^2)$ 
corresponds to the loop corrections to the static potential, 
for which the explicit expressions can be found in 
refs.~\cite{Fischler:1977yf,Schroder:1998vy,Beneke:1999qg,Beneke:2013jia}.
Since the corrections from $\delta \tilde V_C(\bm{q}^2)$ 
to the wavefunctions at the
origin are finite, we do not need
to consider this term in the calculation of $\delta Z$. 
We note that eq.~(\ref{eq:potentials_mom}) reproduces the
position-space expressions in eq.~(\ref{eq:potentials_pert}). 

Now we obtain the $d$-dimensional expression for 
$\delta \tilde {\cal V} (\bm{p})$ from eq.~(\ref{eq:potentials_mom}) 
by repeating the calculations in 
eqs.~(\ref{eq:velpotential_corr}), (\ref{eq:kinetic_corr}),
and (\ref{eq:corr_simplified}) in momentum space in $d$ spacetime dimensions. 
After a straightforward calculation, we obtain 
\begin{eqnarray}
\label{eq:potentials_mom2}
\delta \tilde {\cal V} (\bm{q}) &=& 
\frac{\pi^2 \alpha_s^2 C_F}{m |\bm{q}|} 
\left( \frac{\Lambda^2}{\bm{q}^2} \right)^{\epsilon} 
\left( \frac{C_F}{2} (1-2 \epsilon) - C_A (1-\epsilon) \right)
c_\epsilon 
+ 
\frac{\pi \alpha_s C_F}{m^2} 
(s_\epsilon +2)
\nonumber \\ && 
+ 
\frac{1 }{m} \int_{\bm{k}} \frac{
4 \pi \alpha_s C_F
}{\bm{k}^2} 
\tilde V_{\rm LO} (\bm{k}-\bm{q}) 
- \frac{1}{4 m} \int_{\bm{k}} \tilde V_{\rm LO} (\bm{k}) 
\tilde V_{\rm LO} (\bm{k}-\bm{q}). 
\end{eqnarray}
The first term corresponds to the $1/m$ potential, while
the second term comes from the spin-dependent potential. The third and the
fourth terms correspond to the corrections from the velocity-dependent 
potential and the relativistic correction to the kinetic energy, respectively. 
We see that at $d=4$, the Fourier transform of 
$\delta \tilde {\cal V} (\bm{q})$ is 
exactly the position-space counterpart $\delta {\cal V} (r)$ in
eq.~(\ref{eq:deltav_position}) at short distances.

\subsection{Scheme conversion}

Now we compute $\delta Z$ using the $d$-dimensional momentum-space expressions 
of the Green's function in eq.~(\ref{eq:green_iter}) and 
$\delta \tilde {\cal V}(\bm{q})$ in eq.~(\ref{eq:potentials_mom2}). 
We first compute the contribution to $\delta Z_E$ from the $1/m$ potential, 
which comes from the first term in eq.~(\ref{eq:potentials_mom2}). 
The free propagation term in the formal solution for the
Green's function gives the following contribution in DR:
\begin{eqnarray}
\label{eq:1/mcorr_div1}
&& \hspace{-5ex}
- \frac{\alpha_s^2 C_F \pi^2}{m} 
\left( \frac{C_F}{2} (1-2 \epsilon) - C_A (1-\epsilon) \right)
c_\epsilon 
\int_{\bm{p}} 
\int_{\bm{p}'} 
\frac{(2 \pi)^{d-1} \delta^{(d-1)} (\bm{p}-\bm{p}')}
{E - \bm{p}^2/m} 
\frac{\Lambda^{2 \epsilon}}{|\bm{p}|^{1+2 \epsilon}}
\nonumber \\ 
&=& 
- 
\alpha_s^2 C_F \pi^2 
\left( \frac{C_F}{2} (1-2 \epsilon) - C_A (1-\epsilon) \right) c_\epsilon 
\int_{\bm{p}} \frac{1} {m E - \bm{p}^2} 
\frac{\Lambda^{2 \epsilon}}{|\bm{p}|^{1+2 \epsilon}}
\nonumber \\
&=& 
\frac{\alpha_s^2 C_F}{8} 
\left( \frac{C_F}{2} (1-2 \epsilon) - C_A (1-\epsilon) \right) c_\epsilon 
\left[ \frac{1}{\epsilon} + 2 + 2 \log \left( \frac{- \Lambda^2}{2 m E} \right) 
+ O(\epsilon)
\right], 
\end{eqnarray}
where we used $\int_{\bm{p}'} (2 \pi)^{d-1} \delta^{(d-1)} (\bm{p}-\bm{p}') =1$
and associated a factor of $\left( \Lambda^2 \frac{e^{\gamma_{\rm E}}}{4 \pi}
\right)^\epsilon$ with the integral over $\bm{p}$.
Since the integral in eq.~(\ref{eq:1/mcorr_div1}) is logarithmically divergent, 
the contributions from one or more exchanges of the LO potential in the 
Green's function are finite and do not contribute to $\delta Z$. 
The same quantity in finite-$r$ regularization can be computed using the 
momentum-space expression in eq.~(\ref{eq:div_finiter}), which gives 
\begin{eqnarray}
\label{eq:1/mcorr_div2}
&& \hspace{-5ex}
- \frac{\alpha_s^2 C_F \pi^2}{m} 
\left( \frac{C_F}{2} - C_A \right) 
\int_{\bm{p}} 
\int_{\bm{p}'} 
e^{i \bm{p}' \cdot \hat{\bm{n}} r_0} 
\frac{(2 \pi)^{3} \delta^{(3)} (\bm{p}-\bm{p}')}
{E - \bm{p}^2/m} 
\frac{1}{|\bm{p}|} 
\nonumber \\
&=& 
- \alpha_s^2 C_F \pi^2 
\left( \frac{C_F}{2} - C_A \right) 
\frac{2 \pi^{3/2}}{\Gamma(3/2) (2 \pi)^{3}}
\int_0^\infty dp 
\frac{p^{2} }
{m E - p^2}
\frac{1}{p} \frac{\sin(p r_0)}{p r_0} 
\nonumber \\
&=& 
- \frac{\alpha_s^2 C_F}{8} 
\left( \frac{C_F}{2} - C_A \right)
[ -4 +4 \gamma_{\rm E} + 2 \log (-m E r_0^2) + O(r_0) ].
\end{eqnarray}
The contribution to $\delta Z_E$ from the $1/m$ potential can be found by
subtracting eq.~(\ref{eq:1/mcorr_div2}) from eq.~(\ref{eq:1/mcorr_div1}), and
then subtracting the $1/\epsilon$ pole. Since the dependence on $E$ cancels
between the dimensionally regulated integral and the finite-$r$ regularized
integral, we obtain 
\begin{equation}
\label{eq:conv_1/m}
\delta Z|_{V^{(1)}}
= 
\frac{1}{2} \alpha_s^2 C_F 
\left[ \left(\frac{C_F}{2} - C_A \right)
\log ( \Lambda r_0 e^{\gamma_{\rm E}} ) 
+ \frac{3 C_A}{4} 
- \frac{C_F}{2} 
\right].
\end{equation}
Here, $\Lambda$ is now the $\overline{\rm MS}$ scale associated with the
renormalization of the wavefunction at the origin. 

We now compute the contribution to $\delta Z$ coming from the
velocity-dependent potential, which comes from the third term 
in eq.~(\ref{eq:potentials_mom2}). 
Since $\tilde V_{\rm LO} (\bm{k}) = - 4 \pi \alpha_s C_F/\bm{k}^2$ in 
perturbative QCD, 
the first term in the last line in eq.~(\ref{eq:potentials_mom2}) can be
evaluated as 
\begin{eqnarray}
\label{eq:velcor_mom_evaluated}
\frac{1}{m} 
\int_{\bm{k}} \frac{4 \pi \alpha_s C_F}{\bm{k}^2} 
\tilde V_{\rm LO} (\bm{k}-\bm{p}) 
&=& 
- \frac{1}{m} (4 \pi \alpha_s C_F)^2
\int_{\bm{k}} 
\frac{1}{\bm{k}^2  (\bm{k}-\bm{p})^2} 
\nonumber \\
&=& 
- \frac{1}{m} (4 \pi \alpha_s C_F)^2
\frac{c_\epsilon}{8} 
\frac{\Lambda^{2 \epsilon}}
{| \bm{p} |^{1+2 \epsilon}}, 
\end{eqnarray}
where again we associated a factor of 
$\left( \Lambda^2 \frac{e^{\gamma_{\rm E}}}{4 \pi} \right)^\epsilon$ 
with the integral over $\bm{k}$.
Apart from an $\epsilon$-dependent factor, eq.~(\ref{eq:velcor_mom_evaluated}) 
is the same as the $1/m$ potential in perturbative QCD. Hence,  
\begin{equation}
\label{eq:conv_velocity}
\delta Z|_{V_{p^2}^{(2)}} 
= 
- \alpha_s^2 C_F^2 
\left[ -\frac{1}{2} + \log ( \Lambda r_0 e^{\gamma_{\rm E}} ) \right]. 
\end{equation}
Again, $\Lambda$ is now the $\overline{\rm MS}$ scale associated with the
renormalization of the wavefunction at the origin.

The computation of the contribution from the relativistic correction to the 
kinetic energy is similar to the case of the velocity-dependent potential. 
The last term in eq.~(\ref{eq:potentials_mom2}) can be
computed in perturbative QCD as 
\begin{eqnarray}
\label{eq:relcor_mom_evaluated}
- \frac{1}{4 m} 
\int_{\bm{k}} \tilde V_{\rm LO} (\bm{k})
\tilde V_{\rm LO} (\bm{k}-\bm{q}) 
&=& 
- \frac{1}{4 m} (4 \pi \alpha_s C_F)^2
\int_{\bm{k}} 
\frac{1}{\bm{k}^2  (\bm{k}-\bm{q})^2} 
\nonumber \\
&=& 
- \frac{1}{4 m} (4 \pi \alpha_s C_F)^2
\frac{c_\epsilon}{8} 
\frac{\Lambda^{2 \epsilon}}
{| \bm{q} |^{1+2 \epsilon}}. 
\end{eqnarray}
This is just $1/4$ times the result of the velocity-dependent potential in 
eq.~(\ref{eq:velcor_mom_evaluated}). 
Hence, we obtain 
\begin{equation}
\label{eq:conv_relcor}
\delta Z|_{-\frac{\bm{\nabla}^4}{4 m^3}} 
= 
- \frac{\alpha_s^2 C_F^2 }{4} 
\left[ -\frac{1}{2} + \log ( \Lambda r_0 e^{\gamma_{\rm E}} ) \right]. 
\end{equation}

Finally, we consider the contribution from the spin-dependent potential in
eq.~(\ref{eq:potentials_mom2}).
The free propagation term in the Green's function gives, in DR, 
\begin{eqnarray}
\label{eq:delcor_DR0}
- 
\frac{ \pi \alpha_s C_F (s_\epsilon+2)}{m^2} 
\int_{\bm{p}} 
\int_{\bm{p}'} 
\frac{(2 \pi)^{d-1} \delta^{(d-1)} (\bm{p}-\bm{p}')}
{E - \bm{p}^2/m} 
&=& 
- 
\frac{ \pi \alpha_s C_F (s_\epsilon+2)}{m^2} 
\int_{\bm{p}} \frac{1} {E - \bm{p}^2/m}
\nonumber \\ 
&=& 
- 
\frac{ \alpha_s C_F \bm{S}^2
}{3} \sqrt{-\frac{E}{m}} + O(\epsilon). 
\end{eqnarray}
This integral is power UV divergent; while this is not apparent in the last line 
of eq.~(\ref{eq:delcor_DR0}) because power divergences are subtracted
automatically in DR, the divergence can still be identified in the integrand. 
This implies that the contribution from the second term in
eq.~(\ref{eq:green_iter}) is also divergent. This contribution reads, in DR, 
\begin{eqnarray}
\label{eq:delcor_DR1}
&& \hspace{-10ex} 
- 
\frac{ \pi \alpha_s C_F (s_\epsilon+2)}{m^2} 
\int_{\bm{p}} 
\int_{\bm{p}'} 
\frac{
\tilde V_{\rm LO} (\bm{p}'-\bm{p})
}{(E - \bm{p}'^2/m) 
(E - \bm{p}^2/m)} 
\nonumber \\
&=& 
\frac{4 \pi^2 \alpha_s^2 C_F^2 (s_\epsilon+2)}{m^2} 
\int_{\bm{p}} \frac{1}{E-\bm{p}^2/m} 
\int_{\bm{p}'} 
\frac{1}{(\bm{p}'-\bm{p})^2}
\frac{1}{E-\bm{p}'^2/m} 
\nonumber \\
&=& 
\frac{ \alpha_s^2 C_F^2 (s_\epsilon+2)}{4}
\bigg[ \frac{1}{4 \epsilon} +
\frac{1}{2}- \frac{1}{2} \log \left( - \frac{4 m E}{\Lambda^2} \right)
+ O(\epsilon)
\bigg], 
\end{eqnarray}
where we associated a factor of 
$\left( \Lambda^2 \frac{e^{\gamma_{\rm E}}}{4 \pi} \right)^\epsilon$ 
with each loop integration. 
This integral is logarithmically divergent, and hence, the contributions from
two or more exchanges of the LO potential are finite. 
Now we compute the divergent integrals in finite-$r$ regularization, where the
free propagating contribution gives 
\begin{eqnarray}
\label{eq:delcor_finiter0}
&& \hspace{-5ex}
- 
\frac{4 \pi \alpha_s C_F \bm{S}^2}{3 m^2} 
\int_{\bm{p}} 
\int_{\bm{p}'} 
\frac{(2 \pi)^{3} \delta^{(3)} (\bm{p}-\bm{p}')}
{E - \bm{p}^2/m} 
e^{i \bm{p}' \cdot \hat{\bm{n}} r_0} 
\nonumber \\
&=& 
- 
\frac{4 \pi \alpha_s C_F \bm{S}^2}{3 m^2} 
\int_{\bm{p}} \frac{1} {E - \bm{p}^2/m}
e^{i \bm{p} \cdot \hat{\bm{n}} r_0} 
\nonumber \\ 
&=& 
-  
\frac{4 \pi \alpha_s C_F \bm{S}^2}{3 m^2} 
\int_0^\infty \frac{dp\, p^2}{(2 \pi)^3} 
\frac{1}
{ E - p^2/m}
\frac{4 \pi \sin (p r_0)}{p r_0}
\nonumber \\
&=& 
\frac{\alpha_s C_F \bm{S}^2}{3 m r_0} 
- 
\frac{\alpha_s C_F \bm{S}^2}{3}
\sqrt{-\frac{E}{m}} + O(r_0).
\end{eqnarray}
The contribution from one exchange of the LO potential is 
\begin{eqnarray}
\label{eq:delcor_finiter1}
&& \hspace{-10ex} 
- 
\frac{4 \pi \alpha_s C_F \bm{S}^2}{3 m^2} 
\int_{\bm{p}} 
\int_{\bm{p}'} 
\frac{
\tilde V_{\rm LO} (\bm{p}'-\bm{p})
}{(E - \bm{p}'^2/m) (E - \bm{p}^2/m)} 
e^{i \bm{p}' \cdot \hat{\bm{n}} r_0} 
\nonumber \\
&=& 
\frac{16 \pi^2 \alpha_s^2 C_F^2 \bm{S}^2}{3 m^2} 
\int_{\bm{p}'} \frac{1}{E-\bm{p}'^2/m} 
\int_{\bm{p}} 
\frac{1}{(\bm{p}'-\bm{p})^2}
\frac{1}{E-\bm{p}^2/m} 
e^{i \bm{p}' \cdot \hat{\bm{n}} r_0} 
\nonumber \\
&=& 
\frac{\pi \alpha_s^2 C_F^2 \bm{S}^2}{3}
\int_0^1 dx \int_0^1 dy
\int_0^\infty \frac{dp\,p^2}{(2 \pi)^3} 
\frac{4 \pi \sin(pr_0)}{pr_0} 
\frac{[x y (1-x)] ^{-1/2} }{
[p^2- (1-y+y/x) m E]^{3/2}} 
\nonumber \\
&=& 
\frac{\alpha_s^2 C_F^2 \bm{S}^2}{3} 
\left[ 1 - \gamma_{\rm E} - \frac{1}{2}
\log \left(- 4 m E r_0^2 \right)
+ O(r_0)\right]. 
\end{eqnarray}
The spin-dependent contribution to $\delta Z$ can then be obtained from 
eqs.~(\ref{eq:delcor_DR0}), (\ref{eq:delcor_DR1}), (\ref{eq:delcor_finiter0}),
and (\ref{eq:delcor_finiter1}). 
We see again that the dependences on $E$ cancel between the dimensionally 
regulated integrals and the finite-$r$ regularized integrals. 
After subtracting the $1/\epsilon$ pole, we obtain 
\begin{equation}
\label{eq:conv_spin}
\delta Z |_{V_{S^2}^{(2)}} 
= - \frac{\alpha_s C_F}{3 m r_0} \bm{S}^2 
+ 
\frac{(\alpha_s C_F)^2}{3} \left\{ - \frac{9}{8} 
+ \bm{S}^2  \left[ \frac{1}{6} 
+ \log (\Lambda r_0 e^{\gamma_{\rm E} }) \right] 
\right\}.
\end{equation}
We note that calculation of the spin-dependent contribution 
has been done in ref.~\cite{Kiyo:2010jm}. Equation~(45) of
ref.~\cite{Kiyo:2010jm} can be obtained by subtracting 
eq.~(\ref{eq:delcor_finiter1}) from
eq.~(\ref{eq:delcor_DR1}), dividing by a factor of $\pi \alpha_s C_F
(s_\epsilon +2)/m^2$, subtracting the $1/\epsilon$ pole, and setting 
$\Lambda = e^{-\gamma_{\rm E}}/r_0$. 

The complete result for $\delta Z$, which can be obtained by combining
eqs.~(\ref{eq:conv_1/m}), (\ref{eq:conv_velocity}), (\ref{eq:conv_relcor}), 
and (\ref{eq:conv_spin}), reads 
\begin{equation}
\label{eq:conv_all}
\delta Z
= 
- \frac{\alpha_s C_F}{3 m r_0} \bm{S}^2 
+
\alpha_s^2 C_F \left\{
C_F \left[ - L_\Lambda
+ \frac{\bm{S}^2 }{3} \left( \frac{1}{6} + L_\Lambda
\right)
\right]
- \frac{C_A}{2}  \left( - \frac{3}{4} + L_\Lambda \right)
\right\}, 
\end{equation}
where we use the shorthand 
$L_\Lambda = \log (\Lambda r_0 e^{\gamma_{\rm E}})$.
The scale $\Lambda$ is the $\overline{\rm MS}$ scale associated with the
renormalization of the wavefunction at the origin. 
While the contribution from the spin-dependent potential has been obtained in
ref.~\cite{Kiyo:2010jm}, 
the contributions from the $1/m$ potential, the velocity-dependent
potential, and the relativistic correction to the kinetic energy are new. 
This result is accurate up to order $\alpha_s^2$, and is also sufficiently
accurate to reproduce the divergent small-$r_0$ behavior of the finite-$r$
regularized wavefunctions at the origin $\Psi_n(0)|_{r_0}$ computed to first
order in the QMPT using eq.~(\ref{eq:corr_finiter}). 

We note that in the calculation of $\delta Z_E$, the $E$ dependences cancel
between the $\overline{\rm MS}$-renormalized integrals and the finite-$r$
regularized integrals. This cancellation also occurs in the individual
contributions in eqs.~(\ref{eq:conv_1/m}), (\ref{eq:conv_velocity}),
(\ref{eq:conv_relcor}), and (\ref{eq:conv_spin}).
We argue that this cancellation is not accidental; 
since a shift in $E$ modifies only the finite piece of the divergent integral 
$\int d^3r\, G(\bm{0},\bm{r};E) \delta {\cal V}(r)$, changes in $E$ do not
affect the difference between DR and finite-$r$ regularization. 
Therefore, the scheme conversion coefficient $\delta Z$ is given by 
eq.~(\ref{eq:conv_all}) for all $S$-wave states $n$. 

The cancellation of the $E$ dependence in $\delta Z_E$ follows from the 
fact that $\delta Z_E$ depends only on the behavior of the integrand 
$\tilde G(\bm{p}',\bm{p};E) \delta \tilde{\cal V} (\bm{p})$ at 
large $\bm{p}$ and $\bm{p}'$. This implies that $\delta Z$ is also unaffected
by any modifications to the integrand that preserve the large-momentum
behavior. We note that in position space, inclusion of the nonperturbative 
long-distance contribution to the potential can be done by adding functions of 
$r$ that are regular at $r=0$. In momentum space, this is equivalent to 
modifying $\tilde V_{\rm LO} (\bm{q})$ and $\delta \tilde{\cal V} (\bm{q})$ 
by adding 
functions of $\bm{q}$ that decrease faster than $1/\bm{q}^2$ at large $\bm{q}$. 
Since 
such modifications do not affect the large-momentum behavior of the integrand 
$\tilde G(\bm{p}',\bm{p};E) \delta \tilde{\cal V} (\bm{p})$, they 
do not affect $\delta Z$. As a result, eq.~(\ref{eq:conv_all}) is still valid,
through order-$\alpha_s^2$ accuracy,  
even when the nonperturbative long-distance contributions are included in the
potential.\footnote{There is still a possibility that 
the scheme conversion may depend on nonperturbative
effects, if corrections of even higher orders in $1/m$ and $\alpha_s$ are
included. For example, corrections to the wavefunctions at the origin at
second order in the QMPT may contain subleading divergences that depend on the
nonperturbative contributions in the potentials. 
} 
The same argument can be made in position space: the short-distance
divergence of the integral $\int d^3r \, G(\bm{0},\bm{r};E) \delta{\cal V}(r)$ 
is determined completely in perturbative QCD,
because the short-distance behavior of $\delta{\cal V}(r)$
is unaffected by nonperturbative effects,
and the divergent short-distance behavior of the position-space Green's
function $G(\bm{r}',\bm{r};E)$ is determined only by the short-distance 
behavior of $V_{\rm LO} (r)$.

From the relation $\Psi_n(0)|_{\overline{\rm MS}} = \Psi_n(0)|_{r_0}
- \delta Z \times \Psi_n^{\rm LO} (0)$ we see that the scale dependence of 
$\Psi_n(0)|_{\overline{\rm MS}}$ is determined by $\delta Z$, 
because $\Psi_n(0)|_{r_0}$ and $\Psi_n^{\rm LO} (0)$ do not depend on
$\Lambda$. This allows us to compute the anomalous dimension of $S$-wave 
quarkonium wavefunctions at the origin. For spin triplet, we obtain 
\begin{equation}
\label{eq:RG_wavefunction_triplet}
\frac{d \log \Psi_n(0) |_{\overline{\rm MS}}}{d \log \Lambda} 
\bigg|_{\bm{S}^2=2}
= 
- \frac{d \delta Z}{d \log \Lambda} \bigg|_{\bm{S}^2=2}
= 
\alpha_s^2 C_F \left(\frac{C_F}{3} + \frac{C_A}{2} \right), 
\end{equation}
and for spin singlet, 
\begin{equation}
\label{eq:RG_wavefunction_singlet}
\frac{d \log \Psi_n(0) |_{\overline{\rm MS}}}{d \log \Lambda} 
\bigg|_{\bm{S}^2=0}
= 
- \frac{d \delta Z}{d \log \Lambda} \bigg|_{\bm{S}^2=0}
= 
\alpha_s^2 C_F \left( C_F + \frac{C_A}{2} \right). 
\end{equation}
The anomalous dimensions of the $S$-wave wavefunctions at the origin in
eqs.~(\ref{eq:RG_wavefunction_triplet}) and (\ref{eq:RG_wavefunction_singlet}) 
reproduce the order-$\alpha_s^2$ contributions of the anomalous dimensions of
the NRQCD LDMEs in eqs.~(\ref{eq:RG_vec}) and (\ref{eq:RG_ps}) for spin triplet
and spin singlet, respectively.

\subsection{Unitary transformation}
\label{sec:unitary_transformations}

As we have discussed previously, different forms of the potential can be
obtained by using unitary transformations. 
While the static potential is independent of the matching scheme, the forms of
the $1/m$ and $1/m^2$ potentials depend on the matching scheme used to compute
the potentials, as can be seen in appendix~\ref{appendix:potentials}. 
Since a different form of the potential can lead to a different behavior of the 
finite-$r$ regularized wavefunctions at the origin $\Psi_n(0)|_{r_0}$, 
the expression
for $\delta Z$ in eq.~(\ref{eq:conv_all}) is valid only 
when potentials from on-shell matching are used.
On the other hand, if we want to include corrections from the nonperturbative
long-distance behaviors of the potentials beyond leading order in $1/m$, 
it is necessary
to employ the nonperturbative definitions of the $1/m$ and $1/m^2$ potentials 
from Wilson-loop matching. The wavefunctions computed in Wilson-loop matching
must then be converted to wavefunctions in on-shell matching in order to
compute the $\overline{\rm MS}$-renormalized wavefunctions at the origin using 
the relation in eq.~(\ref{eq:conversion_relation}).\footnote{
In principle, unitary transformations can be avoided if we compute
the NRQCD SDCs that are compatible with Wilson-loop matching 
by using the direct matching procedure in ref.~\cite{Hoang:1997ui}. 
The SDCs in this case will differ from the usual SDCs that are determined from
on-shell matching. Since the differences between the SDCs from Wilson-loop 
matching and the SDCs from on-shell matching are determined in 
perturbative QCD, 
this approach is equivalent to computing the unitary transformation of 
the wavefunctions in perturbative QCD. 
} 

In perturbative QCD, the explicit form of the unitary transformation that is 
necessary to obtain the potentials in Wilson-loop matching
[eq.~(\ref{eq:potentials_wilson_short})] from 
the potentials in on-shell matching [eq.~(\ref{eq:potentials_pert})] 
has been derived in ref.~\cite{Brambilla:2000gk, Peset:2015vvi}.
If we define 
\begin{equation}
U(r) = 
\exp \left( - \frac{i}{m} \{ \bm{W}(\bm{r}), \bm{p} \} \right), 
\end{equation}
with 
\begin{equation}
\bm{W}(\bm{r}) = - \frac{1}{2} 
\left( V^{(1)}(r)|^{\rm OS} - V^{(1)}(r) |^{\rm WL} \right)
\times 
\frac{\bm{\nabla} V^{(0)}(r)}{(\bm{\nabla} V^{(0)}(r))^2},
\end{equation}
then 
\begin{equation}
U^{-1} (r) \left( - \frac{\bm{\nabla}^2}{m} + V (\bm{r},\bm{\nabla}) |^{\rm OS}
\right) U(r)
= 
\left( - \frac{\bm{\nabla}^2}{m} + V (\bm{r},\bm{\nabla}) |^{\rm WL}
\right) + O(1/m^3, \alpha_s^3/m^2), 
\end{equation}
where the superscripts OS and WL imply that the potentials are obtained from
on-shell matching and Wilson-loop matching, respectively. 

Since the differences in the $1/m$ and $1/m^2$ potentials between on-shell
matching and Wilson loop matching are known only at short distances, 
the precise form of $\bm{W}(\bm{r})$ can be obtained only near $r=0$. 
Since the wavefunctions at the origin in finite-$r$ regularization depend only
on the wavefunctions at short distances, it suffices to determine $U(r)$
for small $r$. 
We obtain 
\begin{equation}
\bm{W}(\bm{r}) = -\frac{\alpha_s C_F}{8} \hat{\bm{r}} + O(r),
\end{equation}
where $\hat{\bm{r}} = \bm{r}/|\bm{r}|$. 
Here, we neglect the correction from $\delta V_C(r)$; 
since we consider $\delta V_C(r)$ as perturbations in the QMPT, 
the correction to $\bm{W}(\bm{r})$ from $\delta V_C(r)$ 
corresponds to a piece of the second order correction in QMPT 
from insertions of $\delta V_C(r)$ and $V^{(1)}(r)/m$. 
Since we work at first order in the QMPT, we neglect this correction. 
Hence, to relative order $1/m$,
\begin{equation}
\label{eq:unitran_explicit}
U(r) = 1 + \frac{\alpha_s C_F}{4 m} \left( \frac{1}{r} +
\frac{\partial}{\partial r} \right) + O(1/m^2). 
\end{equation}
If $\Psi_n^{\rm OS}(\bm{r})$ is a solution of the Schr\"odinger equation with 
the potentials from on-shell matching, 
the wavefunction $\Psi_n^{\rm WL} (\bm{r})$ that satisfies the
Schr\"odinger equation from Wilson loop matching is given by 
\begin{equation}
\label{eq:utrans}
\Psi_n^{\rm WL} (\bm{r}) = U^{-1} (r) \Psi_n^{\rm OS} (\bm{r}).
\end{equation}
We note that near $r=0$ this relation also holds for the finite-$r$ regularized
wavefunctions at the origin, 
because the finite-$r$ regularized wavefunction at the origin reproduces the
divergent small $r_0$ behavior of the wavefunction $\Psi_n (\bm{r})$
at $|\bm{r}|=r_0$. While the finite-$r$ regularized wavefunction at the origin
and $\Psi_n (\bm{r})$ at $|\bm{r}| = r_0$ differ by a contribution proportional
to $1/r_0$, this difference does not affect the unitary transformation, because 
$\left( \frac{1}{r} + \frac{\partial}{\partial r} \right) \frac{1}{r}  =0$. 
By using the relation in eq.~(\ref{eq:conversion_relation}), we can compute 
the unitary transformation of $\Psi_n^{\rm OS} (0)|_{r_0}$ as 
\begin{eqnarray}
\label{eq:utrans_schemeconversion}
\Psi_n^{\rm WL}(0)|_{r_0} &=& 
U^{-1}(r_0) \Psi_n^{\rm OS} (0)|_{r_0} = 
U^{-1}(r_0) \left[ 
\Psi_n(0) |_{\overline{\rm MS}} + \delta Z \times \Psi_n^{\rm LO} (0) 
\right] 
\nonumber \\
&=& \Psi_n(0) |_{\overline{\rm MS}} 
+ 
\left[ \delta Z - \frac{\alpha_s C_F}{4 m r_0} \right] \times 
\Psi_n^{\rm LO} (0) + O(1/m^2), 
\end{eqnarray}
where the last equality follows from the fact that the difference between 
$\Psi_n(0)|_{\overline{\rm MS}}$ and $\Psi_n^{\rm LO} (0)$ 
is suppressed by at least $1/m$, 
because the divergent corrections in $\Psi_n^{\rm OS} (0)|_{r_0}$ that are not
suppressed by powers of $1/m$ are subtracted by the scheme conversion. 
Equation~(\ref{eq:utrans_schemeconversion}) implies that 
\begin{equation}
\label{eq:utrans_schemeconversion2}
\Psi_n^{\rm OS}(0)|_{r_0} 
=
\Psi_n^{\rm WL}(0)|_{r_0} 
+ 
\Psi_n^{\rm LO} (0) \times 
\left[ \frac{\alpha_s C_F}{4 m r_0} 
+ O(1/m^2) \right]. 
\end{equation}
We note that the difference between $\Psi_n^{\rm OS}(0)|_{r_0}$ and 
$\Psi_n^{\rm WL}(0)|_{r_0}$ does not depend on the long-distance behavior of
the $1/m$ potential, up to corrections that are suppressed by $1/m^2$. 
This gives rise to the following approximate relation 
\begin{equation}
\label{eq:unitary_diff}
\int d^3r\, \hat G_n (\bm{r}',\bm{r}) 
\left[ 
\frac{\pi \alpha_s C_F \delta^{(3)} (\bm{r}) }{m^2}
- \frac{\alpha_s^2 C_F^2 }{4m r^2} 
\right]
\frac{\Psi_n^{\rm LO} (r)}{\Psi_n^{\rm LO} (0)}
\bigg|_{|\bm{r}'| = r_0}
=
\frac{\alpha_s C_F}{4 m r_0}
+O(1/m^2), 
\end{equation}
which is obtained by dividing eq.~(\ref{eq:utrans_schemeconversion2}) by 
$\Psi_n^{\rm LO} (0)$ and taking the 
$1/m$ and $1/m^2$ potentials given in perturbative QCD. 
We see from eqs.~(\ref{eq:utrans_schemeconversion2}) and 
(\ref{eq:unitary_diff}) that the difference between 
$\Psi_n^{\rm OS}(0)|_{r_0}$ and $\Psi_n^{\rm WL}(0)|_{r_0}$ 
comes only from the difference in the potential at short distances that is
determined in perturbative QCD. 
This implies that, for the purpose of computing 
$\Psi_n^{\rm OS}(0)|_{r_0}$ including the 
nonperturbative long-distance behavior of the $1/m$ potential, 
it suffices to use the following prescription
\begin{equation}
\label{eq:potential_OSWL}
\delta {\cal V} (r)|^{\rm OS} 
=
\delta {\cal V} (r)|^{\rm WL} 
+ \frac{\alpha_s^2 C_F^2}{4 m r^2}  
- \frac{\pi \alpha_s C_F \delta^{(3)} (\bm{r}) }{m^2}, 
\end{equation}
so that while the potential at short distances is given by the expressions
from on-shell matching [eq.~(\ref{eq:potentials_pert})], 
the long-distance behavior is given by Wilson loop matching. 
We use eq.~(\ref{eq:potential_OSWL}) to 
include the nonperturbative long-distance
behavior of the $1/m$ potential in computing 
$\Psi_n^{\rm OS}(0)|_{r_0}$.

If we also wanted to include the long-distance nonperturbative behavior of the
potentials of order $1/m^2$, we would have needed to include order $1/m^2$
contributions in the unitary transformation that we neglected 
in eq.~(\ref{eq:unitran_explicit}). 
This contribution, on the other hand, includes divergences at $r=0$ 
that come from second order corrections to the wavefunctions at the origin. 
In turn, the second order corrections to the wavefunctions at the origin
produce divergences at relative order $\alpha_s^3$, 
which is beyond the accuracy of this paper. 
Hence, we neglect the long-distance nonperturbative behavior of the
potentials of order $1/m^2$.

\section{\boldmath Numerical results}
\label{sec:results}

We now compute wavefunctions at the origin in the $\overline{\rm MS}$ scheme
for $S$-wave charmonia and bottomonia. 
We first compute the wavefunctions at the origin in the finite-$r$
regularization from eq.~(\ref{eq:corr_finiter}), including the nonperturbative
long-distance contribution to the static potential. We also consider the
nonperturbative long-distance contribution to the $1/m$ potential by using the 
prescription given in eq.~(\ref{eq:potential_OSWL}). 
Then, the $\overline{\rm MS}$-renormalized wavefunctions at the origin can be
obtained from the relation in eq.~(\ref{eq:conversion_relation}), where 
$\delta Z$ is given by eq.~(\ref{eq:conv_all}). 
The validity of this numerical procedure is verified by comparing with 
the known analytical results from perturbative QCD calculations 
in appendix~\ref{appendix:perttest}. 

We compute decay constants and electromagnetic decay rates of 
$S$-wave charmonia and bottomonia based on the values of the 
wavefunctions at the origin that we obtain. 
The decay constants that we consider are defined in QCD by 
\begin{equation}
\langle 0 | \bar Q
\bm{\gamma} Q | V \rangle
= f_V  m_V \bm{\epsilon},
\end{equation}
for a vector quarkonium $V$, 
where the Dirac spinor $Q$ is a heavy quark field in QCD, $m_V$ is the mass
of the quarkonium $V$, and $\bm{\epsilon}$ is the polarization vector for the
state $|V\rangle$. In the QCD definition of $f_V$, the state $|V\rangle$ is 
normalized relativistically. 
We also consider the decay constants of pseudoscalar quarkonium $P$, 
which are defined by 
\begin{equation}
\langle 0 | \bar Q \gamma_\mu \gamma_5 Q | P \rangle
= f_P p_\mu,
\end{equation}
where $p_\mu$ is the 4-momentum of the quarkonium $P$,
and the state $|P\rangle$ is normalized relativistically.
The decay constants $f_V$ and $f_P$ are renormalization scheme and scale
independent. 
We list the NRQCD factorization formulas and the SDCs 
for the decay constants in appendix~\ref{appendix:sdcs}. 
We note that the decay constants are given at leading order in $\alpha_s$ and
$v$ by 
\begin{subequations}
\begin{eqnarray}
f_V^{\rm LO} &=& \sqrt{\frac{4 N_c}{m_V}} \Psi_V^{\rm LO} (0), 
\\
f_P^{\rm LO} &=& \sqrt{\frac{4 N_c}{m_P}} \Psi_P^{\rm LO} (0), 
\end{eqnarray}
\end{subequations}
where $m_P$ is the mass of the quarkonium $P$.
The decay constant $f_V$ is related to the leptonic decay rate of $V$ by
\begin{equation}
\label{eq:leptonic_rate}
\Gamma(V \to e^+ e^-) = \frac{4 \pi}{3} \alpha^2 e_Q^2
\frac{f_V^2}{m_V},
\end{equation}
where $\alpha$ is the QED coupling constant, and $e_Q$ is the
fractional charge of the heavy quark $Q$.
For pseudoscalar quarkonium, $f_P$ cannot be compared directly with an
experimentally measurable quantity, although it appears in hard exclusive
production rates of pseudoscalar quarkonium $P$~\cite{Brodsky:1989pv,
Chernyak:1983ej, Jia:2008ep, Chung:2019ota}, 
and also have been studied in lattice QCD~\cite{Davies:2010ip, Hatton:2020qhk}.
We also compute the two-photon decay rates of pseudoscalar quarkonia, 
which can be compared with measurements. 
We list the NRQCD factorization formula and the SDCs 
for the two-photon decay rate in appendix~\ref{appendix:sdcs}. 

Because the decay constants are quarkonium-to-vacuum matrix elements, they can
develop imaginary parts when there are contributions from cut diagrams. In the
NRQCD factorization formulas, the cut diagrams that involve momentum
transfers of the scale $m$ manifest as imaginary parts in the SDCs, 
as can be seen in the SDC for the decay constant $f_P$ in 
eq.~(\ref{eq:sdcs_pscurrent1}) and also in the SDC for the two-photon decay
amplitude in eq.~(\ref{eq:sdcs_psdecay1}). The cut diagrams that involve
momentum transfers of scales that are much less than $m$, such as transitions
between quarkonium states, can affect the quarkonium-to-vacuum LDMEs such as 
$\langle 0 | \chi^\dag \bm{\epsilon} \cdot \bm{\sigma} \psi | V \rangle$
and $\langle 0 | \chi^\dag \psi | P \rangle$. The sizes of these contributions
are at most of order $v^2$, because they are induced by insertions of the
dimension-5 operators in the NRQCD Lagrangian. 
In practice, we are only interested in the size of the decay constants, 
and imaginary parts in the SDCs are tiny, so we assume that the 
quarkonium-to-vacuum LDMEs at leading order in $v$ are real and positive, 
by utilizing the freedom to choose the phase of the quarkonium states, and
neglect the small imaginary parts of the SDCs in computing the decay constants.

\subsection{Numerical inputs}

We list the numerical inputs that we use in 
the numerical calculations in this section.

\subsubsection{Heavy quark mass and the strong coupling}
\label{sec:renormalon}

The heavy quark mass $m$ that appears in the Schr\"odinger equation, as well as
the scheme conversion coefficient in eq.~(\ref{eq:conv_all}) is the 
heavy quark pole mass, which suffers from renormalon
ambiguity~\cite{Beneke:1998ui}. 
In order to avoid this issue, we use the modified renormalon subtracted ($\rm
RS'$) mass $m_{\rm RS'}$, which is related to the pole mass $m$
by~\cite{Pineda:2001zq}
\begin{equation}
m = m_{\rm RS'} (\nu_f) + \delta m_{\rm RS'}(\nu_f), 
\end{equation}
where $\nu_f$ is the scale associated with the renormalon subtraction, and 
$\delta m_{\rm RS'}(\nu_f)$ is given as a series in $\alpha_s$. At leading
order in $\alpha_s$ (order $\alpha_s^2$), $\delta m_{\rm RS'}(\nu_f)$ is given
by 
\begin{equation}
\label{eq:delmrsp}
\delta m_{\rm RS'}(\nu_f) =
\frac{\alpha_s^2}{2 \pi} 
N_m \nu_f \beta_0
\sum_{k=0}^\infty c_k 
(b-k+1) 
+ O(\alpha_s^3). 
\end{equation}
Here, the constant $N_m$ is known numerically as $N_m =
0.5626(260)$~\cite{Peset:2018ria}, 
and the constants $b$ and $c_k$ are determined by the QCD $\beta$ function. 
Explicit formulas for $b$ and $c_k$ are given in ref.~\cite{Pineda:2001zq}. 
We truncate the series in eq.~(\ref{eq:delmrsp}) by including terms up to
$k=2$, which is equivalent to considering the running of $\alpha_s$ at $4$-loop
accuracy. 
In principle, the QCD renormalization scale at which $\alpha_s$ is computed in 
eq.~(\ref{eq:delmrsp}) can be different from $\nu_f$, and the scale dependence
is compensated by corrections of higher orders in $\alpha_s$.

Because the leading renormalon ambiguity of order $\Lambda_{\rm QCD}$ 
in the pole mass is not present in the $\rm RS'$ mass,
accurate values of the $\rm RS'$ masses have been obtained for charm and 
bottom. In order to use the $\rm RS'$ masses in calculation of the 
wavefunctions, we replace the heavy quark pole mass $m$ by 
$m_{\rm RS'} + \delta m_{\rm RS'}$, and expand perturbatively in powers of 
$\delta m_{\rm RS'}$. Since $\delta m_{\rm RS'}$ begins at order
$\alpha_s^2$, it suffices to consider only the correction coming from the
kinetic energy in the Schr\"odinger equation.\footnote{
The advantage of the use of the $\rm RS'$ mass, compared to other renormalon
subtraction schemes~\cite{Beneke:1998rk, Pineda:2001zq, Brambilla:2017hcq}, 
is that the renormalon subtraction term 
$\delta m_{\rm RS'}(\nu_f)$ begins at order $\alpha_s^2$, rather than order
$\alpha_s$. This makes it easier to organize the corrections from 
$\delta m_{\rm RS'}(\nu_f)$ in powers of $\alpha_s$. For example, 
in the $\rm RS'$ scheme, the corrections to the binding energies from 
$\delta m_{\rm RS'}(\nu_f)$ 
contribute to decay constants and decay rates from order
$\alpha_s^2 v^2$, which we ignore at the current level of accuracy. 
}
The correction to the wavefunctions at the origin from the $\rm RS'$
subtraction term is given by 
\begin{equation}
\label{eq:delmrspcorr}
\frac{\delta m_{\rm RS'}}{m_{\rm RS'}} 
\int d^3 r \, \hat G_n (\bm{0},\bm{r}) 
\left( - \frac{\bm{\nabla}^2}{m_{\rm RS'}} \right)
\Psi_n^{\rm LO} (r)
= 
- \frac{\delta m_{\rm RS'}}{m_{\rm RS'}} 
\int d^3 r \, \hat G_n (\bm{0},\bm{r}) 
V_{\rm LO} (r) 
\Psi_n^{\rm LO} (r), 
\end{equation}
which is finite. The corrections associated with the $1/m$ and
$1/m^2$ potentials can be neglected, because they are suppressed by higher
powers of $\alpha_s$. 
Hence, in computing the wavefunctions at the origin, 
it suffices to replace $m$ by
$m_{\rm RS'}(\nu_f)$ everywhere, and then compensate for the difference by
adding the correction term in eq.~(\ref{eq:delmrspcorr}) to the 
wavefunctions at the origin,
where $\delta m_{\rm RS'}$ is truncated at order $\alpha_s^2$. 
We adopt the values for the $\rm RS'$ charm and bottom masses determined in
ref.~\cite{Peset:2018ria} for $\nu_f = 2$~GeV, which are given by 
\begin{subequations}
\begin{eqnarray}
\label{eq:charmmassrsprime}
m_{c, {\rm RS'}} (2\textrm{~GeV}) &=& 1316(41)\textrm{~MeV},
\\
\label{eq:bottommassrsprime}
m_{b, {\rm RS'}} (2\textrm{~GeV}) &=& 4743(41)\textrm{~MeV}.
\end{eqnarray}
\end{subequations}

We compute $\alpha_s$ in the $\overline{\rm MS}$ scheme at a fixed QCD
renormalization scale $\mu_R$, except when we consider resummation of
logarithms in the loop corrections to the potentials. 
This is in order to facilitate exact order by order cancellation of 
the dependence on the factorization scale $\Lambda$ in 
NRQCD factorization formulas, which requires the same $\alpha_s$ to be used in
the SDCs and in the calculations of the wavefunctions at the origin. 
Reference~\cite{Peset:2018ria} gives ranges of $\mu_R$ in which
the theoretical
determinations of $\eta_c$ and $\eta_b$ masses have mild dependences 
on the scale; these are given by $\mu_R = 2.5{}^{+1.5}_{-1.0}$~GeV 
for charm, and $\mu_R = 5{}^{+3}_{-3}$~GeV for bottom.
We compute the numerical values of $\alpha_s$ 
in the $\overline{\rm MS}$ scheme at these scales using 
{\tt RunDec} at 4-loop accuracy~\cite{Chetyrkin:2000yt,Herren:2017osy}.

\subsubsection{Static potential from lattice QCD}

Precise nonperturbative determinations of the static potential $V^{(0)} (r)$ 
have been done in unquenched lattice QCD. 
We take the following parameterization of the
static potential in ref.~\cite{Bali:2000vr} given by 
\begin{equation}
\label{eq:staticpotential_lattice}
V^{(0)} (r) \big|_{\rm lattice} 
= V_0 + \sigma r - \frac{e}{r} + g \left( \frac{1}{r}- \left[ \frac{1}{\bm{r}}
\right] \right), 
\end{equation}
where $V_0$, $\sigma$, $e$, and $g$ are determined in fits to lattice
measurements. The term proportional to $g$ is included to quantify short
distance lattice artifacts, where $\left[ \frac{1}{\bm{r}} \right]$ is the 
tree level lattice propagator in position space~\cite{Bali:2000gf,Bali:2000vr}. 
We take the results in the physical limit given in 
table~V of ref.~\cite{Bali:2000vr}, where the central values read 
$V_0 = 0.760/a$, $\sigma = (0.171/a)^2$, $e = 0.368$, with $a^{-1} = 2.68$~GeV. 
We ignore the term proportional to $g$ in 
eq.~(\ref{eq:staticpotential_lattice}), because we only need the lattice QCD
result at long distances in the continuum limit.
We note that only the slope in $r$ is physically meaningful in the lattice QCD 
result for the static potential, because in
eq.~(\ref{eq:staticpotential_lattice}) an overall constant has been subtracted
to make $V^{(0)} (r) \big|_{\rm lattice}$ vanish at $a r^{-1} =
0.1485$~\cite{Bali:2000vr}. 

\begin{figure}[tbp]
\centering
\includegraphics[width=.95\textwidth]{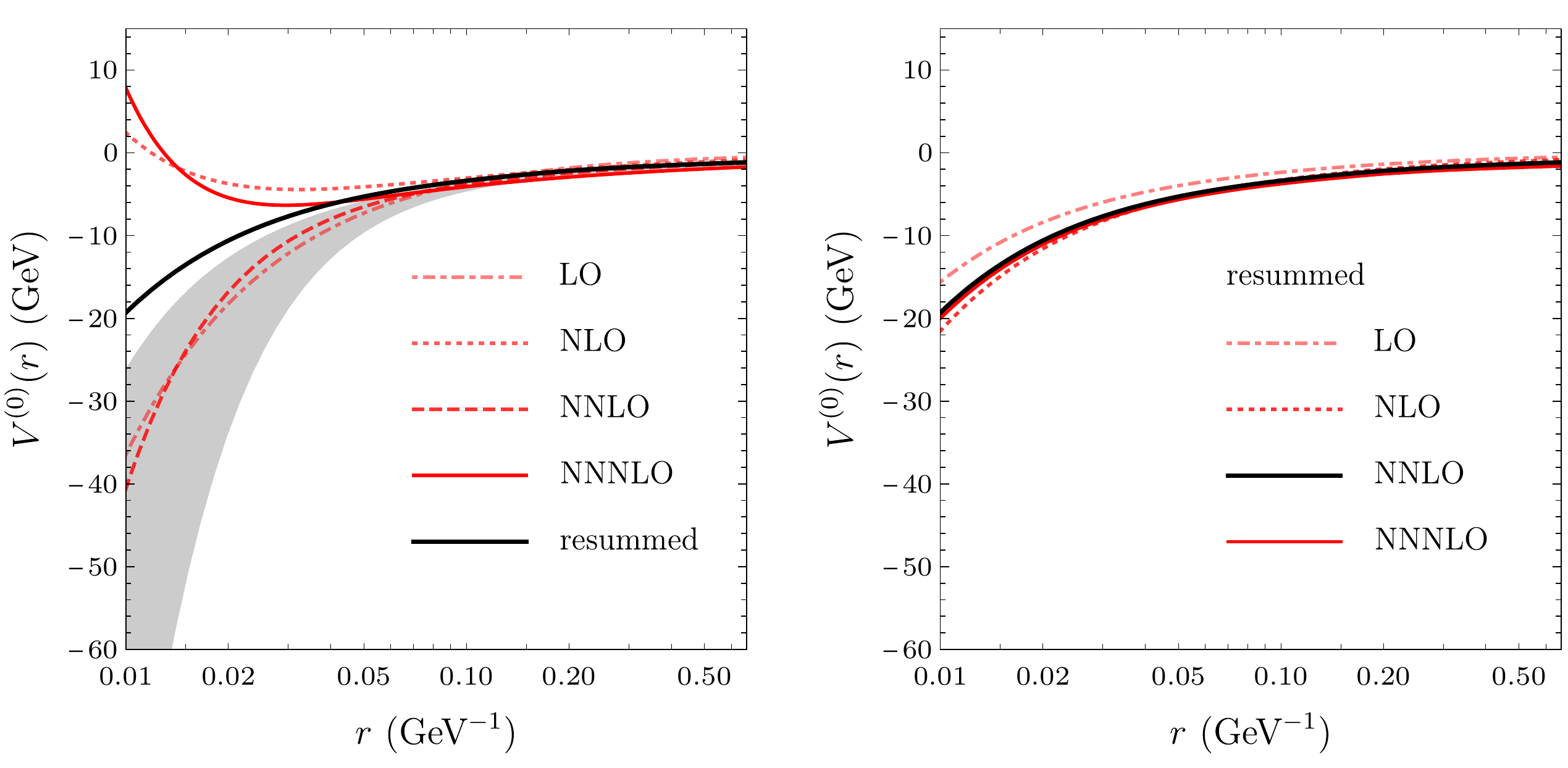}
\caption{\label{fig:static_pert}
Left panel: perturbative QCD results for the static potential 
at a fixed renormalization scale $\mu_R = 2.5$~GeV with $n_f =3$ 
at LO (dot-dashed line), NLO (dotted line), NNLO (dashed line), and NNNLO
(red solid line) accuracies, 
compared with the resummed expression at NNLO accuracy (black solid line) 
given in eq.~(\ref{eq:staticpotential_short_resummed}). 
The gray band shows the effect of varying the fixed renormalization scale 
$\mu_R$ between $1.5$~GeV and 4~GeV on the NNLO expression. 
Right panel: resummed perturbative QCD results for the static potential 
at the $r$-dependent renormalization scale $\mu_r = (r^{-2}+\mu_R^2)^{1/2}$ 
with $n_f =3$ at LO (dot-dashed line), NLO (dotted line), NNLO (black solid
line), and NNNLO (red solid line) accuracies. 
The position-space expression at NNNLO accuracy 
has been taken from ref.~\cite{Pineda:2011dg}. 
}
\end{figure}

We match eq.~(\ref{eq:staticpotential_lattice}) with the perturbative QCD
expression in eq.~(\ref{eq:staticpotential_pert}) at $r=r_{\rm match}$,
where the $r$ dependences of the two expressions agree well. 
While the renormalization scale dependence in $V^{(0)} (r) \big|_{\rm pert}$
cancels order by order in perturbative QCD, it is known that the convergence 
of the $r$ dependence of $V^{(0)} (r) \big|_{\rm pert}$ is poor when the 
renormalization scale is fixed~\cite{Kiyo:2010jm}. 
The convergence can be improved if we resum the logarithms that are
associated with the running of $\alpha_s$, which can be done by choosing the
renormalization scale to be proportional to $1/r$ at short distances. 
In order to avoid the renormalization scale being too small, 
we set the $r$-dependent renormalization scale to be
$\mu_r = (r^{-2}+\mu_R^2)^{1/2}$, so that 
$\mu_r > \mu_R$, while $\mu_r \approx 1/r$ at short distances. 
That is, we write 
\begin{equation}
\label{eq:staticpotential_short_resummed}
V^{(0)} (r) \big|_{\textrm{pert, resum}} 
= 
- \frac{\alpha_s(\mu_r) C_F}{r} 
\left[ 1 + \sum_{n=1}^2 \left( \frac{\alpha_s(\mu_r)}{4 \pi} \right)^n
a_n (r; \mu_r) \right], 
\end{equation}
where the $a_n$ are given in eqs.~(\ref{eq:coulombcorr}). 
This is just the perturbative QCD expression for the static potential in 
eq.~(\ref{eq:staticpotential_pert}), 
computed at the renormalization scale $\mu_r$. 
This choice of the renormalization scale may also help smoothen the matching
between the perturbative QCD expression at short distances and the
nonperturbative long-distance determination from lattice QCD. 
We compare the resummed expression for the static potential in
eq.~(\ref{eq:staticpotential_short_resummed}) with expressions at fixed
renormalization scale at LO, NLO, NNLO, and NNNLO accuracies 
in fig.~\ref{fig:static_pert}.

\begin{figure}[tbp]
\centering 
\includegraphics[width=.55\textwidth]{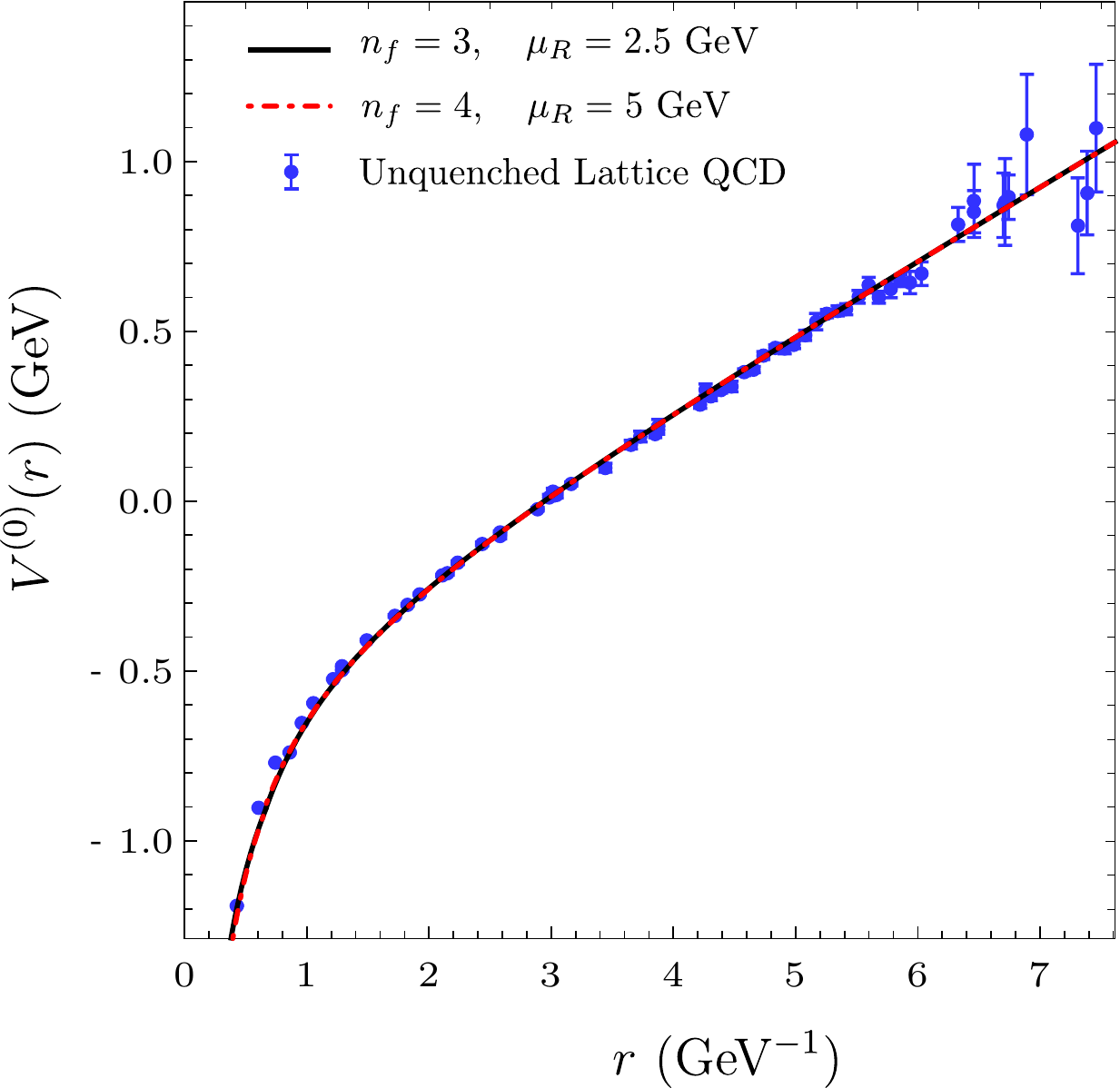}
\caption{\label{fig:staticpotential} 
The static potential $V^{(0)} (r)$ 
including the nonperturbative long-distance contribution 
[eq.~(\ref{eq:staticpotential_longandshort})] 
for $n_f =3$ (black solid line) 
and for $n_f =4$ (red dot-dashed line), 
shown with unquenched lattice QCD results from ref.~\cite{Bali:2000vr},
shifted vertically to match eq.~(\ref{eq:staticpotential_longandshort}). 
The renormalization scale for $\alpha_s$ has been chosen to be 
$\mu_r = (r^{-2}+\mu_R^2)^{1/2}$, 
as described in text, with $\mu_R = 2.5$~GeV for $n_f = 3$, and 
$\mu_R = 5$~GeV for $n_f = 4$. 
}
\end{figure}

We define the nonperturbative long-distance contribution to the static
potential as 
\begin{equation}
\label{eq:staticpotential_long}
V^{(0)} (r) \big|_{\rm long}
= \theta(r-r_{\rm match}) \times 
\left[ V^{(0)} (r) \big|_{\rm lattice} - 
V^{(0)} (r) \big|_{\textrm{pert, resum}} - \Delta V^{(0)} \right], 
\end{equation}
where $V^{(0)} (r) \big|_{\textrm{pert, resum}}$ is 
given by eq.~(\ref{eq:staticpotential_short_resummed}), 
and $\Delta V^{(0)}$ is chosen so that the right-hand side 
vanishes at $r = r_{\rm match}$, which removes the
unphysical constant shift in the lattice QCD parametrization $V^{(0)} (r)
\big|_{\rm lattice}$. 
We choose $r_{\rm match}^{-1} = 1.5$~GeV, 
which is where the slopes of $V^{(0)} (r) \big|_{\rm lattice}$ and 
$V^{(0)} (r) \big|_{\textrm{pert, resum}}$ are approximately same. 
Since $V^{(0)} (r) \big|_{\rm long}$ vanishes for $r< r_{\rm match}$, 
we obtain the following expression for the static potential that is valid
for both short and long distances: 
\begin{equation}
\label{eq:staticpotential_longandshort}
V^{(0)} (r) =  V^{(0)} (r) \big|_{\textrm{pert, resum}} 
+ V^{(0)} (r) \big|_{\rm long}, 
\end{equation}
so that $V^{(0)}(r)$ coincides with the perturbative QCD expression for 
$r<r_{\rm match}$, while it reproduces the lattice QCD determination for 
$r>r_{\rm match}$. Again, the perturbative QCD expression 
$V^{(0)} (r) \big|_{\textrm{pert, resum}}$ is computed at the renormalization
scale $\mu_r$, so that logarithms associated with the running of $\alpha_s$ 
are resummed. 
In fig.~\ref{fig:staticpotential} 
we compare the unquenched lattice QCD results in ref.~\cite{Bali:2000vr} 
with the expression for $V^{(0)}(r)$ in 
eq.~(\ref{eq:staticpotential_longandshort}). 
The perturbative QCD expressions of the static potential depends on the number
of light quark flavors $n_f$, which we take to be $n_f = 3$ for charm, 
and $n_f = 4$ for bottom. 
We note that the matching of perturbative QCD and lattice QCD for the pNRQCD
potentials have been done in a similar way in refs.~\cite{Laschka:2011zr,
Laschka:2012cf} for heavy quarkonium spectroscopy.

Eq.~(\ref{eq:staticpotential_longandshort}) implies that the 
leading-order potential is given by 
\begin{equation}
\label{eq:staticpotential_leadingorder}
V_{\rm LO} (r) = -\frac{\alpha_s(\mu_R) C_F}{r} + V^{(0)} (r) \big|_{\rm long},
\end{equation}
where in the first term on the right-hand side, $\alpha_s$ is evaluated at a
fixed renormalization scale $\mu_R$. 
The Coulombic correction term $\delta V_{\rm C} (r)= 
V^{(0)} (r) - V_{\rm LO} (r)$ that appears in the second line of
eq.~(\ref{eq:corr_finiter}) is given by 
\begin{equation}
\label{eq:coulombiccorr}
\delta V_{\rm C} (r) = 
V^{(0)} (r) \big|_{\textrm{pert, resum}}
+ \frac{\alpha_s(\mu_R) C_F}{r}, 
\end{equation}
where in the first term, $\alpha_s$ is evaluated at the scale $\mu_r$, 
while in the last term, $\alpha_s$ is evaluated at a fixed renormalization 
scale $\mu_R$, so that $V_{\rm LO} (r) + \delta V_C (r)$ reproduces the 
expression for $V^{(0)}(r)$ in eq.~(\ref{eq:staticpotential_longandshort}). 
The dependence on $\mu_R$ in $V_{\rm LO} (r)$ is cancelled explicitly by 
the order-$\alpha_s$ piece in $\delta V_{\rm C} (r)$, 
which is given by $- [\alpha_s (\mu_r) - \alpha_s (\mu_R)] C_F/r$. 
We note that $\delta V_C(r)$ contain contributions 
whose net effect is to shift the Coulomb strength of the LO potential at 
relative order $\alpha_s$, and so, the correction to the wavefunctions at 
the origin from $\delta V_C(r)$ begins at relative order $\alpha_s$. 
Although we work through first order in the QMPT, second order corrections 
from $\delta V_C(r)$ might be important, because this is of order $\alpha_s^2$.
Computing the second order Coulombic correction can also be useful in testing
the convergence of the Coulombic corrections. 
The second order Coulombic correction to the wavefunction at the origin 
$\Psi_n (0)$ can be computed by using the usual formula for the second order 
correction in the Rayleigh-Schr\"odinger perturbation theory, which reads 
\begin{eqnarray}
&&
\sum_{k \neq n} \sum_{\ell \neq n} 
\Psi_k^{\rm LO} (0) 
\frac{
\int d^3r_1 d^3r_2 
\Psi_k^{\rm LO}{}^* ({r}_2) \delta V_C(r_2) \Psi_\ell^{\rm LO} ({r}_2)
\Psi_\ell^{\rm LO}{}^* ({r}_1) \delta V_C(r_1) \Psi_n^{\rm LO} ({r}_1)
} 
{( E_n^{\rm LO} - E_k^{\rm LO} ) (E_n^{\rm LO} - E_\ell^{\rm LO} ) } 
\nonumber \\ 
&& - 
\left(\int d^3r \Psi_n^{\rm LO}{}^* ({r}) \delta V_C(r) \Psi_n^{\rm LO} ({r})
\right)
\times 
\sum_{k \neq n} 
\Psi_k^{\rm LO} (0)
\frac{
\int d^3r_1 
\Psi_k^{\rm LO}{}^* ({r}_1) \delta V_C(r_1) \Psi_n^{\rm LO} ({r}_1)
}
{( E_n^{\rm LO} - E_k^{\rm LO}) ^2}
\nonumber \\
&& - \frac{ \Psi_n^{\rm LO} (0) }{2} 
\sum_{k \neq n} 
\frac{\left| \int d^3r \Psi_k^{\rm LO}{}^* ({r}) \delta V_C (r) 
\Psi_n^{\rm LO} ({r}) \right|^2}{(E_n^{\rm LO} - E_k^{\rm LO})^2}. 
\end{eqnarray}
This expression can be rewritten in terms of the reduced Green's functions as
\begin{eqnarray}
\label{eq:coulombcorr_2ndorder}
&&
\int d^3r_1 d^3r_2 \, \hat G_n(\bm{0},\bm{r}_2) \delta V_C(r_2) 
\hat G_n(\bm{r}_2,\bm{r}_1) \delta V_C(r_1) \Psi_n^{\rm LO} (r_1) 
\nonumber \\
&& - 
\left( \int d^3r |\Psi_n^{\rm LO} (r)|^2 \delta V_C(r) \right) 
\times 
\int d^3r_1 d^3r_2 \, \hat G_n(\bm{0},\bm{r}_2) 
\hat G_n(\bm{r}_2,\bm{r}_1) \delta V_C(r_1) \Psi_n^{\rm LO} (r_1) 
\nonumber \\
&&
- \frac{\Psi_n^{\rm LO} (0)}{2} 
\int d^3r_1 d^3r_2 d^3r_3 \, 
\Psi_n^{\rm LO} (r_3) \delta V_C(r_3) 
\hat G_n (\bm{r}_3,\bm{r}_2) 
\hat G_n (\bm{r}_2,\bm{r}_1) 
\delta V_C(r_1) \Psi_n^{\rm LO} (r_1). 
\quad 
\end{eqnarray}
The equivalence between the two expressions can be verified by using 
the identity 
\begin{eqnarray}
\int d^3r \, \hat G_n (\bm{r}_2,\bm{r})\hat G_n (\bm{r},\bm{r}_1)
=
\sum_{k \neq n} 
\frac{\Psi_k^{\rm LO} (\bm{r}_2) 
\Psi_k^{\rm LO}{}^* (\bm{r}_1)
}{(E_n-E_k)^2}. 
\end{eqnarray}
In computing the second order Coulombic corrections, we neglect the 
$a_2$ term in the static potential [eq.~(\ref{eq:staticpotential_pert})], 
because at second order in the QMPT, this
term contributes at relative order $\alpha_s^3$.

\subsubsection[$1/m$ potential from lattice QCD]
{\boldmath $1/m$ potential from lattice QCD}

We use a similar strategy as the previous section to 
determine the nonperturbative long-distance contribution to the $1/m$
potential. Unlike the static potential, nonperturbative determinations of the
$1/m$ potential are available only from quenched lattice QCD. 
We use the parametrization in ref.~\cite{Koma:2012bc} given by 
\begin{equation}
\label{eq:1mpotential_lattice}
V^{(1)} (r) \big|_{\textrm{lattice}}^{\rm WL}
=
- \frac{9 A^2}{8 r^2} + \sigma^{(1)} \log r, 
\end{equation}
where $A = 0.297$ and $\sigma^{(1)} = 0.142$~GeV$^2$. 
This parametrization, which is based on the long-distance behavior expected 
from effective string theory in ref.~\cite{PerezNadal:2008vm},
is obtained in ref.~\cite{Koma:2012bc} from quenched lattice QCD results at 
lattice coupling $\beta=6.0$. Similarly to the lattice QCD determination of the
static potential in eq.~(\ref{eq:staticpotential_lattice}), only the slope in
$r$ is meaningful in the lattice QCD result in 
eq.~(\ref{eq:1mpotential_lattice}). 

\begin{figure}[tbp]
\centering
\includegraphics[width=.95\textwidth]{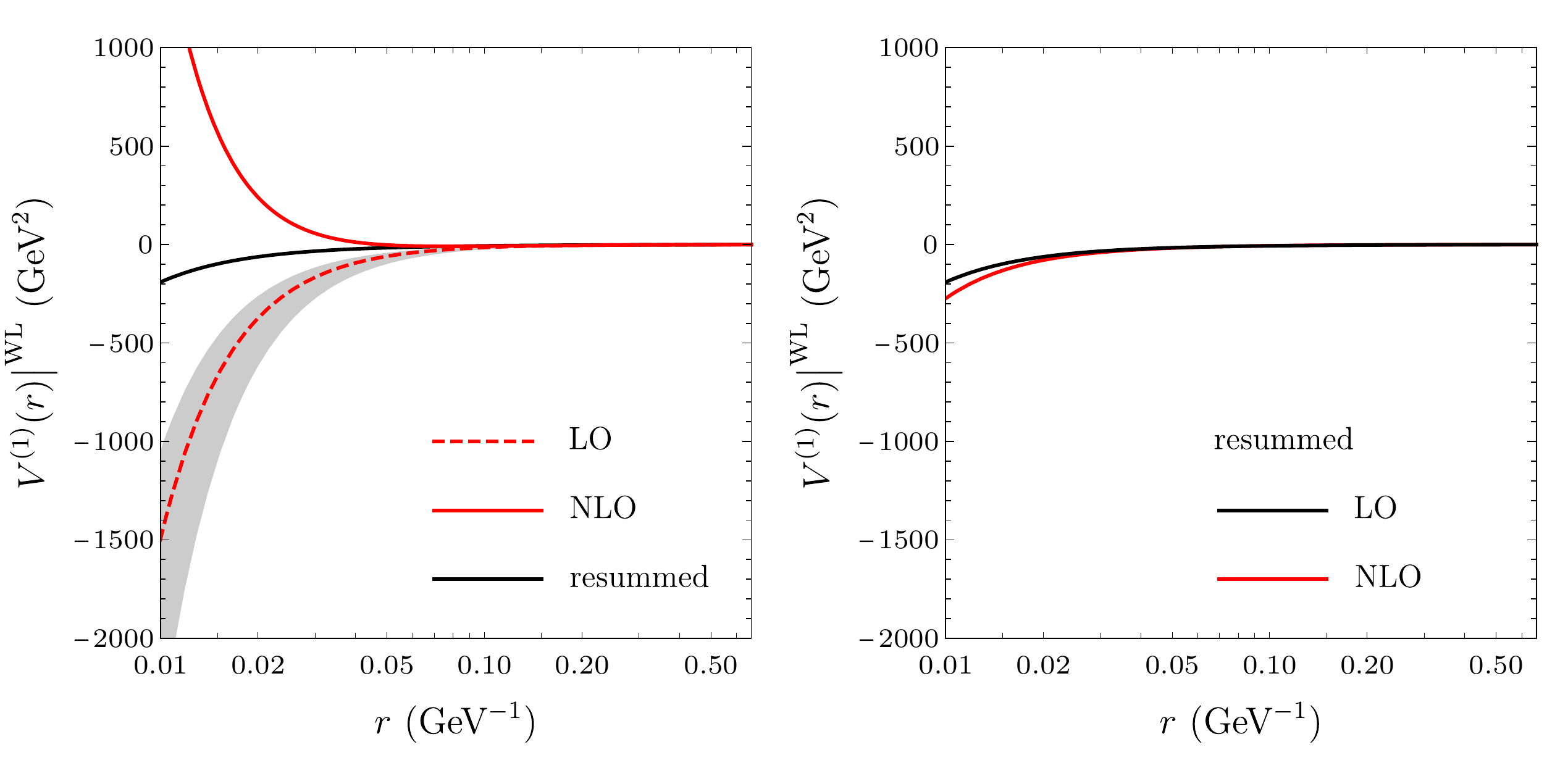}
\caption{\label{fig:mpot_pert}
Left panel: perturbative QCD results for the $1/m$ potential
in Wilson loop matching at a fixed renormalization scale $\mu_R = 2.5$~GeV
at LO (dashed line) and NLO (red solid line) accuracies,
compared with the resummed expression at LO accuracy (black solid line)
given in eq.~(\ref{eq:1mpotential_resum}). The gray band shows the
effect of varying the fixed renormalization scale $\mu_R$ between $1.5$~GeV 
and 4~GeV on the LO expression. 
Right panel: resummed perturbative QCD results for the $1/m$ potential
in Wilson loop matching 
at the $r$-dependent renormalization scale $\mu_r = (r^{-2}+\mu_R^2)^{1/2}$ at
LO (black solid line) and NLO (red solid line) accuracies. 
The position-space expression at NLO accuracy has been taken from 
ref.~\cite{Peset:2015vvi}, 
which is renormalized in the ${\overline {\rm MS}}$ scheme at scale 1~GeV. 
}
\end{figure}

We match eq.~(\ref{eq:1mpotential_lattice}) with the perturbative QCD 
expression at $r=r_{\rm match}$. 
Since we do not include loop corrections to the $1/m$ potential in our
calculations of the wavefunctions at the origin, 
the expression for the $1/m$ potential $V^{(1)} (r) \big|_{\rm pert}^{\rm WL}$
at leading order in $\alpha_s$ depends on the
choice of the renormalization scale. 
Similarly to our treatment of the static potential, we choose the
renormalization scale to be $\mu_r = (r^{-2} + \mu_R^2)^{1/2}$, so that the
logarithms associated with the running of $\alpha_s$ are resummed, which may
help smoothen the matching between the short-distance perturbative QCD
expression and the nonperturbative lattice QCD parametrization at long
distances. 
That is, we write 
\begin{equation}
\label{eq:1mpotential_resum}
V^{(1)} (r) \big|_{\textrm{pert, resum}}^{\rm WL}
= - \frac{\alpha_s^2 (\mu_r) C_F C_A}{2 r^2}. 
\end{equation}
We compare this resummed expression with expressions at a fixed
renormalization scale at LO and NLO accuracies in fig.~\ref{fig:mpot_pert}.

\begin{figure}[tbp]
\centering
\includegraphics[width=.55\textwidth]{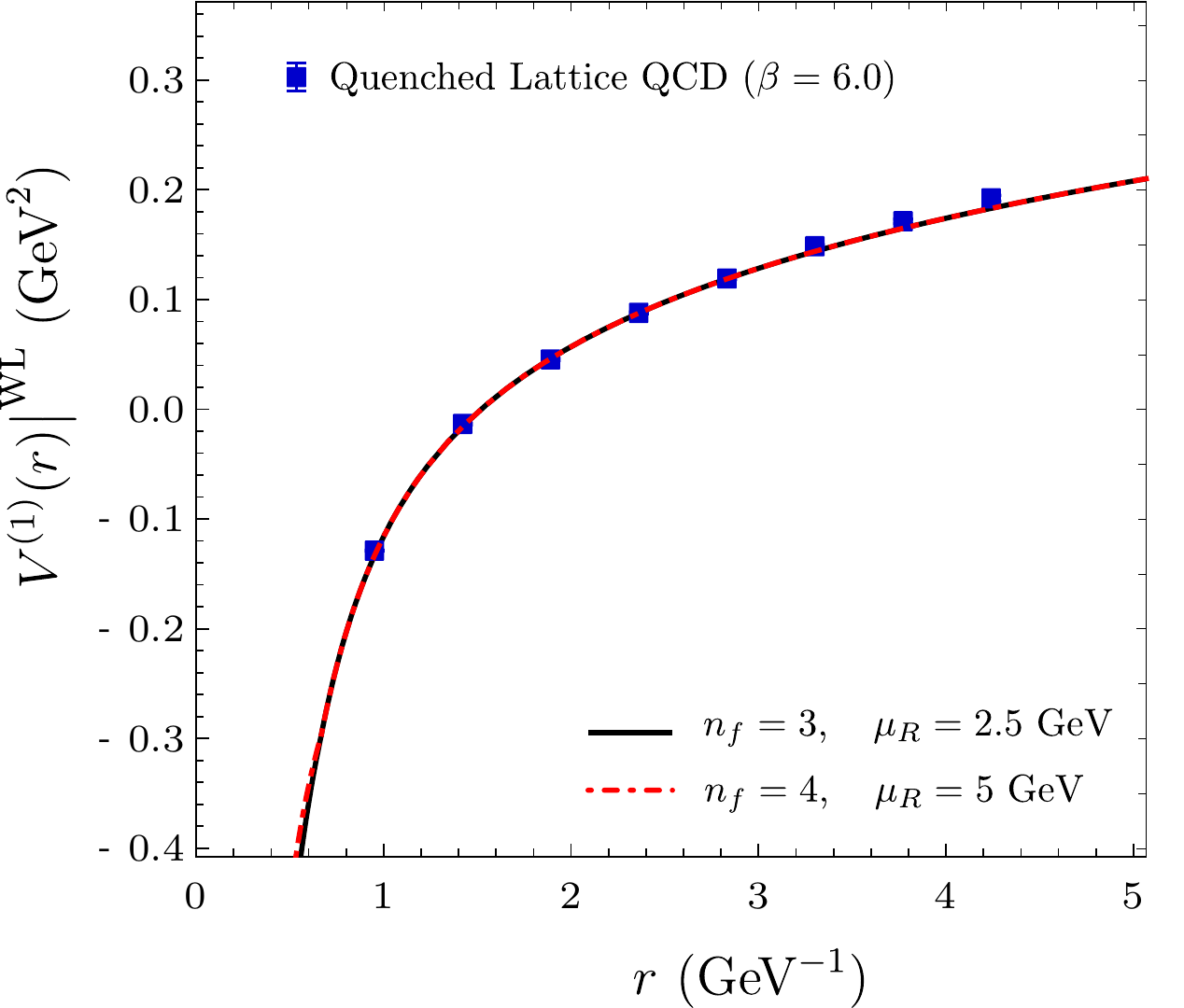}
\caption{\label{fig:1mpotential}
The $1/m$ potential $V^{(1)} (r)$ in Wilson loop matching, 
for $n_f =3$ and $\mu_R = 2.5$~GeV (black solid line),
and for $n_f =4$ and $\mu_R = 5$~GeV (red dot-dashed line),
shown with quenched lattice QCD results with lattice coupling $\beta=6.0$ in
ref.~\cite{Koma:2012bc}, shifted vertically to match
eq.~(\ref{eq:1mpotential_longandshort}). 
}
\end{figure}

We define the nonperturbative long-distance contribution to the $1/m$ potential by 
\begin{equation}
\label{eq:1mpotential_long}
V^{(1)} (r) \big|_{\textrm{long}}^{\rm WL}
=
\theta(r-r_{\rm match}) \times \left[ 
V^{(1)} (r) \big|_{\textrm{lattice}}^{\rm WL}
-V^{(1)} (r) \big|_{\textrm{pert, resum}}^{\rm WL} 
- \Delta V^{(1)} \right], 
\end{equation}
where $\Delta V^{(1)}$ is chosen so that the right-hand side 
vanishes at $r = r_{\rm match}$, which removes the unphysical constant shift in
the lattice QCD parametrization $V^{(1)} (r) \big|_{\textrm{lattice}}^{\rm
WL}$.
We choose $r_{\rm match}^{-1} = 1.5$~GeV. 
Since $V^{(1)} (r) \big|_{\textrm{long}}^{\rm WL}$ vanishes for 
$r<r_{\rm match}$, we obtain an expression for the $1/m$ potential that is
valid for both short and long distances given by 
\begin{equation}
\label{eq:1mpotential_longandshort}
V^{(1)} (r) \big|^{\rm WL} = 
V^{(1)} (r) \big|_{\textrm{pert, resum}}^{\rm WL}
+ V^{(1)} (r) \big|_{\rm long}^{\rm WL}.
\end{equation}
We compare the lattice QCD determination in 
eq.~(\ref{eq:1mpotential_lattice}) with the expression for 
$V^{(1)}(r) \big|^{\rm WL}$ in eq.~(\ref{eq:1mpotential_longandshort}) 
in fig.~\ref{fig:1mpotential}. 

Based on the argument given in section~\ref{sec:unitary_transformations}, we
obtain the expression for the $1/m$ potential in on-shell matching that is
valid for computation of wavefunctions at the origin, given by  
\begin{equation}
\label{eq:1mpotential_longandshort_onshell}
V^{(1)} (r) \big|^{\rm OS} =
\frac{\alpha_s^2(\mu_R) C_F (\frac{1}{2} C_F- C_A)}{2 r^2}
+ V^{(1)} (r) \big|_{\rm long}^{\rm WL},
\end{equation}
where in the first term on the right-hand side, $\alpha_s$ is computed at a
fixed renormalization scale $\mu_R$. 
We use this form of the $1/m$ potential in
the calculation of the wavefunctions at the origin.

\subsubsection{Reduced Green's function}

We compute the reduced Green's function $\hat G_n (\bm{r}',\bm{r})$ numerically
by using two different methods, which are valid in different regimes of 
$r$ and $r'$. In the first method, which is valid for small $r$ and $r'$, 
we compute the Green's function in position space numerically 
by using the method given in ref.~\cite{Strassler:1990nw}. 
We only need to compute the $S$-wave contribution, which is defined by
including only the $S$-wave states in the sum in eq.~(\ref{eq:greenfunction1}). 
This contribution can be written as
\begin{equation}
\label{eq:greenfunc_swave}
G^S (\bm{r}',\bm{r};E) 
= \frac{m}{4 \pi} \frac{u_<(r_<)}{r_<} \frac{u_> (r_>)}{r_>},
\end{equation}
where $r_< = {\rm min}(|\bm{r}|,|\bm{r}'|)$, 
$r_> = {\rm max}(|\bm{r}|,|\bm{r}'|)$, and the superscript $S$ denotes the
$S$-wave contribution. 
The functions $u_<$ and $u_>$ are two independent solutions of the
differential equation
\begin{equation}
\label{eq:greenequations}
\left[ \frac{d^2}{dr^2} + m (E-V_{\rm LO} (r)) \right] 
u (r) = 0,
\end{equation}
with the following boundary condition
\begin{subequations}
\begin{eqnarray}
&&
u_<(0) = 0, \quad u'_< (0) = 1,
\\
&&
u_>(\infty) = 0, \quad u_> (0) = 1,
\end{eqnarray}
\end{subequations}
so that $u_<(r)/r$ is regular at $r=0$, while $u_>(r)$ is square integrable. 
We determine the functions $u_<$ and $u_>$ by numerically solving the
differential equation for a given $E$. 
The reduced Green's function can then be obtained by using the relation in
eq.~(\ref{eq:redgreen_relation}), where we take the limit numerically. 
We note that, if $E$ coincides with an eigenenergy of the LO Schr\"odinger
equation $E_n^{\rm LO}$, then the corresponding wavefunction 
$\Psi_n^{\rm LO} (r)$ is proportional to $u_<(r)/r$. 
This means that $u_<(r)$ is square integrable if $E = E_n^{\rm LO}$, 
and in such case, the square-integrable solution $u_>(r)$ does not exist. 
Hence, the limit in eq.~(\ref{eq:redgreen_relation}) must be taken with care, 
because the numerical solution for $u_>(r)$ becomes unstable if 
$E$ is too close to $E_n^{\rm LO}$. 
When we compute the reduced Green's functions numerically using
eq.~(\ref{eq:redgreen_relation}), we set $\eta = 10^{-3}$~GeV. 

Since the first method involves computing $u_<(r_<)$ by solving a 
differential equation with initial conditions at $r_<=0$, the method becomes
unreliable when $r$ and $r'$ are both large. For large $r$ and $r'$, we 
compute the reduced Green's function by using the formal solution in
eq.~(\ref{eq:redgreen_definition}), where we truncate the series by including
only a limited number of the lowest eigensolutions of the LO 
Schr\"odinger equation. In the numerical calculations, we include the 9 
lowest $S$-wave states in the calculation of the reduced Green's function. 
This method in turn becomes unreliable at small $r$ and $r'$. For example, 
if the LO potential $V_{\rm LO} (r)$ is linear in $r$ at long distances, 
the eigenenergies of highly excited $S$-wave states 
increase linearly with increasing principal quantum number, 
and the LO wavefunctions at the origin 
are constant in the principal quantum number. 
Hence, the series in eq.~(\ref{eq:redgreen_definition}) diverges like 
$\sum_n^\infty 1/n$ at $r=r'=0$. This implies that the truncated series becomes
unreliable at small $r$ and $r'$.

We combine the reduced Green's function at long and short distances by 
\begin{equation}
\label{eq:redgreen_blend}
\hat G_n (\bm{r}',\bm{r}) 
= 
b(r_<) \times 
\hat G_n (\bm{r}',\bm{r}) |_{\rm short} 
+ 
\left[1-b(r_<) \right] \times 
\hat G_n (\bm{r}',\bm{r}) |_{\rm long}, 
\end{equation}
where $\hat G_n (\bm{r}',\bm{r}) |_{\rm short}$ is computed by using 
eqs.~(\ref{eq:greenfunc_swave}) and (\ref{eq:redgreen_relation}), 
$\hat G_n (\bm{r}',\bm{r}) |_{\rm long}$ is
computed by truncating the series in eq.~(\ref{eq:redgreen_definition}), and 
$b(r)$ is a smooth function that satisfies $b(0)=1$ and $b(\infty) = 0$,
so that eq.~(\ref{eq:redgreen_blend}) is reliable for all $r$ and $r'$.
We define $b(r)$ by 
\begin{equation}
b(r) = \frac{1}{\pi} 
\left[ \tan^{-1} ( 4 m (r_b-r)) - \tan^{-1} (4 m r_b) \right] +1, 
\end{equation}
with $r_b = 1$~GeV$^{-1}$. 
The validity of the reduced Green's function obtained in 
eq.~(\ref{eq:redgreen_blend}) can be tested 
by numerically checking the relations
\begin{subequations}
\begin{eqnarray}
\left(E_k^{\rm LO} - E_n^{\rm LO} \right)
\int d^3r\, \hat G_n(\bm{r}',\bm{r}) \Psi_k^{\rm LO} (r)
&=& 
\Psi_k^{\rm LO} (r'), 
\\
\int d^3r\, \hat G_n(\bm{r}',\bm{r}) \Psi_n^{\rm LO} (r) &=& 0, 
\end{eqnarray}
\end{subequations}
for $k \neq n$. 

We note that, due to the boundary condition $u_>(0) =1$, 
it is evident that $G^S (\bm{0},\bm{r};E)$ develops a power divergence given
by $m/(4 \pi r)$ near $r=0$. 
It has been shown in ref.~\cite{Kiyo:2010jm}
that if the LO potential is given by $V_{\rm LO}(r) =
-\alpha_s C_F/r$ at short distances, $u_>(r)/r$ also contains a 
logarithmic divergence given by $-\alpha_s C_F m \log r$. 
Therefore, near $r=0$, the Green's function behaves like 
\begin{equation}
G^S (\bm{0},\bm{r};E) = 
\frac{m}{4 \pi r} - \frac{\alpha_s C_F m^2}{4 \pi} \log r 
+ \cdots, 
\end{equation}
where the ellipsis represent contributions that are finite at $r=0$. 
This shows that the divergent small $r$ behavior of 
$G^S (\bm{0},\bm{r};E)$ depends only on the short-distance behavior of the LO
potential, which is determined in perturbative QCD. 

\subsubsection{Gluonic correlators}

The pNRQCD expressions of the NRQCD LDMEs in eqs.~(\ref{eq:pNRQCD_vec})
and (\ref{eq:pNRQCD_ps}) depend on gluonic correlators that scale with powers
of $\Lambda_{\rm QCD}$. Also, corrections to the wavefunctions at the origin 
from the velocity-dependent potential involve $V_{p^2}^{(2)} (0)$, which
in DR, is proportional to the correlator $i {\cal E}_2$. 
While the gluonic correlators of mass dimension two contribute to the NRQCD
LDMEs at relative order $\Lambda_{\rm QCD}^2/m^2$, the dimensionless correlator 
${\cal E}_3$ contributes to 
$\langle 0 | \chi^\dag \bm{\epsilon} \cdot \bm{\sigma} \psi | V \rangle$ 
and $\langle 0 | \chi^\dag \psi | P \rangle$ at relative order $v^2$, 
and the correlator $i {\cal E}_2$ contributes to the wavefunctions at the origin
at relative order $\Lambda_{\rm QCD}/m$. 

The dimensionless correlator ${\cal E}_3$ in the $\overline{\rm MS}$ scheme 
has been determined in ref.~\cite{Brambilla:2020xod} 
from measured decay rates of $P$-wave
charmonia. At the $\overline{\rm MS}$ scale $\Lambda = 1$~GeV, 
\begin{equation}
{\cal E}_3 (1\textrm{~GeV}) = 2.05^{+0.94}_{-0.65}. 
\end{equation}
The correlator ${\cal E}_3$ depends logarithmically on the scale. We 
compute ${\cal E}_3$ at other scales by using the one-loop renormalization
group improved expression~\cite{Brambilla:2001xy, Brambilla:2020xod}
\begin{equation}
\label{eq:e3_evolve}
{\cal E}_3 (\Lambda) = 
{\cal E}_3 (\Lambda') + \frac{24 C_F}{\beta_0} 
\log \frac{\alpha_s (\Lambda')}{\alpha_s (\Lambda)}. 
\end{equation}

Reference~\cite{Brambilla:2020xod} also provides a determination of 
$i{\cal E}_2$ from measured
electromagnetic decay and production rates of $P$-wave charmonia. However, the
determination in ref.~\cite{Brambilla:2020xod} has uncertainties that 
are larger than the typical size
of the correlator that is expected from its power counting. 
For this reason, instead of taking the determination in
ref.~\cite{Brambilla:2020xod}, we consider the effect of $V_{p^2}^{(2)}(0)$ 
to the wavefunctions at the origin in the uncertainties by assuming 
$|V_{p^2}^{(2)}| \lesssim 500$~MeV, which corresponds to the typical size of 
$\Lambda_{\rm QCD}$.

Since the gluonic correlators of mass dimension two contribute to the NRQCD 
LDMEs at relative order $\Lambda_{\rm QCD}^2/m^2$, we neglect them in 
calculations of the LDMEs compared to corrections of relative order 
$\Lambda_{\rm QCD}/m$ and $v^2$.

\subsection[Numerical results for $S$-wave charmonia]
{\boldmath Numerical results for $S$-wave charmonia}

In this section, 
we compute the $\overline{\rm MS}$-renormalized wavefunctions at the origin for
the $1S$ and $2S$ charmonium states. 
We identify the $J/\psi$ and $\eta_c$ as the $1S$ charmonium states in 
spin-triplet and spin-singlet states, respectively, while the 
$\psi(2S)$ and $\eta_c(2S)$ states are the $2S$ charmonium states in 
spin-triplet and spin-singlet states, respectively. 

As we discussed in previous sections, we solve the Schr\"odinger equation
numerically with
the LO potential in eq.~(\ref{eq:staticpotential_leadingorder}) 
and the charm quark mass in eq.~(\ref{eq:charmmassrsprime}) to determine
$\Psi_n^{\rm LO} (r)$, $E_n^{\rm LO}$, and $\hat G_n (\bm{r}',\bm{r})$. 
For this purpose, 
it suffices to solve the differential equation in eq.~(\ref{eq:greenequations}) 
and obtain the solutions $u_<(r)$ and $u_>(r)$ for a range of $E$, 
because $u_<(r)$ becomes square integrable when $E = E_n^{\rm LO}$, and 
the corresponding eigenfunction $\Psi_n^{\rm LO} (r)$ is then proportional to
$u_<(r)/r$. We obtain the solution $u_<(r)$ by solving the differential 
equation in eq.~(\ref{eq:greenequations}) numerically in {\scshape Mathematica} 
using the {\tt NDSolve} command with the initial conditions $u_<(0) = 0$ and
$u'_<(0)=1$. Instead of obtaining directly the  $u_>(r)$ with the boundary 
conditions $u_>(0) = 1$ and $u_>(\infty) = 0$, we find a linearly independent
second solution $v(r)$ which is in general a linear combination of $u_<(r)$ and
$u_>(r)$. Similarly to what has been done in ref.~\cite{Kiyo:2010jm}, 
we find a solution $v(r)$ that satisfies  $v(0)=1$, and let $v'(r)$ be nonzero
at small $r$ (in general $v'(r)$ is singular at 
$r=0$~\cite{Strassler:1990nw, Kiyo:2010jm}).
Then, the solution $u_>(r)$ that satisfies the boundary conditions $u_>(0) = 1$ 
and $u_>(\infty)=0$ is given by 
$u_>(r) = v(r) - \frac{v(\infty)}{u_<(\infty)} u_<(r)$. 

Then, we compute the corrections to the wavefunctions at the origin in the
finite-$r$ regularization using eq.~(\ref{eq:corr_finiter}). 
In order to compensate for the use of the $\rm RS'$ mass, we add 
to eq.~(\ref{eq:corr_finiter}) the finite
correction from the $\rm RS'$ subtraction term 
in eq.~(\ref{eq:delmrspcorr}). 
We also add to eq.~(\ref{eq:corr_finiter}) the Coulombic correction at 
second order in QMPT in eq.~(\ref{eq:coulombcorr_2ndorder}). 
In the calculation of the corrections to the wavefunctions at the origin, 
we use the $1/m$ potential given by 
eq.~(\ref{eq:1mpotential_longandshort_onshell}), 
while we take the perturbative QCD expressions of the $1/m^2$ potentials given
by eq.~(\ref{eq:potentials_pert}). 
When we compute the central values of the wavefunctions at the origin, 
we set $V_{p^2}^{(2)} (0) = 0$ in eq.~(\ref{eq:corr_finiter}), 
and consider the effect of the correction from $V_{p^2}^{(2)} (0)$ 
in the uncertainties. 
We then use eq.~(\ref{eq:conversion_relation}) to obtain the 
$\overline{\rm MS}$-renormalized wavefunctions at the origin. 

In computing the finite-$r$ regularized wavefunctions at the origin, the
regulator $r_0$ must be chosen to be as small as possible, as long as the
numerical calculation is stable. 
We determine an optimal choice of $r_0$ by numerically testing the 
approximate relation in eq.~(\ref{eq:unitary_diff}). 
We find that the relation is 
well reproduced numerically at $1\%$ level for $r_0 \gtrsim 0.1$~GeV$^{-1}$. 
Hence, we choose $r_0 = 0.2$~GeV$^{-1}$, and vary $r_0$ between
$0.1$~GeV$^{-1}$ and $0.3$~GeV$^{-1}$. 
We set the $\overline{\rm MS}$ scale $\Lambda$ to be the charm quark mass $m$,
and choose the central value of the QCD renormalization scale $\mu_R$ to be
2.5~GeV, as discussed in sec.~\ref{sec:renormalon}. 

\begin{figure}[tbp]
\centering
\includegraphics[width=.55\textwidth]{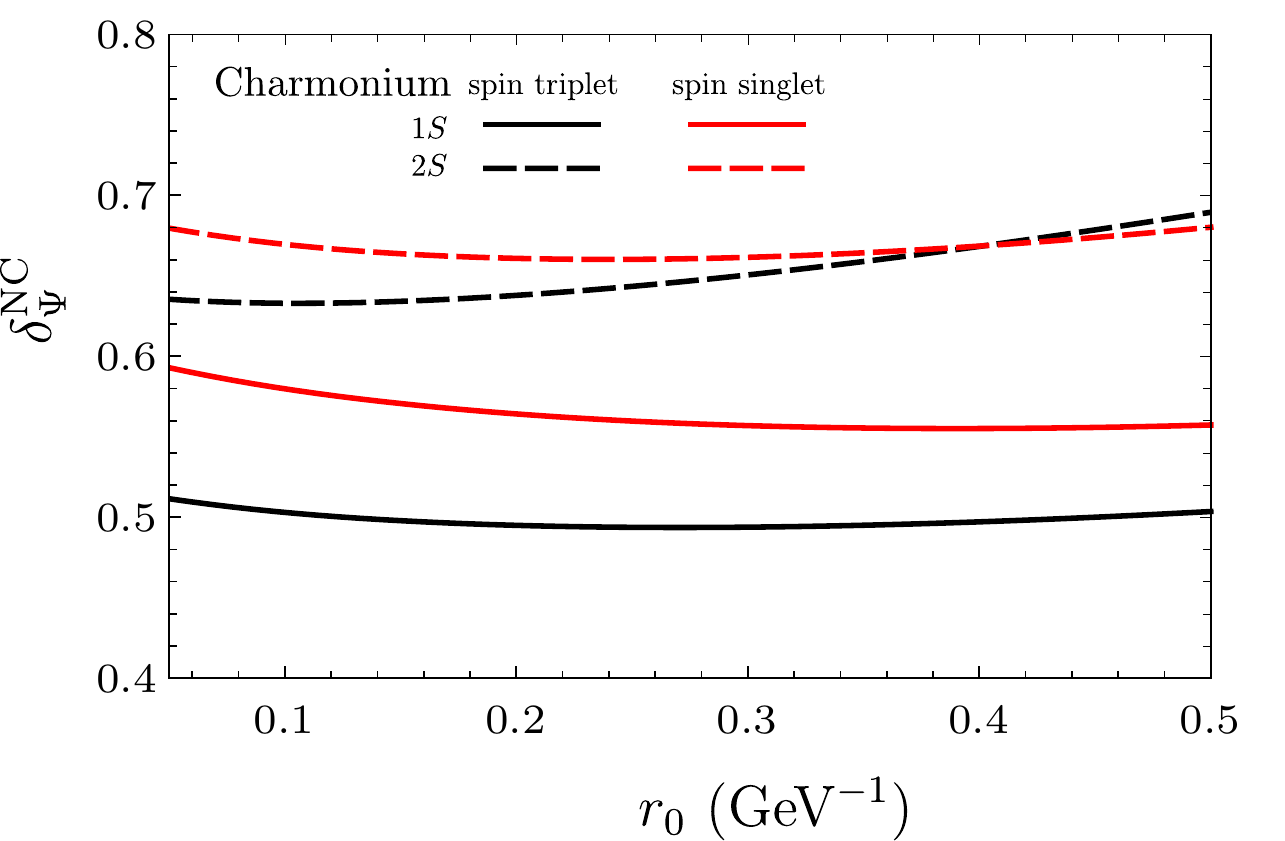}
\caption{\label{fig:charmr0}
The non-Coulombic corrections $\delta_\Psi^{\rm NC}$ at finite $r_0$ 
for the charmonium $1S$ (solid lines) and $2S$ (dashed lines) states, 
for spin triplet (black) and spin singlet (red). The $r_0$ dependences are 
mild for the range 0.1~GeV$^{-1}< r_0 <$ 0.3~GeV$^{-1}$ that we consider. 
}
\end{figure}

We list the central values of the LO wavefunctions at the origin and 
the LO binding energies in table~\ref{tab:charmcorrections}. 
We also list the 
corrections to the wavefunctions at the origin relative to 
$\Psi_n^{\rm LO}(0)$ in table~\ref{tab:charmcorrections}. 
We classify the corrections by their origins in the following way: 
the non-Coulombic correction $\delta_\Psi^{\rm NC}$ 
comes from the $1/m$ and $1/m^2$ potentials, 
the Coulombic corrections $\delta_\Psi^{\rm C1}$ and $\delta_\Psi^{\rm C2}$
come from $\delta V_C(r)$ 
at first and second order in the QMPT, respectively, 
and the correction $\delta_\Psi^{\rm RS'}$ 
comes from the $\rm RS'$ subtraction term. 
The explicit expressions for $\delta_{\Psi}^{\rm NC}$ and 
$\delta_\Psi^{\rm C1}$ are given by 
\begin{subequations}
\begin{eqnarray}
\label{eq:NCcorr}
\delta_{\Psi}^{\rm NC} 
&=&
-\delta Z 
- \frac{1}{\Psi_n^{\rm LO} (0)}
\int d^3r \, \hat G_n (\bm{r}',\bm{r}) \delta {\cal V} (r)
\Psi_n^{\rm LO} (r) \Big|_{|\bm{r}'| = r_0}
\nonumber \\ &&
- 
\frac{1}{\Psi_n^{\rm LO} (0)}
\frac{E_n^{\rm LO}}{m}
\int d^3r \, \hat G_n (\bm{0},\bm{r})
\left[ V_{p^2}^{(2)} (r) + \frac{1}{2} V_{\rm LO}(r)
\right]
\Psi_n^{\rm LO} (r)
\nonumber \\ &&
+ \frac{1}{2 m}  \int d^3 r \,
\left[ V_{p^2}^{(2)} (r) + \frac{1}{2} V_{\rm LO} (r) \right]
\left|\Psi_n^{\rm LO} (r) \right|^2
- \frac{
V_{p^2}^{(2)} (0)
}{2 m}, 
\\
\delta_\Psi^{\rm C1}
&=&
- \frac{1}{\Psi_n^{\rm LO} (0)}
\int d^3r \, \hat G_n (\bm{0},\bm{r})
\delta V_C(r)
\Psi_n^{\rm LO} (r), 
\end{eqnarray}
\end{subequations}
while $\delta_\Psi^{\rm C2}$ is given by dividing
eq.~(\ref{eq:coulombcorr_2ndorder}) by $\Psi_n^{\rm LO} (0)$, 
and $\delta_\Psi^{\rm RS'}$ is given by dividing 
eq.~(\ref{eq:delmrspcorr}) by $\Psi_n^{\rm LO} (0)$. 
The $\overline{\rm MS}$-renormalized wavefunctions at the origin are then given
by 
\begin{equation}
\Psi_n(0)|_{\overline{\rm MS}} = \Psi_n^{\rm LO}(0) \times \left( 1+ 
\delta_\Psi^{\rm NC} + \delta_\Psi^{\rm C1} + \delta_\Psi^{\rm C2} +
\delta_\Psi^{\rm RS'} \right). 
\end{equation}
We note that the $r_0$ dependence cancels in $\delta_\Psi^{\rm NC}$ between
$\delta Z$ and the finite-$r$ regularized integral for small $r_0$. 
We demonstrate this cancellation of the $r_0$ dependence in
fig.~\ref{fig:charmr0}. 

\begin{table}[tbp]
\centering
\begin{tabular}{|c||c|c||c|c|c|c|c|}
\hline
State & $\Psi^{\rm LO} (0)$~(GeV$^{3/2}$) & $E^{\rm LO}$~(GeV) &
$\delta_\Psi^{\rm NC}|_{\bm{S}^2=2}$ & $\delta_\Psi^{\rm NC}|_{\bm{S}^2=0}$ & 
$\delta_\Psi^{\rm C1}$ & $\delta_\Psi^{\rm C2}$ & $\delta_\Psi^{\rm RS'}$ \\
\hline
\hline
$1S$ & 0.183 & 0.233 & 0.495 & 0.564 & 0.173 & $-0.010$ & 0.080
\\
\hline
$2S$ & 0.177 & 0.769 & 0.638 & 0.661 & 0.079 & $-0.004$ & 0.072
\\
\hline
\end{tabular}
\caption{\label{tab:charmcorrections}
LO wavefunctions at the origin, LO binding energies 
and relative corrections to the wavefunctions at the
origin 
in the $\overline{\rm MS}$ scheme at scale $\Lambda=m$ for $1S$ and $2S$
charmonium states. 
$\delta_\Psi^{\rm NC}$ is the correction from the $1/m$ and $1/m^2$ potentials,
$\delta_\Psi^{\rm C1}$ and $\delta_\Psi^{\rm C2}$ are Coulombic corrections at
first and second order in QMPT, respectively, 
and $\delta_\Psi^{\rm RS'}$ is the correction from the $\rm RS'$ 
subtraction term. 
The $\delta_\Psi^{\rm NC}$, $\delta_\Psi^{\rm C1}$, $\delta_\Psi^{\rm C2}$,
and $\delta_\Psi^{\rm RS'}$ are dimensionless. 
}
\end{table}

The results for the LO binding energies for the $1S$ and $2S$ states in
table~\ref{tab:charmcorrections} are roughly compatible with the mass
difference between $J/\psi$ and $\psi(2S)$. 
We see that the non-Coulombic corrections from $1/m$ and $1/m^2$ potentials,
given by $\delta_\Psi^{\rm NC}$ in table~\ref{tab:charmcorrections}, 
are large and positive for both $1S$ and $2S$ states. 
This is in contrast with the order-$\alpha_s^2$ corrections to the NRQCD SDCs
in appendix~\ref{appendix:sdcs}, which are large and negative at $\Lambda = m$. 
This implies that if we combine the pNRQCD expressions of the LDMEs with
the NRQCD SDCs, large cancellations will occur between the order-$\alpha_s^2$ 
corrections to the SDCs and the corrections to the wavefunctions at the origin. 
The contribution from the long-distance part of the $1/m$ potential, which is
given by the second term in eq.~(\ref{eq:1mpotential_longandshort_onshell}),
amounts to about $+10\%$ of the LO wavefunction at the origin for the $1S$
state, and about $+6\%$ of the LO wavefunction at the origin for the $2S$ 
state. 
We note that while the Coulombic corrections at first order are 
positive, the Coulombic corrections at second order are small and negative,
signaling good convergence of the Coulombic corrections. The corrections from
the renormalon subtraction term $\delta_\Psi^{\rm RS'}$ are mild for both $1S$
and $2S$ states. 

The LO wavefunctions at the origin in table~\ref{tab:charmcorrections} are 
much larger than what we would obtain if we neglect the long-distance 
nonperturbative part of the static potential, for example, by using the
analytical solution of the Schr\"odinger equation in perturbative QCD (see
appendix~\ref{appendix:perttest}). For the $1S$ state, neglecting
the long-distance nonperturbative part of the static potential reduces the
wavefunction at the origin by more than a factor of 2, and for the $2S$ state,
the wavefunction at the origin reduces by more than a factor of 7. At the
squared amplitude level, neglecting the long-distance part of the static
potential can reduce the $1S$ charmonium decay rates by almost an order of
magnitude, and $2S$ charmonium decay rates by more than an order of magnitude. 
Hence, the long-distance nonperturbative part of the static potential has 
a significant effect on charmonium wavefunctions at the origin and charmonium 
decay rates. 

We use the results for the $\overline{\rm MS}$ wavefunctions at the origin 
in table~\ref{tab:charmcorrections}
to compute decay constants and electromagnetic decay rates of $S$-wave
charmonium states. 
We first compute the decay constants $f_V$ of $V = J/\psi$ and
$\psi(2S)$. By using the pNRQCD expressions of the LDMEs in
eqs.~(\ref{eq:pNRQCD_vec}) and (\ref{eq:pNRQCD_vec_v2}) and the SDCs in 
appendix~\ref{appendix:sdcs}, and expanding the corrections to the SDCs and to
the wavefunctions at the origin, we obtain 
\begin{eqnarray}
\label{eq:veccurrent_pNRQCD}
f_V &=& 
\sqrt{\frac{4 N_c}{m_V} }
\Psi_V^{\rm LO} (0) 
\bigg[ 1 
+ \alpha_s c_v^{(1)} 
+ \delta_\Psi^{\rm C1} + \delta_\Psi^{\rm C2} + \delta_\Psi^{\rm RS'} 
+ \delta_\Psi^{\rm NC}|_{\bm{S}^2=2} + \alpha_s^2 c_v^{(2)} 
\nonumber \\ && \hspace{16ex}
+ \alpha_s c_v^{(1)} \delta_\Psi^{\rm C1} 
+ \frac{2 E_V^{\rm LO}}{m_V} \left(d_v 
- \frac{{\cal E}_3}{9}
\right) 
+O(\alpha_s^3, v^3, \Lambda_{\rm QCD}^2/m^2)
\bigg], 
\quad 
\end{eqnarray}
where $c_v = 1+ \alpha_s c_v^{(1)} + \alpha_s^2 c_v^{(2)} +
O(\alpha_s^3)$, and $\alpha_s = \alpha_s (\mu_R)$. 
This expression is valid up to corrections of relative order $\alpha_s^3$, 
$v^3$, and $\Lambda_{\rm QCD}^2/m^2$. 
We set $n_f=3$ in the SDCs. 
Since we assume $\delta_\Psi^{\rm C1}$ to be of order $\alpha_s$, 
we keep the 
cross term $\alpha_s c_v^{(1)} \delta_\Psi^{\rm C1}$. 
The dependence on the $\overline{\rm MS}$ scale $\Lambda$ in 
$\delta_\Psi^{\rm NC}|_{\bm{S}^2=2}$ cancels completely 
with the $\Lambda$ dependence in $\alpha_s^2 c_v^{(2)}$, 
while the $\Lambda$ dependence in the order-$\alpha_s$ correction to 
$d_v$ cancels with the scale dependence of the correlator ${\cal E}_3$. 
Hence, variation of the factorization scale $\Lambda$ has almost no effect 
in eq.~(\ref{eq:veccurrent_pNRQCD}). 
We set the scale $\Lambda = m$ in the one-loop correction to $d_v$, 
and compute the correlator ${\cal E}_3$ at the same scale 
using the renormalization group improved expression in 
eq.~(\ref{eq:e3_evolve}). 
We take the measured quarkonium masses from ref.~\cite{Tanabashi:2018oca}.

The numerical result for the $J/\psi$ decay constant is 
\begin{equation}
\label{eq:fJpsi_result}
f_{J/\psi} = 
0.363 
{}^{+0.015}_{-0.003}
{}^{+0.003}_{-0.000}
\pm 0.069 
\pm 0.054
\textrm{~GeV} 
= 0.363{}^{+0.089}_{-0.088} 
\textrm{~GeV}, 
\end{equation}
where the first uncertainty comes from varying $\mu_R$ between $1.5$~GeV and
$4$~GeV, and the second uncertainty comes from varying $r_0$ between
0.1~GeV$^{-1}$ and 0.3~GeV$^{-1}$. The third uncertainty comes from 
the neglect of the correction $-V_{p^2}^{(2)} (0)/(2 m)$ to the wavefunction at
the origin, which we take to be $\pm 500\textrm{~MeV}/(2 m)$ times the central
value. 
The last uncertainty comes from the uncalculated corrections of order $v^3$, 
which we take to be 15\% of the central value, based on the
typical estimate $v^2 \approx 0.3$ for charmonium states. 
In the last equality, we add the uncertainties in quadrature. 

We note that the central value for $f_{J/\psi}$ that we obtain
is very close to the leading-order
value $f_{J/\psi}^{\rm LO} 
= 0.360$~GeV. 
The order-$\alpha_s$ terms in eq.~(\ref{eq:veccurrent_pNRQCD}) 
from $\alpha_s c_v^{(1)}$ and $\delta_\Psi^{\rm C1}$ amount to about $-6\%$
of the central value, and the corrections proportional to $E_V^{\rm LO}$ is
about $-7\%$ of the central value. 
The remaining corrections from $\alpha_s^2 c_v^{(2)}$, 
$\delta_\Psi^{\rm NC}$, 
$\delta_\Psi^{\rm C2}$, 
$\delta_\Psi^{\rm RS'}$, 
and the cross term $\alpha_s c_v^{(1)} \delta_\Psi^{\rm C1}$ add up to about 
$+14\%$ of the central value, so that the numerical result for $f_{J/\psi}$ 
in eq.~(\ref{eq:fJpsi_result})
is only about $1\%$ larger than the leading-order value. 
If we had ignored the corrections to the wavefunctions at the origin, the
one-loop correction would have been $-23\%$ of the leading order result, 
while the two-loop correction would have been $-39\%$ of the leading order
value, so that the loop corrections add up to $-62\%$ at two-loop accuracy. 
The inclusion of the corrections to the wavefunction at the origin 
in the calculation of the decay constant $f_{J/\psi}$ has 
substantially improved the convergence of the expansion in $\alpha_s$ and $v$. 

The result for $f_{J/\psi}$ agrees within uncertainties with the 
lattice QCD determination using relativistic charm quarks in
ref.~\cite{Hatton:2020qhk},
which gives $f_{J/\psi} = 0.4104(17)$~GeV. 
In order to compare with measurements, 
we compute the leptonic decay rate of $J/\psi$ from $f_{J/\psi}$ 
by using eq.~(\ref{eq:leptonic_rate}). We obtain 
\begin{equation}
\label{eq:Jpsi_decayrate}
\Gamma(J/\psi \to e^+ e^-) = 4.5 {}^{+2.5}_{-1.9} \textrm{~keV},
\end{equation}
where we used $\alpha = 1/133$, which is computed at the scale of the $J/\psi$
mass. This result agrees with the experimental value 
$\Gamma(J/\psi \to e^+ e^-) = 5.53 \pm 0.10$~keV 
in ref.~\cite{Tanabashi:2018oca} within uncertainties.

The leptonic decay rate can also be computed by using the NRQCD factorization
formula at the decay rate level, which is obtained by squaring the
amplitude-level factorization formula (\ref{eq:fac_vectordecayconstant}), 
and expanding in powers of $\alpha_s$ and $v$. 
In order to facilitate exact order-by-order cancellation of the
NRQCD factorization scale dependence, we also expand the pNRQCD expressions
for the NRQCD LDMEs at the squared amplitude level, as well as the square of
the wavefunction at the origin in powers of $\alpha_s$, $v$, and 
$\Lambda_{\rm QCD}/m$. That is, we square the expression for the decay constant
in eq.~(\ref{eq:veccurrent_pNRQCD}) and then expand the corrections in powers 
of $\alpha_s$, $v$, and $\Lambda_{\rm QCD}/m$.
In this case, we obtain $\Gamma(J/\psi \to e^+ e^-) = 4.5^{+1.9}_{-1.8}$~keV, 
which agrees well with the result in eq.~(\ref{eq:Jpsi_decayrate}) within 
uncertainties. This agreement is due to the fact that the convergence
of the expansion in powers of $\alpha_s$, $v$, and $\Lambda_{\rm QCD}/m$ 
have improved significantly in both the decay constant and the leptonic 
decay rate, thanks to the corrections to the wavefunctions at the origin that
we have included. 

The numerical result for the $\psi(2S)$ decay constant is 
\begin{equation}
\label{eq:psi2S_result}
f_{\psi(2S)} =
0.309
{}^{+0.011}_{-0.010}
{}^{+0.004}_{-0.002}
\pm 0.059
\pm 0.046
\textrm{~GeV}
= 0.309{}^{+0.076}_{-0.076}
\textrm{~GeV},
\end{equation}
where the uncertainties are as in eq.~(\ref{eq:fJpsi_result}). 
Again, the central value for $f_{\psi(2S)}$ that we obtain 
is very close to the leading-order value $f_{\psi(2S)}^{\rm LO} 
= 0.318$~GeV. This follows from the improvement of
the convergence of the corrections of higher orders in $\alpha_s$ and $v$ 
by the inclusion of the corrections to the wavefunction at the origin. 
We compute the leptonic decay rate of $\psi(2S)$ 
by using eq.~(\ref{eq:leptonic_rate}). We obtain
\begin{equation}
\label{eq:psi2S_decayrate}
\Gamma(\psi(2S) \to e^+ e^-) = 2.7 {}^{+1.5}_{-1.2} \textrm{~keV},
\end{equation}
where we used $\alpha = 1/133$, which is computed at the scale of the 
$\psi(2S)$ mass. This result agrees with the experimental value 
$\Gamma(\psi(2S) \to e^+ e^-) = 2.33 \pm 0.04$~keV 
in ref.~\cite{Tanabashi:2018oca} within uncertainties.
If we use the expression for the decay rate expanded in powers of $\alpha_s$,
$v$, and $\Lambda_{\rm QCD}/m$ at the squared amplitude level, we obtain 
$\Gamma(\psi(2S) \to e^+ e^-) = 2.7 \pm 1.1$~keV, which agrees well within
uncertainties with eq.~(\ref{eq:psi2S_decayrate}). 

Now we compute the decay constants $f_P$ of $P = \eta_c$ and $\eta_c(2S)$. 
Although $f_P$ cannot be obtained directly from experimental measurements, 
this decay constant appears in exclusive production cross sections of
pseudoscalar quarkonia at high energies~\cite{Jia:2008ep, Chung:2019ota}. 
We obtain the following expression for $f_P$ by expanding the corrections to
the SDCs and the corrections to the wavefunctions at the origin: 
\begin{eqnarray}
\label{eq:pseudoveccurrent_pNRQCD}
f_P &=&
\sqrt{\frac{4 N_c}{m_P} }
\Psi_P^{\rm LO} (0)
\bigg[ 1
+ \alpha_s c_p^{(1)}
+ \delta_\Psi^{\rm C1} + \delta_\Psi^{\rm C2} + \delta_\Psi^{\rm RS'}
+ \delta_\Psi^{\rm NC}|_{\bm{S}^2=0} + \alpha_s^2 c_p^{(2)}
\nonumber \\ && \hspace{16ex}
+ \alpha_s c_p^{(1)} \delta_\Psi^{\rm C1}
+ \frac{2 E_P^{\rm LO}}{m_P} \left(d_p 
- \frac{{\cal E}_3}{9}
\right)
+O(\alpha_s^3, v^3, \Lambda_{\rm QCD}^2/m^2)
\bigg],
\end{eqnarray}
where $c_p = 1+ \alpha_s c_p^{(1)} + \alpha_s^2 c_p^{(2)} +
O(\alpha_s^3)$, and $\alpha_s = \alpha_s (\mu_R)$. 
We neglect the small imaginary part in $c_p^{(2)}$, 
which amounts to less than $0.02$. 
This expression is valid up to corrections of relative order $\alpha_s^3$, 
$v^3$, and $\Lambda_{\rm QCD}^2/m^2$. 
We set $n_f=3$ in the SDCs.
We note that the $\Lambda$ dependence in 
$\delta_\Psi^{\rm NC}|_{\bm{S}^2=0}$ cancels exactly
with $\alpha_s^2 c_p^{(2)}$. 
Since the order-$\alpha_s$ correction to $d_p$ is not available, 
our expression for $f_P$ in 
eq.~(\ref{eq:pseudoveccurrent_pNRQCD}) depends mildly on $\Lambda$ through 
the scale dependence of ${\cal E}_3$.\footnote{
It is expected from NRQCD factorization that 
$d_p$ will have the same logarithmic dependence on $\Lambda$ 
at order $\alpha_s$ 
as $d_{\gamma \gamma}$ in eq.~(\ref{eq:sdcs_psdecay2}), 
so that the dependence on $\Lambda$ cancels between $d_p$ and ${\cal E}_3$
in eq.~(\ref{eq:pseudoveccurrent_pNRQCD}). 
}
Nevertheless, variation of the factorization scale $\Lambda$ has a very
small effect in eq.~(\ref{eq:pseudoveccurrent_pNRQCD}). 
We compute ${\cal E}_3$ at the scale $\Lambda=m$. 
We take the measured quarkonium masses from ref.~\cite{Tanabashi:2018oca}.

The numerical result for $f_{\eta_c}$ is 
\begin{equation}
\label{eq:etacf_result}
f_{\eta_c} =
0.385 
{}^{+0.013}_{-0.000}
{}^{+0.006}_{-0.003}
\pm 0.073
\pm 0.057
\textrm{~GeV}
= 0.385{}^{+0.094}_{-0.093}
\textrm{~GeV},
\end{equation}
where the uncertainties are as in eq.~(\ref{eq:fJpsi_result}). 
We add the uncertainties in quadrature. 
We neglect the uncertainty from the scale dependence of ${\cal E}_3$, 
which is small compared to other uncertainties. 
This result for $f_{\eta_c}$ agrees with the lattice QCD determination 
using relativistic charm quarks in refs.~\cite{Hatton:2020qhk},
which gives $f_{\eta_c} = 0.3981(10)$~MeV. 

Similarly to the case of $f_{J/\psi}$, 
the central value for $f_{\eta_c}$ that we obtain 
is very close to the leading-order value 
$f_{\eta_c}^{\rm LO} 
= 0.367$~GeV. 
The order-$\alpha_s$ corrections in eq.~(\ref{eq:pseudoveccurrent_pNRQCD})
from $\alpha_s c_p^{(1)}$ and $\delta_\Psi^{\rm C1}$ amount to about $1\%$
of the central value, and the corrections proportional to $E_P^{\rm LO}$ is
about $-13\%$ of the central value.
The remaining corrections from $\alpha_s^2 c_v^{(2)}$,
$\delta_\Psi^{\rm NC}$, $\delta_\Psi^{\rm C2}$, $\delta_\Psi^{\rm RS'}$,
and the cross term $\alpha_s c_v^{(1)} \delta_\Psi^{\rm C1}$ add up to about
$+17\%$ of the central value, so that the central value for $f_{\eta_c}$
in eq.~(\ref{eq:etacf_result})
is only about $4\%$ larger than the leading-order value.
In contrast, if we had ignored the corrections to the wavefunctions
at the origin, the
one-loop correction would have been $-16\%$ of the leading order result,
while the two-loop correction would have been $-44\%$ of the leading order
value, so that the loop corrections add up to $-60\%$ at 
two-loop accuracy. 
Just like the case of the $J/\psi$ and $\psi(2S)$ decay constants, 
the inclusion of the corrections to the wavefunction at the origin
in the calculation of the decay constant $f_{\eta_c}$ 
greatly improves the convergence of the expansion in $\alpha_s$ and $v$.

The numerical result for $f_{\eta_c(2S)}$ is 
\begin{equation}
\label{eq:etac2Sf_result}
f_{\eta_c(2S)} =
0.275
{}^{+0.010}_{-0.019}
{}^{+0.003}_{-0.000}
\pm 0.052
\pm 0.041
\textrm{~GeV}
= 0.271{}^{+0.068}_{-0.069}
\textrm{~GeV},
\end{equation}
where the uncertainties are as in eq.~(\ref{eq:etacf_result}). 
For the case of $\eta_c(2S)$, 
the central value for $f_{\eta_c(2S)}$ that we obtain is about $18\%$ 
smaller than the leading-order value $f_{\eta_c(2S)}^{\rm LO} = 0.321$~GeV.
The difference is larger than the case of $\eta_c$, 
because the binding energy of the $2S$ state is larger than the 
binding energy of the $1S$ state, and so, 
the correction proportional to $E^{\rm LO}_{\eta_c(2S)}$ 
is larger compared to the $\eta_c$ case. 

Finally, we compute the two-photon decay rate of $P = \eta_c$ and $\eta_c(2S)$. 
The NRQCD factorization formula for the decay rate is given in 
appendix~\ref{appendix:sdcs}. 
The following expression for the decay rate is obtained by expanding the 
corrections to the SDCs and the corrections to the wavefunctions at the origin
at the amplitude level:
\begin{eqnarray}
\label{eq:twophotondecay_pNRQCD}
\Gamma(P\to \gamma \gamma) 
&=&
\frac{16 N_c \pi \alpha^2 e_Q^4}{m_P^2} 
|\Psi_P^{\rm LO} (0)|^2
\bigg| 
\bigg[
1
+ \alpha_s c_{\gamma \gamma}^{(1)}
+ \delta_\Psi^{\rm C1} + \delta_\Psi^{\rm C2} + \delta_\Psi^{\rm RS'}
+ \delta_\Psi^{\rm NC}|_{\bm{S}^2=0} 
\nonumber \\ && 
+ \alpha_s^2 c_{\gamma \gamma}^{(2)}
+ \alpha_s c_{\gamma \gamma}^{(1)} \delta_\Psi^{\rm C1}
+ \frac{2 E_P^{\rm LO}}{m_P} \left(d_{\gamma \gamma} 
- \frac{{\cal E}_3}{9}
\right)
+O(\alpha_s^3, v^3,\Lambda_{\rm QCD}^2/m^2)
\bigg]
\bigg|^2,
\quad
\quad 
\end{eqnarray}
where $c_{\gamma \gamma} = 1+ \alpha_s c_{\gamma \gamma}^{(1)} 
+ \alpha_s^2 c_{\gamma \gamma}^{(2)} + O(\alpha_s^3)$, and 
$\alpha_s = \alpha_s (\mu_R)$. 
This expression is valid up to corrections of relative order $\alpha_s^3$, 
$v^3$, and $\Lambda_{\rm QCD}^2/m^2$. 
We set $n_f=3$ in the SDCs, and use 
$e_Q = 2/3$ for charm. We choose $\alpha = 1/137$, because the QED coupling 
constant in eq.~(\ref{eq:twophotondecay_pNRQCD}) is associated with on-shell
photons in the final state. 
The dependence on the $\overline{\rm MS}$ scale $\Lambda$ in
$\delta_\Psi^{\rm NC}|_{\bm{S}^2=0}$ cancels completely
with the $\Lambda$ dependence in $\alpha_s^2 c_{\gamma \gamma}^{(2)}$,
while the $\Lambda$ dependence in the order-$\alpha_s$ correction to
$d_{\gamma \gamma}$ cancels with the scale dependence of the correlator 
${\cal E}_3$. 
Similarly to the case of decay constants, 
variation of the factorization scale $\Lambda$ has almost no effect 
in eq.~(\ref{eq:twophotondecay_pNRQCD}). 
We set the scale $\Lambda = m$ in the one-loop correction to
$d_{\gamma \gamma}$,
and compute the correlator ${\cal E}_3$ at scale $m$ using the renormalization
group improved expression in eq.~(\ref{eq:e3_evolve}).

The numerical result for the two-photon decay rate of $\eta_c$ is
\begin{equation}
\label{eq:etacdecay_result}
\Gamma(\eta_c \to \gamma \gamma) =
6.8 {}^{+0.4}_{-0.0} {}^{+0.2}_{-0.1} 
{}^{+2.8}_{-2.3} 
\pm 1.0
\textrm{~keV}
= 6.8 
{}^{+3.0}_{-2.5} 
\textrm{~keV},
\end{equation}
where the uncertainties are as in eq.~(\ref{eq:fJpsi_result}). 
This result is compatible within uncertainties with the PDG value of the
two-photon decay rate 
$\Gamma(\eta_c \to \gamma \gamma) = 5.06 \pm 0.34$~keV 
in ref.~\cite{Tanabashi:2018oca}. 
The central value for the decay rate that we obtain
is not very different from the leading-order value 
$\Gamma(\eta_c \to \gamma \gamma)|^{\rm LO} 
= 16 N_c \pi \alpha^2 e_Q^4 |\Psi_{\eta_c}^{\rm LO} (0)|^2/m_{\eta_c}^2
= 6.0$~keV. 
If we had ignored the corrections to the wavefunctions at the origin, 
the one-loop correction would have been $-15\%$ of the leading order result
at the amplitude level, 
while the two-loop correction would have been $-46\%$ of the leading order
amplitude, 
so that the effect of the loop corrections would add up to $-61\%$ at
two-loop accuracy at the amplitude level.
In contrast, the 
order-$\alpha_s$ corrections in eq.~(\ref{eq:twophotondecay_pNRQCD})
from $\alpha_s c_{\gamma \gamma}^{(1)}$ and $\delta_\Psi^{\rm C1}$ amount to 
about $3\%$, 
and the corrections proportional to $E_{\eta_c}^{\rm LO}$ is
about $-10\%$ compared to the leading-order amplitude.
The remaining corrections from $\alpha_s^2 c_v^{(2)}$,
$\delta_\Psi^{\rm NC}$, $\delta_\Psi^{\rm C2}$, $\delta_\Psi^{\rm RS'}$,
and the cross term $\alpha_s c_v^{(1)} \delta_\Psi^{\rm C1}$ add up to about
$+14\%$ of the leading-order amplitude, 
so that the central value for the decay rate that we obtain 
is about $7\%$ larger than the leading-order value at the amplitude level,
or about $13\%$ larger at the squared amplitude level. 
Inclusion of the corrections to the wavefunctions at the origin in
eq.~(\ref{eq:twophotondecay_pNRQCD}) has improved the convergence of the 
corrections of higher orders in $\alpha_s$ and $v$. 
If we compute the decay rate by expanding the expression in
eq.~(\ref{eq:twophotondecay_pNRQCD}) at the squared amplitude level, as we have done
for the leptonic decay rates of $J/\psi$ and $\psi(2S)$, we obtain 
$\Gamma(\eta_c \to \gamma \gamma) = 6.7^{+3.0}_{-2.5}$~keV, which agrees well
with eq.~(\ref{eq:etacdecay_result}) within uncertainties.

The numerical result for the two-photon decay rate of $\eta_c(2 S)$ is 
\begin{equation}
\label{eq:etac2Sdecay_result}
\Gamma(\eta_c(2S) \to \gamma \gamma) =
3.0 {}^{+0.4}_{-0.7} {}^{+0.1}_{-0.0}
{}^{+1.3}_{-1.0}
\pm 0.6
\textrm{~keV}
= 3.0
{}^{+1.4}_{-1.3}
\textrm{~keV},
\end{equation}
where the uncertainties are as in eq.~(\ref{eq:fJpsi_result}). 
The central value of the decay rate is about $19\%$ smaller than the 
leading-order value 
$\Gamma(\eta_c(2S) \to \gamma \gamma)|^{\rm LO}
= 16 N_c \pi \alpha^2 e_Q^4 |\Psi_{\eta_c(2S)}^{\rm LO} (0)|^2/m_{\eta_c(2S)}^2
= 3.7$~keV.
If we use the expression for the decay rate obtained by expanding
eq.~(\ref{eq:etacdecay_result}) at the squared amplitude level, we obtain 
$\Gamma(\eta_c(2S) \to \gamma \gamma) = 3.0 ^{+1.4}_{-1.5}$~keV, which agrees 
well with eq.~(\ref{eq:etac2Sdecay_result}) within uncertainties. 
We note that the result for the decay rate in 
eq.~(\ref{eq:etac2Sdecay_result})
disagrees with existing experimental values of the decay
rate in refs.~\cite{Asner:2003wv, Xu:2018uye}, which disagree with each other.

\subsection[Numerical results for $S$-wave bottomonia]
{\boldmath Numerical results for $S$-wave bottomonia}

Now we compute the $\overline{\rm MS}$-renormalized wavefunctions at the origin
for the $1S$, $2S$, and $3S$ bottomonium states. We identify the spin-triplet
states as $\Upsilon(nS)$, while the spin-singlet states correspond to
$\eta_b(nS)$, where $n=1,2,$ and 3. 
The calculations for bottomonia are done similarly as the calculations for 
charmonium states, except that we use the bottom quark $\rm RS'$ mass for $m$, 
set $n_f = 4$, and choose the central value of the QCD renormalization scale 
to be $\mu_R = 5$~GeV. The range for $r_0$ is again determined from numerically
checking the approximate relation in eq.~(\ref{eq:unitary_diff}). We choose the
central value for $r_0$ to be $r_0 = 0.1$~GeV$^{-1}$, and vary $r_0$ between 
$0.05$~GeV$^{-1}$ and $0.2$~GeV$^{-1}$. We set the $\overline{\rm MS}$
scale $\Lambda$ to be the bottom quark mass $m$. 
We use the measured masses of the $\Upsilon(nS)$ and $\eta_b(nS)$ states from 
ref.~\cite{Tanabashi:2018oca}. 
Because the mass of the $\eta_b(3S)$ state has not been measured, 
we estimate $m_{\eta_b(3S)}$
by $m_{\Upsilon(3S)}-(m_{\Upsilon(2S)}-m_{\eta_b(2S)})$, 
assuming that the hyperfine splitting is same for the $2S$ and $3S$ states.

\begin{figure}[tbp]
\centering
\includegraphics[width=.55\textwidth]{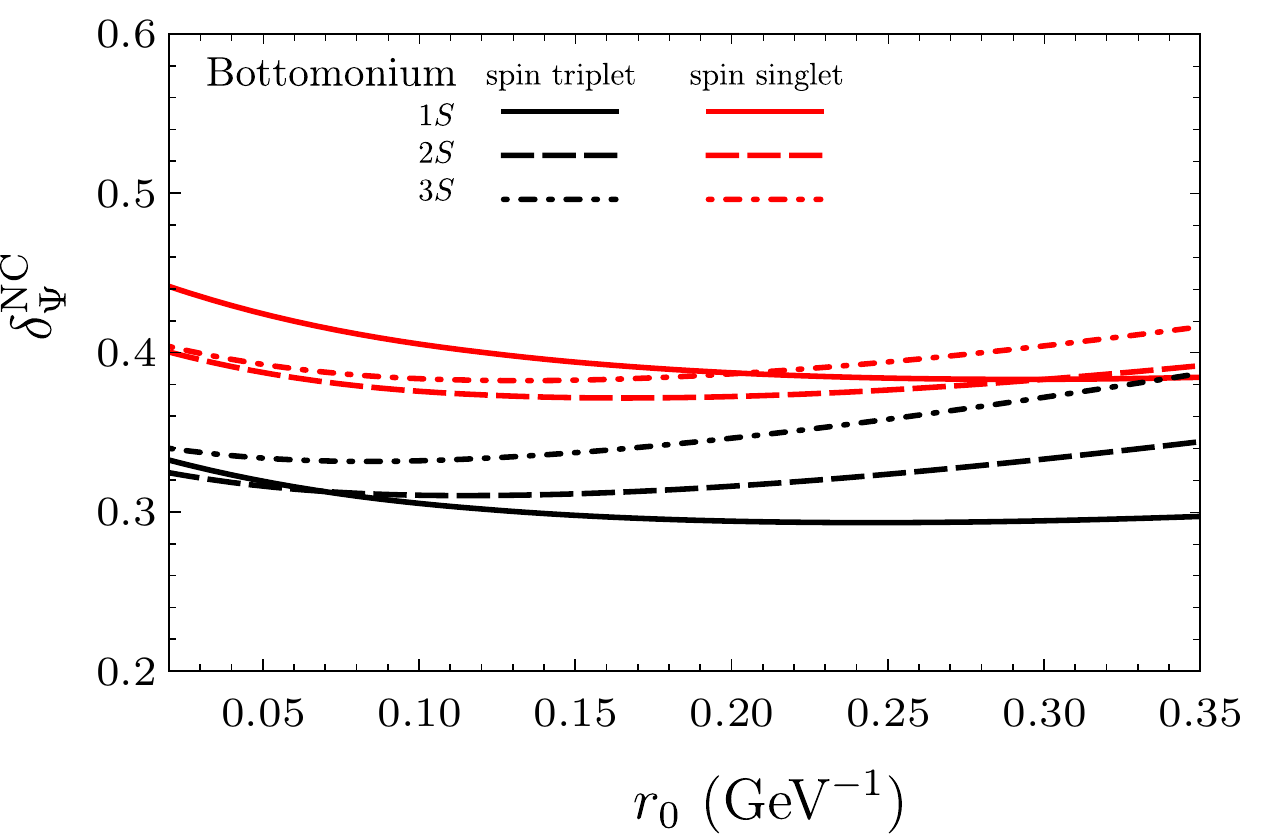}
\caption{\label{fig:bottomr0}
The non-Coulombic corrections $\delta_\Psi^{\rm NC}$ at finite $r_0$ 
for the bottomonium $1S$ (solid lines), $2S$ (dashed lines), 
and $3S$ (dot-dashed lines) states,
for spin triplet (black) and spin singlet (red). The $r_0$ dependences are
mild for the range 0.05~GeV$^{-1}< r_0 <$ 0.2~GeV$^{-1}$ that we consider.
}
\end{figure}

We list the numerical results for the LO wavefunctions at the origin,
the LO binding energies,
and the corrections to the wavefunctions at the origin relative to the
LO wavefunctions at the origin in table~\ref{tab:bottomcorrections}.
The relative corrections $\delta_\Psi^{\rm NC}$, $\delta_\Psi^{\rm C1}$,
$\delta_\Psi^{\rm C2}$, and $\delta_\Psi^{\rm RS'}$ are defined in the previous
section. 
We show the $r_0$ dependence of the non-Coulombic correction
$\delta_\Psi^{\rm NC}$ in fig.~\ref{fig:bottomr0}.

\begin{table}[tbp]
\centering
\begin{tabular}{|c||c|c||c|c|c|c|c|}
\hline
State & $\Psi^{\rm LO} (0)$~(GeV$^{3/2}$) & $E^{\rm LO}$~(GeV) &
$\delta_\Psi^{\rm NC}|_{\bm{S}^2=2}$ & $\delta_\Psi^{\rm NC}|_{\bm{S}^2=0}$ &
$\delta_\Psi^{\rm C1}$ & $\delta_\Psi^{\rm C2}$ & $\delta_\Psi^{\rm RS'}$ \\
\hline
\hline
$1S$ & 0.496 & 0.023 & 0.305 & 0.405 & 0.241 & $-0.018$ & 0.016
\\
\hline
$2S$ & 0.423 & 0.417 & 0.311 & 0.376 & 0.108 & $-0.009$ & 0.013
\\
\hline
$3S$ & 0.400 & 0.723 & 0.332 & 0.384 & 0.068 & $-0.003$ & 0.012
\\
\hline
\end{tabular}
\caption{\label{tab:bottomcorrections}
LO wavefunctions at the origin, LO binding energies
and the relative corrections to the wavefunctions at the origin 
in the $\overline{\rm MS}$ scheme at scale $\Lambda=m$ for $1S$, $2S$, and $3S$
bottomonium states.
$\delta_\Psi^{\rm NC}$ is the correction from the $1/m$ and $1/m^2$ potentials,
$\delta_\Psi^{\rm C1}$ and $\delta_\Psi^{\rm C2}$ are Coulombic corrections at
first and second order in QMPT, respectively,
and $\delta_\Psi^{\rm RS'}$ is the correction from the $\rm RS'$
subtraction term.
The $\delta_\Psi^{\rm NC}$, $\delta_\Psi^{\rm C1}$, $\delta_\Psi^{\rm C2}$,
and $\delta_\Psi^{\rm RS'}$ are dimensionless.
}
\end{table}

The results for the LO binding energies for the $1S$, $2S$ and $3S$ states in
table~\ref{tab:bottomcorrections} are roughly compatible with the mass
differences between $\Upsilon(1S)$, $\Upsilon(2S)$, and $\Upsilon(3S)$ states.
We see that the non-Coulombic corrections from $1/m$ and $1/m^2$ potentials,
given by $\delta_\Psi^{\rm NC}$ in table~\ref{tab:bottomcorrections},
are large and positive for the $1S$, $2S$, and $3S$ states,
although the relative sizes of the corrections are smaller than the case of
charmonium $1S$ and $2S$ states. 
As it was in the case of charmonia, the order-$\alpha_s^2$ corrections to 
the SDCs in appendix~\ref{appendix:sdcs} are large and negative at
$\Lambda = m$, so that if we combine the pNRQCD expressions of the LDMEs
with the SDCs, large cancellations will occur between the
order-$\alpha_s^2$ corrections to the SDCs and the corrections to the
wavefunctions at the origin.
The contribution from the long-distance part of the $1/m$ potential, 
originating from the second term in the expression for the $1/m$ 
potential in eq.~(\ref{eq:1mpotential_longandshort_onshell}),
amounts to about $+4\%$, $+2\%$, and $+2\%$ of the 
LO wavefunction at the origin for the $1S$, $2S$, and $3S$ states,
respectively, which are less than half of the corresponding corrections to the 
charmonium wavefunctions at the origin. 
We note that while the Coulombic corrections at first order are 
positive, 
the Coulombic corrections at second order are small and negative,
signaling good convergence of the Coulombic corrections. The corrections from
the renormalon subtraction term $\delta_\Psi^{\rm RS'}$ are small.

As it was in the case of charmonium, the LO wavefunctions at the origin in
table~\ref{tab:bottomcorrections} are larger than what we obtain if we neglect
the long-distance nonperturbative part of the static potential. 
Neglecting the long-distance part of
the static potential reduces the wavefunctions at the origin for the $2S$ and
$3S$ states by more than factors of 3 and 6, respectively, while for the $1S$
state, the wavefunction at the origin reduces by a factor of about $1.5$. 
At the squared amplitude level, neglecting the long-distance part of the
static potential can reduce the $2S$ bottomonium decay rates by almost an order
of magnitude, and the $3S$ bottomonium decay rates by more than an order of 
magnitude.  Hence, even for the
bottomonium states, the nonperturbative long-distance part of the static
potential is important, especially for the $2S$ and $3S$ states. 

Now we compute the decay constants $f_{\Upsilon(nS)}$, $f_{\eta_b(nS)}$, and 
the electromagnetic decay rates of $\Upsilon(nS)$ and
$\eta_b(nS)$ based on the
bottomonium wavefunctions at the origin that we obtained. 
We use the same pNRQCD expressions for these quantities in
eqs.~(\ref{eq:veccurrent_pNRQCD}), (\ref{eq:pseudoveccurrent_pNRQCD}), and
(\ref{eq:twophotondecay_pNRQCD}) that we used in the previous section for the
charmonium states, except that we set $n_f=4$ in the SDCs, and use 
$e_Q = -1/3$ for bottom. 
We note that the correction terms in the pNRQCD expressions of the LDMEs
in eqs.~(\ref{eq:pNRQCD_vec}) and (\ref{eq:pNRQCD_ps}) that come from the
gluonic correlators may not be valid for $1S$ bottomonium states, because the
assumption $mv \gtrsim \Lambda_{\rm QCD} \gg mv^2$ may not hold for these
states. 
Hence, when we make predictions for the bottomonium $1S$ states, 
we assume that eqs.~(\ref{eq:veccurrent_pNRQCD}), 
(\ref{eq:pseudoveccurrent_pNRQCD}), and (\ref{eq:twophotondecay_pNRQCD}) are 
valid up to corrections of order $v^2$.

The numerical results for the decay constants $f_{\Upsilon(nS)}$ are
\begin{subequations}
\label{eq:Upsilonf_result}
\begin{eqnarray}
f_{\Upsilon(1S)} &=& 
0.621 {}^{+0.045}_{-0.000} {}^{+0.008}_{-0.006} \pm 0.033 \pm 0.062
\textrm{~GeV}
= 0.621 {}^{+0.084}_{-0.070} \textrm{~GeV},
\\
f_{\Upsilon(2S)} &=& 
0.447 {}^{+0.002}_{-0.000} {}^{+0.003}_{-0.003} \pm 0.024 \pm 0.013
\textrm{~GeV}
= 0.447 {}^{+0.028}_{-0.027} \textrm{~GeV},
\\
f_{\Upsilon(3S)} &=&
0.395 {}^{+0.001}_{-0.000} {}^{+0.006}_{-0.000} \pm 0.021 \pm 0.012
\textrm{~GeV}
= 0.395 {}^{+0.025}_{-0.024} \textrm{~GeV},
\end{eqnarray}
\end{subequations}
where the first uncertainties come from varying $\mu_R$ between 2~GeV and
8~GeV, and the second uncertainties come from varying $r_0$ between
0.05~GeV$^{-1}$ and 0.2~GeV$^{-1}$. The third uncertainties take into account
the neglect of the correction $-V_{p^2}^{(2)} (0)/(2 m)$ to the wavefunctions 
at the origin, which we take to be $\pm 500\textrm{~MeV}/(2m)$ times the 
central value. For $f_{\Upsilon(1S)}$, the final uncertainty comes from the
uncalculated order-$v^2$ corrections to the LDME, 
which we take to be 10\% of the central value.
This is based on the typical estimate $v^2 \approx 0.1$ for bottomonium states. 
For $f_{\Upsilon(2S)}$ and $f_{\Upsilon(3S)}$, the final uncertainties come
from the uncalculated corrections of order $v^3$, which we take to be 3\% of 
the central value, based on the typical estimate $v^2 \approx 0.1$. 
We add the uncertainties in quadrature. 

Compared to the LO values $f_{\Upsilon(nS)}^{\rm LO}$, 
the central values in eq.~(\ref{eq:Upsilonf_result}) are $12$\% larger for
$\Upsilon(1S)$, $3\%$ smaller for $\Upsilon(2S)$, and $8\%$ smaller for
$\Upsilon(3S)$. If we had ignored the corrections to the wavefunctions at the
origin, the order-$\alpha_s$ correction would have been $-18$\% of the central
value, while the order-$\alpha_s^2$ correction would have been $-20$\% of the
central value, so that the perturbative corrections to two-loop accuracy would
add up to $-38\%$ of the central value. Similarly to the case of charmonia, 
inclusion of the corrections to the
wavefunctions at the origin reduces the sizes of the corrections considerably,
significantly improving the convergence of the corrections. 

The results for $f_{\Upsilon(1S)}$ and $f_{\Upsilon(2S)}$ 
that we obtain 
agree well within uncertainties with the lattice NRQCD determinations 
$f_{\Upsilon(1S)} = 0.639(31)$~GeV and 
$f_{\Upsilon(2S)} = 0.481(39)$~GeV 
from ref.~\cite{Colquhoun:2014ica}, where the SDCs and the LDMEs are both 
obtained in lattice regularization, avoiding the use of the $\overline{\rm MS}$
scheme. In order to compare  with experimental measurements, 
we compute the leptonic decay rates of $\Upsilon(nS)$ from $f_{\Upsilon(nS)}$
by using eq.~(\ref{eq:leptonic_rate}). We obtain
\begin{subequations}
\label{eq:Upsilon_decayrate}
\begin{eqnarray}
\Gamma(\Upsilon(1S) \to e^+ e^-) &=& 1.11 {}^{+0.32}_{-0.24} \textrm{~keV},
\\
\Gamma(\Upsilon(2S) \to e^+ e^-) &=& 0.54 {}^{+0.07}_{-0.06} \textrm{~keV},
\\
\Gamma(\Upsilon(3S) \to e^+ e^-) &=& 0.41 {}^{+0.05}_{-0.05} \textrm{~keV},
\end{eqnarray}
\end{subequations}
where we used $\alpha = 1/131$, which is computed at the scale of the
$\Upsilon(nS)$ mass. These results agree within uncertainties
with the experimental values 
$\Gamma(\Upsilon(1S) \to e^+ e^-) = 1.340 \pm 0.018$~keV, 
$\Gamma(\Upsilon(2S) \to e^+ e^-) = 0.612 \pm 0.011$~keV, 
and $\Gamma(\Upsilon(3S) \to e^+ e^-) = 0.443 \pm 0.008$~keV 
in ref.~\cite{Tanabashi:2018oca}. 
If we use the expressions for the decay rates expanded at the squared 
amplitude level, we obtain 
$\Gamma(\Upsilon(1S) \to e^+ e^-) = 1.10 {}^{+0.23}_{-0.16}$~keV, 
$\Gamma(\Upsilon(2S) \to e^+ e^-) = 0.54 \pm 0.06$~keV, 
and 
$\Gamma(\Upsilon(3S) \to e^+ e^-) = 0.41 \pm 0.05$~keV, 
which agree well with the results in eq.~(\ref{eq:Upsilon_decayrate})
within uncertainties.

We note that the result for $\Gamma(\Upsilon(1S) \to e^+ e^-)$ 
that we obtain also agrees well with the perturbative QCD prediction 
at third order in ref.~\cite{Beneke:2014qea}.
However, the convergence of the perturbative expansion in the 
perturbative QCD calculation is poor; according to ref.~\cite{Beneke:2014qea},
the size of the corrections at first and second order, when combined, 
exceeds the leading-order result, 
while the third order correction is moderate. 
In contrast, in the calculation of the decay rate 
$\Gamma(\Upsilon(1S) \to e^+ e^-)$ in this work, 
the loop corrections to the SDCs and 
the corrections to the wavefunction at the origin combine to be 
$25$\% of the leading order result at the squared amplitude level. 
It seems that 
the improvement of the convergence has mostly to do with the 
Coulombic corrections, because the non-Coulombic correction 
$\delta_{\Psi}^{\rm NC}$ does not change much from the result in
table~\ref{tab:bottomcorrections} when we neglect the 
long-distance, nonperturbative part of the LO potential given by the 
second term in eq.~(\ref{eq:staticpotential_leadingorder}). 
Hence, the convergence of the perturbative QCD calculation may improve if 
the logarithms associated with the loop corrections to the static potential 
are resummed, as we have done in computing the Coulombic corrections. 

The numerical results for the decay constants $f_{\eta_b(nS)}$ are
\begin{subequations}
\label{eq:etabf_result}
\begin{eqnarray}
f_{\eta_b(1S)} &=&
0.691 {}^{+0.117}_{-0.015} {}^{+0.010}_{-0.010} 
\pm 0.037 \pm 0.069
\textrm{~GeV}
= 0.691{}^{+0.141}_{-0.080} \textrm{~GeV},
\\
f_{\eta_b(2S)} &=&
0.471 {}^{+0.006}_{-0.004} {}^{+0.005}_{-0.002} 
\pm 0.025 \pm 0.014 \textrm{~GeV}
= 0.471 {}^{+0.030}_{-0.029} \textrm{~GeV},
\\
f_{\eta_b(3S)} &=&
0.403 {}^{+0.000}_{-0.002} {}^{+0.004}_{-0.000} 
\pm 0.021 \pm 0.012 \textrm{~GeV}
= 0.403 {}^{+0.026}_{-0.025} \textrm{~GeV},
\end{eqnarray}
\end{subequations}
where the uncertainties are as in eq.~(\ref{eq:Upsilonf_result}). 
We neglect the small uncertainty from the scale dependence of the correlator
${\cal E}_3$. 
We add the uncertainties in quadrature.
Compared to the LO results $f_{\eta_b(nS)}^{\rm LO}$, 
the central values in eq.~(\ref{eq:etabf_result}) are $23$\% larger for
$\eta_b(1S)$, $1\%$ larger for $\eta_b(2S)$, and $7\%$ smaller for
$\eta_b(3S)$. If we had ignored the corrections to the wavefunctions at the
origin, the order-$\alpha_s$ correction would have been $-14$\% of the central
value, while the order-$\alpha_s^2$ correction would have been $-24$\% of the
central value, so that the perturbative corrections to two-loop accuracy would
add up to $-38\%$ of the central value. Inclusion of the corrections to the
wavefunctions at the origin reduces the sizes of the corrections considerably,
especially for $\eta_b(2S)$ and $\eta_b(3S)$, 
greatly improving the convergence of the corrections.

The numerical results for the decay rates 
$\Gamma(\eta_b(nS) \to \gamma \gamma)$ are 
\begin{subequations}
\label{eq:etabdecay_result}
\begin{eqnarray}
\Gamma(\eta_b(1S) \to \gamma \gamma) &=&
0.433 {}^{+0.142}_{-0.016} {}^{+0.013}_{-0.012} {}^{+0.047}_{-0.045} \pm 0.043
\textrm{~keV}
= 0.433{}^{+0.165}_{-0.065} \textrm{~keV},
\\
\Gamma(\eta_b(2S) \to \gamma \gamma) &=&
0.194 {}^{+0.003}_{-0.002} {}^{+0.004}_{-0.001} {}^{+0.021}_{-0.020} 
\pm 0.006
\textrm{~keV}
= 0.194 {}^{+0.022}_{-0.021} \textrm{~keV},
\\
\Gamma(\eta_b(3S) \to \gamma \gamma) &=&
0.141 {}^{+0.000}_{-0.005} {}^{+0.003}_{-0.001} {}^{+0.015}_{-0.014} 
\pm 0.004
\textrm{~keV}
= 0.141 {}^{+0.016}_{-0.015} \textrm{~keV},
\end{eqnarray}
\end{subequations}
where the uncertainties are as in $f_{\Upsilon(nS)}$. 
We add the uncertainties in quadrature.
Compared to the LO calculation of the decay rates, the corrections from loop
corrections to the SDCs and the corrections to the wavefunctions at the origin
combine to be about 58\%, 15\%, and 3\% for the $\eta_b(1S)$, $\eta_b(2S)$,
and $\eta_b(3S)$ states, respectively. 
At the amplitude level, the corrections amount
to about 26\%, 7\%, and 2\% for the $\eta_b(1S)$, $\eta_b(2S)$,
and $\eta_b(3S)$ states, respectively. If we had ignored the corrections to the
wavefunctions at the origin, the order-$\alpha_s$ correction would have been 
$-11$\%, and the order-$\alpha_s^2$ correction would have been $-26$\% of the
central value at the amplitude level, so that the loop corrections to two-loop
accuracy would add up to $-38\%$ of the leading-order amplitude. 
By the inclusion of the corrections to the
wavefunctions at the origin, the sizes of the corrections are reduced
considerably for the $\eta_b(2S)$ and $\eta_b(3S)$ states, while the
improvement is moderate for the $\eta_b(1S)$ state. 
If we use the expressions for the decay rates expanded at the squared amplitude
level, we obtain 
$\Gamma(\eta_b(1S) \to \gamma \gamma) = 0.422^{+0.155}_{-0.064}$~keV, 
$\Gamma(\eta_b(2S) \to \gamma \gamma) = 0.196 \pm 0.022$~keV, and 
$\Gamma(\eta_b(3S) \to \gamma \gamma) = 0.142 \pm 0.016$~keV, 
which agree well with the results in eq.~(\ref{eq:etabdecay_result}).

\section{\boldmath Summary and discussion}
\label{sec:summary}

In this paper, we have computed the wavefunctions at the origin of $S$-wave
heavy quarkonia in the $\overline{\rm MS}$ renormalization scheme. 
We include the nonperturbative long-distance contributions to the potential, 
which are neglected in perturbative QCD calculations. 
We compute corrections to the wavefunctions at the origin at subleading orders
in $1/m$ in position space, where the ultraviolet divergences are regulated
by using finite-$r$ regularization. The position-space expressions for the
corrections to the wavefunctions at the origin are given in
section~\ref{sec:wavefunctions}.
The wavefunctions at the origin in 
finite-$r$ regularization is then converted to the 
$\overline{\rm MS}$ scheme by computing the scheme conversion
in perturbative QCD. The result for the scheme conversion coefficient 
is given in section~\ref{sec:conversion}. 
We use the results for the wavefunctions at the origin to
make first-principles based, model-independent predictions of 
decay constants and electromagnetic decay rates of 
$S$-wave charmonium and bottomonium states in section~\ref{sec:results}. 

The predictions for the electromagnetic decay rates of $J/\psi$, $\psi(2S)$, 
$\eta_c$, and $\Upsilon(nS)$ states in this work 
agree with experimental measurements within uncertainties. 
The predictions for the $J/\psi$ and $\eta_c$ decay constants 
agree within uncertainties with lattice QCD calculations in 
ref.~\cite{Hatton:2020qhk}, 
which make use of relativistic charm quarks. 
The predictions for the $\Upsilon(1S)$ and $\Upsilon(2S)$ decay 
constants agree
with the lattice NRQCD determinations in ref.~\cite{Colquhoun:2014ica}, 
where lattice regularization is used to compute both the 
short-distance coefficients and the NRQCD matrix elements. 

The calculation of the wavefunctions at the origin in this work 
contains several
improvements compared to existing model dependent methods. 
First of all, in this work, we include potentials at leading and
subleading orders in $1/m$, which are determined by perturbative QCD at 
short distances, while their nonperturbative behaviors at long distances
are fixed by lattice QCD. Secondly, the ultraviolet divergences that appear in
corrections to the wavefunctions at the origin are properly renormalized 
in the $\overline{\rm MS}$ scheme, so that the wavefunctions at
the origin that we obtain have the correct scale dependences that 
are expected from perturbative QCD. 
Finally, the ambiguity in the heavy quark pole mass is removed by the 
use of the modified renormalon subtracted mass, 
whose numerical values are accurately known. 
These improvements are generally not possible in potential-model calculations.

Because the wavefunctions at the origin that we have computed have the correct
scale dependence that reproduce the anomalous dimensions of NRQCD LDMEs, 
the dependences on the NRQCD factorization scale cancel completely 
through two-loop order in the 
pNRQCD expressions for the decay constants and electromagnetic decay rates. 
Together with the order-by-order cancellation of the dependence on the QCD
renormalization scale, this greatly reduces the uncertainty associated with 
scale dependences. 

A novel feature of the calculation of the decay constants and decay rates 
in this work is
that large cancellations occur between 
the corrections to the wavefunctions at the origin and the perturbative
corrections to the short-distance coefficients. 
These cancellations substantially improve 
the convergence of the expansion in $\alpha_s$ and $v$. 
This may have important implications in understanding the appearance of 
large perturbative corrections in calculations of 
short-distance coefficients in the $\overline{\rm MS}$
scheme. A possible explanation of the cancellations is that, 
due to the confining nature of the nonperturbative potentials, 
including the long-distance contributions to the potentials 
in calculating the wavefunctions at the origin 
may have the effect of introducing an infrared cutoff, 
so that the renormalon ambiguities associated with the
infrared contributions of loop corrections in perturbative QCD are resolved. 

The pNRQCD expressions of the wavefunctions at the origin, as well as the 
decay constants and decay rates, depend on gluonic correlators, whose values 
are in general not well known. Especially, the correction from the
velocity-dependent potential at zero distance, which is given by a gluonic
correlator whose size is of order $\Lambda_{\rm QCD}$, 
is the largest source of uncertainties, 
with the exception of the bottomonium $1S$ states. 
Improved determinations of the gluonic correlators, 
which can in principle be done in lattice QCD, 
will be necessary in further reducing the uncertainties. 

The calculation of the renormalization of the wavefunctions at the 
origin in this work is 
accurate to two-loop accuracy. In principle, the calculation in this
work can be extended to
three-loop accuracy, by computing the divergent corrections to the
wavefunctions at the origin to second order in the quantum-mechanical
perturbation theory, and computing the scheme conversion from finite-$r$
regularization to the $\overline{\rm MS}$ scheme to order-$\alpha_s^3$
accuracy. Such a calculation will make possible the inclusion of the
long-distance nonperturbative contributions to the potentials of order $1/m^2$,
because the second-order correction in the quantum-mechanical perturbation
theory is necessary in extending the calculation of the 
unitary transformation between on-shell matching and Wilson-loop matching 
in section~\ref{sec:conversion} to order-$1/m^2$ accuracy. 

Finally, we note that the calculation in this paper may be extended to states
with nonzero orbital angular momentum. This necessarily involves considering
the orbital angular momentum dependent terms in the potential, as well as
the angular dependence of the wavefunctions in dimensional regularization, 
which were not present in this work thanks to the rotational symmetry of the
$S$-wave states. Such calculations will allow us to make accurate predictions 
of production and decay rates of $P$-wave heavy quarkonium states.

\acknowledgments

The author is grateful to Nora Brambilla and Antonio Vairo for fruitful
discussions and their encouragement in completing this work. 
This work is supported by Deutsche Forschungsgemeinschaft 
(DFG, German Research Foundation) cluster of excellence ``ORIGINS'' under
Germany's Excellence Strategy - EXC-2094 - 390783311.

\appendix

\section{Anomalous dimensions}
\label{sec:anomalous_dimensions}

In this appendix, we compute the anomalous dimensions of the NRQCD LDMEs 
that are given in eqs.~(\ref{eq:RG_vec}) and (\ref{eq:RG_ps}).
Although the results 
are available in refs.~\cite{Bodwin:1994jh, Czarnecki:1997vz,Beneke:1997jm,
Czarnecki:2001zc,Kniehl:2006qw}, 
it can be useful to compute them through loop calculations
in NRQCD, because such calculations can reveal the origins of the anomalous
dimensions which are obscure in perturbative QCD calculations of SDCs. 

The anomalous dimensions can be computed as perturbation series in $\alpha_s$ 
by replacing the quarkonium states in the definitions of the LDMEs 
by perturbative $Q \bar Q$ states, and
computing loop corrections to the LDMEs, with the vertices coming from the
operators in the 
NRQCD Lagrangian. The $Q$ and $\bar Q$ in the $Q \bar Q$ states are on shell,
which have 
nonrelativistic 4-momenta $(E, \bm{q})$ and $(E, -\bm{q})$, respectively,
where $E = \bm{q}^2/(2 m)$. 
We work in Coulomb gauge, and use the NRQCD Feynman rules given 
in ref.~\cite{Bodwin:1998mn}. 
We use DR in $d=4-2 \epsilon$ spacetime dimensions, where the 
anomalous dimensions are simply given by the coefficients of the $1/\epsilon$ 
poles that are associated with UV divergences. 

The NRQCD loop integrals are evaluated in the following way.  First, we
integrate over the temporal components of the loop momenta, using contour
integration. Then, we expand the integrand in powers of $1/m$, which is
necessary in preserving the nonrelativistic power counting in DR. 
Finally, we integrate over the spatial components of the loop momenta, 
regulating the resulting divergences in DR. 
The anomalous dimension is given by the coefficients of the single UV poles, 
after
differentiating and multiplying by $g_s = \sqrt{4 \pi \alpha_s}$.

\subsection[One-loop anomalous dimension at relative order $v^2$]
{\boldmath One-loop anomalous dimension at relative order $v^2$}

\begin{figure}[tbp]
\centering
\includegraphics[width=.80\textwidth]{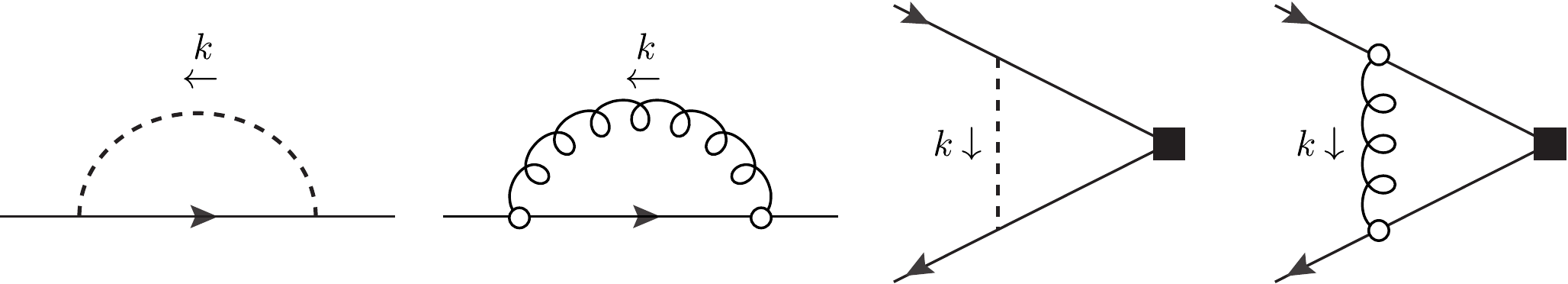}
\caption{\label{fig:oneloop}
Feynman diagrams for one-loop corrections to the NRQCD LDMEs. 
Solid lines are heavy quarks and antiquarks, dashed lines are temporal gluons, 
and curly lines are transverse gluons. 
Open circles represent insertions of the $\bm{p} \cdot \bm{A}$ vertex, 
and filled squares represent 
the operator $\chi^\dag \bm{\epsilon} \cdot \bm{\sigma} \psi$ for spin triplet,
and $\chi^\dag \psi$ for spin singlet. 
}
\end{figure}

We first consider the one-loop corrections to the NRQCD LDMEs 
$\langle 0 | \chi^\dag \bm{\epsilon} \cdot \bm{\sigma} \psi | Q \bar Q \rangle$
and $\langle 0 | \chi^\dag \psi | Q \bar Q \rangle$, which 
come from the Feynman diagrams in fig.~\ref{fig:oneloop}. 
Because the one-loop integrals that we compute only involve single poles in
$\epsilon$, we may set $\epsilon=0$ in the loop integrands 
without affecting the $1/\epsilon$ poles. 

The one-loop heavy quark self energy from the first two diagrams in 
fig.~\ref{fig:oneloop} reads
\begin{equation}
\Sigma(E,\bm{q}) = 
2 \pi \alpha_s C_F \int_{\bm{k}} \frac{1}{\bm{k}^2}
+ \frac{4 \pi \alpha_s C_F}{m^2} 
\int_{\bm{k}} \frac{\bm{q}^2-(\bm{q} \cdot \hat{\bm{k}})^2}
{(2 |\bm{k}|-i \varepsilon)
[E-|\bm{k}|-(\bm{q}+\bm{k})^2/(2 m) + i \varepsilon]},
\end{equation}
where the first and second terms come from exchange of temporal and 
spatial gluons, respectively. 
The first term is scaleless power divergent, and therefore can be discarded. 
Since the second term already has a factor of $\bm{q}^2/m^2$, we can expand in
powers of $1/m$ and keep only the leading contribution. 
To find the one-loop correction to the quark field renormalization factor 
$Z_Q^{\rm NRQCD}$, 
we differentiate $\Sigma(E,\bm{q})$ by $E$ and 
set $E = \bm{q}^2/(2 m)$ to obtain
\begin{equation}
\label{eq:oneloopgam1}
Z_Q^{\rm NRQCD} = 
1 - \frac{2 \pi \alpha_s C_F}{m^2} 
\int_{\bm{k}} \frac{\bm{q}^2-(\bm{q} \cdot \hat{\bm{k}})^2}
{|\bm{k}|^3}  
= 
1 - \frac{\alpha_s C_F}{3 \pi} \frac{\bm{q}^2}{m^2} 
\frac{1}{\epsilon_{\rm UV}} + \cdots, 
\end{equation}
where we only keep the UV pole in the last equality. We use the subscript UV to
make clear that the pole is associated with a UV divergence. 

The vertex correction diagram from exchange of a temporal gluon gives 
\begin{equation}
4 \pi \alpha_s C_F m \int_{\bm{k}} 
\frac{1}{\bm{k}^2 (\bm{k}^2 + 2 \bm{q} \cdot \bm{k} -i \varepsilon)}, 
\end{equation}
which does not have a UV divergence, and hence does not contribute to the
anomalous dimension. The transverse-gluon exchange diagram gives
\begin{equation}
\label{eq:vertex_spatial}
\frac{2 \pi \alpha_s C_F}{m^2} 
\int_{\bm{k}} \frac{\bm{q}^2 - (\bm{q} \cdot \hat{\bm{k}})^2}
{A (\bm{k}^2- A^2 -i \varepsilon)}
- 
\frac{2 \pi \alpha_s C_F}{m^2} 
\int_{\bm{k}} \frac{\bm{q}^2 - (\bm{q} \cdot \hat{\bm{k}})^2}
{ (|\bm{k}| -i \varepsilon) (|\bm{k}| + A-i \varepsilon) (|\bm{k}| -A-i
\varepsilon)}, 
\end{equation}
where $A = (\bm{k}+\bm{q})^2/(2 m) - \bm{q}^2/(2 m) -i \varepsilon$.
Here, the first term comes from the residue of the pole from the quark
propagator, and the second term comes from the residue of the pole from the
gluon propagator. Since eq.~(\ref{eq:vertex_spatial}) already has a factor of
$\bm{q}^2/m^2$, we can keep only the leading contribution in the $1/m$
expansion, which gives 
\begin{equation}
\label{eq:oneloopgam2}
\frac{2 \pi \alpha_s C_F}{m^2}
\int_{\bm{k}} \frac{\bm{q}^2 - (\bm{q} \cdot \hat{\bm{k}})^2}
{A (\bm{k}^2 -i \varepsilon)}
-
\frac{2 \pi \alpha_s C_F}{m^2}
\int_{\bm{k}} \frac{\bm{q}^2 - (\bm{q} \cdot \hat{\bm{k}})^2}
{ |\bm{k}|^3} 
= - \frac{\alpha_s C_F}{3 \pi} \frac{\bm{q}^2}{m^2} \frac{1}{\epsilon_{\rm UV}}
+ \cdots,
\end{equation}
where we only keep the UV pole. We note that the first term on the left-hand
side does not have a UV divergence, and the UV pole comes only from the second
term. It can be shown that if we replace the $\bm{p} \cdot \bm{A}$
vertices with $\bm{\sigma} \cdot \bm{B}$ vertices, the transverse-gluon 
exchange diagram does not produce logarithmic UV divergences. 

Since the diagrams in fig.~\ref{fig:oneloop} 
give rise to logarithmic UV divergences at relative order $\alpha_s v^2$ 
already at leading power in the $1/m$ expansion, 
it is not necessary to consider vertices from higher dimensional
operators in the NRQCD Lagrangian. 
We combine eqs.~(\ref{eq:oneloopgam1}) and (\ref{eq:oneloopgam2})
to find the UV pole in the one-loop correction to the NRQCD LDME 
$\langle 0 | \chi^\dag \bm{\epsilon} \cdot \bm{\sigma} \psi | Q \bar Q
\rangle$, which reads 
\begin{eqnarray}
\langle 0 | \chi^\dag \bm{\epsilon} \cdot \bm{\sigma} \psi | Q \bar Q \rangle
|_{\textrm{one loop}}
&=& - \frac{2 \alpha_s C_F}{3 \pi} 
\frac{\bm{q}^2}{m^2} \frac{1}{\epsilon_{\rm UV}}
\langle 0 | \chi^\dag \bm{\epsilon} \cdot \bm{\sigma} \psi | Q \bar Q \rangle
|_{\textrm{tree}}
+ \cdots
\nonumber \\
&=& - \frac{4 \alpha_s C_F}{3 \pi m^2} 
\frac{1}{2\epsilon_{\rm UV}}
\langle 0 | \chi^\dag \bm{\epsilon} \cdot \bm{\sigma} 
(- \tfrac{i}{2} \overleftrightarrow{\bm{D}})^2
\psi | Q \bar Q \rangle
|_{\textrm{tree}}
+ \cdots, 
\end{eqnarray}
where we used $\langle 0 | \chi^\dag \bm{\epsilon} \cdot \bm{\sigma}
(- \tfrac{i}{2} \overleftrightarrow{\bm{D}})^2
\psi | Q \bar Q \rangle
= \bm{q}^2 \langle 0 | \chi^\dag \bm{\epsilon} \cdot \bm{\sigma} \psi | Q \bar
Q \rangle$ at tree level. 
This confirms the order-$\alpha_s v^2$ term of the anomalous dimension in
eq.~(\ref{eq:RG_vec}). 
Similarly, the UV pole in the one-loop correction to the LDME
$\langle 0 | \chi^\dag \psi | Q \bar Q \rangle$ is
\begin{equation}
\langle 0 | \chi^\dag \psi | Q \bar Q \rangle
|_{\textrm{one loop}}
= - \frac{4 \alpha_s C_F}{3 \pi m^2}
\frac{1}{2\epsilon_{\rm UV}}
\langle 0 | \chi^\dag 
(- \tfrac{i}{2} \overleftrightarrow{\bm{D}})^2
\psi | Q \bar Q \rangle
|_{\textrm{tree}}
+ \cdots,
\end{equation}
which confirms the order-$\alpha_s v^2$ term of the anomalous dimension in
eq.~(\ref{eq:RG_ps}).

Since the logarithmic UV divergences in the vertex correction diagrams come
from the contribution from the gluon pole, there are no contributions to the
anomalous dimension at relative order $\alpha_s v^2$ 
that comes from gluon exchanges between the quark 
and the antiquark when the virtual quark or the virtual antiquark is on 
shell. Hence, at one loop level, there is no
contribution to the anomalous dimension that comes from exchanges of potentials
between the $Q$ and $\bar Q$. 
This is consistent with the pNRQCD expressions of the LDMEs in
eqs.~(\ref{eq:pNRQCD_vec}) and (\ref{eq:pNRQCD_ps}), where the
one-loop anomalous dimension comes from the gluonic correlator ${\cal E}_3$,
and not from corrections to the wavefunctions at the origin.

\subsection[Two-loop anomalous dimension at leading order in $v$]
{\boldmath Two-loop anomalous dimension at leading order in $v$}

\begin{figure}[tbp]
\centering
\includegraphics[width=1.0\textwidth]{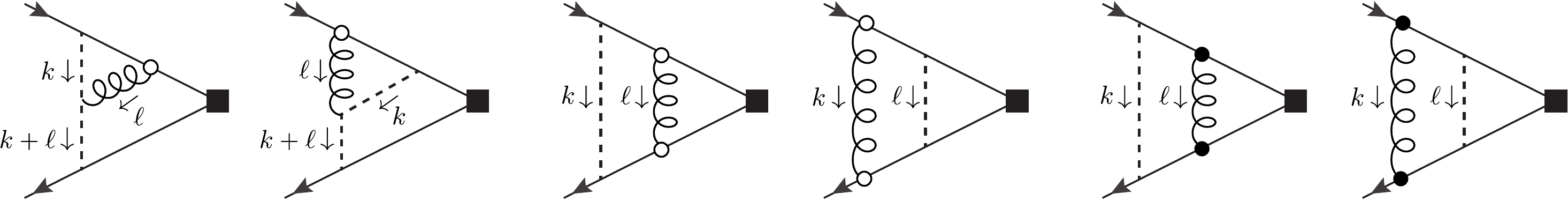}
\caption{\label{fig:twoloop}
Feynman diagrams for two-loop corrections to the NRQCD LDMEs that produce
logarithmic UV divergences. 
There are additional diagrams that can be obtained
from charge conjugation, which we do not show here. 
Solid lines are heavy quarks and antiquarks, dashed lines are temporal gluons,
and curly lines are transverse gluons. 
Open and filled circles represent insertions of the 
$\bm{p} \cdot \bm{A}$ vertex and the spin-dependent 
$\bm{\sigma} \cdot \bm{B}$ vertex, respectively. 
Filled squares represent
the operator $\chi^\dag \bm{\epsilon} \cdot \bm{\sigma} \psi$ for spin triplet,
and $\chi^\dag \psi$ for spin singlet.
}
\end{figure}

Now let us consider the UV divergences in the two-loop corrections to the NRQCD
LDMEs. Since we work at leading order in $v$, we can set the relative
momentum $\bm{q}$ between the quark and antiquark to zero. 
The two-loop diagrams that contain logarithmic UV divergences are shown in
fig.~\ref{fig:twoloop}. We neglect diagrams that do not contain logarithmic UV
divergences, most of which are scaleless power divergent, and hence vanish in
DR.
As we will see later, in the Coulomb gauge the diagrams only involve single 
poles in $\epsilon$, and hence, we can set $\epsilon=0$ in the numerators of 
loop integrands without affecting the $1/\epsilon$ poles. 
The non-Abelian diagrams yield 
\begin{equation}
\label{eq:twoloopcorr_nonabelian}
16 \pi^2 \alpha_s^2
C_A
C_F 
\int_{\bm{k}} \int_{\bm{\ell}}
\frac{
\bm{k}^2 - (\bm{k} \cdot \hat{\bm{\ell}})^2
}{\bm{\ell}^2  \bm{k}^2 (\bm{k}+\bm{\ell})^4} 
= 
\frac{\alpha_s^2 C_A C_F }{8} 
\frac{1}{\epsilon_{\rm UV}} + \cdots, 
\end{equation}
where we only keep the UV pole. 
The spin-independent ladder diagrams yield 
\begin{equation}
\label{eq:twoloopcorr_spinindep}
16 \pi^2 \alpha_s^2
C_F^2 
\int_{\bm{k}} \int_{\bm{\ell}}
\frac{
\bm{k}^2 - (\bm{k} \cdot \hat{\bm{\ell}})^2 
}{\bm{k}^4 \bm{\ell}^2 (\bm{k}+\bm{\ell})^2}
=
\frac{\alpha_s^2 C_F^2 }{8}
\frac{1}{\epsilon_{\rm UV}} + \cdots,
\end{equation}
where again we only keep the UV pole. 
The spin-dependent ladder diagrams give 
\begin{equation}
\label{eq:twoloopcorr_spindep}
4 \pi^2 \alpha_s^2
C_F^2
\int_{\bm{k}} \int_{\bm{\ell}}
\left( 
\frac{
\delta^{il} - \hat{\ell}^i \hat{\ell}^l 
}{\bm{k}^4 \bm{\ell}^2 (\bm{k}+\bm{\ell})^2}
+ \frac{ \delta^{il} - \hat{\ell}^i \hat{\ell}^l 
}{\bm{k}^2 \bm{\ell}^4 (\bm{k}+\bm{\ell})^2}
\right)
\epsilon_{ijk} \ell_j \sigma_k \otimes \sigma_n \epsilon_{lmn} \ell_m ,
\end{equation}
where we use the notation $\otimes$ to make clear that the 
Pauli matrix on the left acts on the quark line, 
while the one on the right acts on the antiquark line. 
We first integrate over $\bm{k}$, and average over the angles of $\bm{\ell}$ 
to obtain the following UV-divergent 
contribution from the spin-dependent ladder diagrams:
\begin{equation}
\label{eq:twoloopcorr_spindepresult}
\pi^2 \alpha_s^2
C_F^2
\int_{\bm{\ell}}
\frac{1}{|\bm{\ell}|^{3+2 \epsilon}}
\frac{\sigma_i \otimes \sigma_i }{3} 
= 
\frac{\alpha_s^2 C_F^2}{8}
\frac{1}{\epsilon_{\rm UV}} 
\frac{\sigma_i \otimes \sigma_i}{3} 
+ \cdots, 
\end{equation}
where again we keep only the UV pole. 
The spin-dependent factor $\sigma_i \otimes \sigma_i$ yields, for spin triplet, 
\begin{equation}
\frac{1}{3} 
\sigma_i \bm{\epsilon} \cdot \bm{\sigma} \sigma_i
= - \frac{1}{3} \bm{\epsilon} \cdot \bm{\sigma}, 
\end{equation}
and for spin singlet, 
\begin{equation}
\frac{1}{3} \sigma_i \sigma_i
= 1. 
\end{equation}
We collect the results in eqs.~(\ref{eq:twoloopcorr_nonabelian}),
(\ref{eq:twoloopcorr_spinindep}), and 
(\ref{eq:twoloopcorr_spindepresult}) to obtain 
the logarithmic UV divergence in the two-loop correction to the LDME 
$\langle 0 | \chi^\dag \bm{\epsilon} \cdot \bm{\sigma} \psi | Q \bar Q
\rangle$, which reads 
\begin{equation}
\langle 0 | \chi^\dag \bm{\epsilon} \cdot \bm{\sigma} \psi | Q \bar Q \rangle
|_{\textrm{two loop}}
= 
\alpha_s^2 C_F \left(\frac{C_F}{3} + \frac{C_A}{2}  \right) 
\frac{1}{4 \epsilon_{\rm UV}}
\langle 0 | \chi^\dag \bm{\epsilon} \cdot \bm{\sigma} \psi | Q \bar Q \rangle
|_{\textrm{tree}}
+ \cdots.
\end{equation}
This reproduces the order-$\alpha_s^2$ term of the anomalous dimension in
eq.~(\ref{eq:RG_vec}). 
Similarly, the logarithmic UV divergence in the two-loop correction to the LDME
$\langle 0 | \chi^\dag \psi | Q \bar Q \rangle$ is
\begin{equation}
\langle 0 | \chi^\dag \psi | Q \bar Q \rangle
|_{\textrm{two loop}}
= 
\alpha_s^2 C_F \left( C_F + \frac{C_A}{2}  \right)
\frac{1}{4 \epsilon_{\rm UV}}
\langle 0 | \chi^\dag
\psi | Q \bar Q \rangle
|_{\textrm{tree}}
+ \cdots,
\end{equation}
which agrees with the order-$\alpha_s^2$ term of the anomalous dimension in
eq.~(\ref{eq:RG_ps}).

We note that in the calculation of the two-loop diagrams in 
fig.~\ref{fig:twoloop}, 
at least one of the integrations over the temporal components of the loop 
momenta must involve residues of the poles from the quark or antiquark
propagators in order to produce logarithmic UV divergences. 
This is clear in the ladder diagrams, because the temporal-gluon propagator 
does not have a pole in the temporal components of loop momenta.
For the non-Abelian diagrams, it can be shown that if we neglect the
pole that comes from the transverse gluon propagator in the integration over
$\ell_0$, we obtain the same UV pole as in 
eq.~(\ref{eq:twoloopcorr_nonabelian}). 
This shows that the two-loop anomalous dimensions come solely from exchanges 
of potentials between the $Q$ and $\bar Q$. 
This is consistent with the pNRQCD expressions of the NRQCD LDMEs in
eqs.~(\ref{eq:pNRQCD_vec}) and (\ref{eq:pNRQCD_ps}), which imply that the 
two-loop anomalous dimension can only come from the wavefunctions at the
origin.

\section{Potentials in perturbative QCD} 
\label{appendix:potentials}

In this appendix, we list the short-distance behaviors of the potentials, which
are obtained from perturbative QCD. 
In perturbative QCD, the static potential is given through relative order
$\alpha_s^2$ by~\cite{Fischler:1977yf, Schroder:1998vy} 
\begin{equation}
\label{eq:staticpotential_pert}
V^{(0)} (r) \big|_{\rm pert}
= - \frac{\alpha_s (\mu) C_F}{r} 
\left[ 1 + \sum_{n=1}^2 \left( \frac{\alpha_s(\mu)}{4 \pi} \right)^n 
a_n(r;\mu) \right] + O(\alpha_s^3), 
\end{equation}
where $\alpha_s = \alpha_s(\mu)$ is the $\overline{\rm MS}$-renormalized QCD
coupling constant at scale $\mu$, and the functions $a_n (r;\mu)$ are defined 
by
\begin{subequations}
\label{eq:coulombcorr}
\begin{eqnarray}
a_1 (r;\mu) &=& \frac{31 C_A - 20 T_F n_f}{9} + 
2 \beta_0 \log (\mu e^{\gamma_{\rm E}} r), 
\\
a_2 (r;\mu) &=& \frac{400 n_f^2 T_F^2}{81} 
- C_F T_F n_f \left( \frac{55}{3} - 16 \, \xi (3) \right) \nonumber \\
&& + C_A^2 \left( \frac{4343}{162} + \frac{16 \pi^2-\pi^4}{4} 
+ \frac{22 \, \xi(3)}{3} \right) 
-C_A T_F n_f \left( \frac{1798}{81} + \frac{56 \, \xi(3)}{3} \right) 
\nonumber \\
&& + \frac{\pi^2}{3} \beta_0^2 + 
\left( 4 \bar a_1 \beta_0+2 \beta_1 \right) 
\log (\mu e^{\gamma_{\rm E}} r)
+ 4 \beta_0^2 \log^2 (\mu e^{\gamma_{\rm E}} r),
\end{eqnarray}
\end{subequations}
with $\beta_0 = \frac{11}{3} C_A - \frac{4}{3} T_F n_f$, 
$\beta_1 = \frac{34}{3} C_A^2- \frac{20}{3} C_A T_F n_f - 4 C_F T_F n_f$, 
$T_F = \frac{1}{2}$, 
$\gamma_{\rm E}$ is the Euler-Mascheroni constant, 
$n_f$ is the number of light quark flavors, 
and $\bar a_1 = a_1 (r=e^{-\gamma_{\rm E}}/\mu;\mu)$.
We note that the dependence on $\mu$ cancels order by order 
in eq.~(\ref{eq:staticpotential_pert}). 
The corrections at relative order $\alpha_s^3$ have been computed in
refs.~\cite{Brambilla:1999qa, Kniehl:1999ud, Smirnov:2008pn, Anzai:2009tm,
Smirnov:2009fh}, and the position-space expression can be found in
ref.~\cite{Pineda:2011dg}.  

The forms of the $1/m$ and $1/m^2$ potentials generally depend on the 
matching scheme in which the potentials are determined. 
In on-shell matching, where we match on-shell $S$-matrix elements in
NRQCD and pNRQCD in momentum space, we obtain~\cite{Gupta:1982kp, 
Pantaleone:1987qh, Titard:1993nn, Manohar:2000hj, 
Kniehl:2001ju, Kniehl:2002br,Beneke:1999qg,Beneke:2013jia}
\begin{subequations}
\label{eq:potentials_pert}
\begin{eqnarray}
V^{(1)} (r) \big|_{\rm pert}^{\rm OS}
&=& \frac{\alpha_s^2 C_F (\tfrac{1}{2} C_F - C_A)}{2 r^2}
 + O(\alpha_s^3), 
\\
V_r^{(2)} (r) \big|_{\rm pert}^{\rm OS}
&=& 0 + O(\alpha_s^2), \\
\label{eq:velpotential_pert}
V_{p^2}^{(2)} (r) \big|_{\rm pert}^{\rm OS}
&=& - \frac{\alpha_s C_F}{r} + O(\alpha_s^2), \\
\label{eq:spinpotential_pert}
V_{S^2}^{(2)} (r) \big|_{\rm pert}^{\rm OS}
&=& \frac{4 \pi \alpha_s C_F}{3} \delta^{(3)} (\bm{r}) 
+ O(\alpha_s^2). 
\end{eqnarray}
\end{subequations}
We use the superscript OS to denote the on-shell matching scheme. 

In Wilson-loop matching, the potentials are given in terms of the rectangular
Wilson loop $W_{r \times T}$ with spatial size $r$ and time extension $T$, with
insertions of the gluon fields. 
The nonperturbative expressions for the $1/m$ potential and the
velocity-dependent potential in Wilson loop matching are given
by~\cite{Brambilla:2000gk, Pineda:2000sz}  
\begin{subequations}
\label{eq:potentials_wilson_def}
\begin{eqnarray}
V^{(1)} (r) \big|^{\rm WL} &=&
- \lim_{T \to \infty} \int_0^T
dt \, t \left( \langle\!\langle g_s \bm{E}_1^i (t) g_s \bm{E}_1^j (0) \rangle\! \rangle
- \langle\!\langle g_s \bm{E}_1^i (t)  \rangle\!\rangle
\langle\!\langle g_s \bm{E}_1^j (0) \rangle\!\rangle
\right),
\\
V_{p^2}^{(2)} (r) \big|^{\rm WL} &=&
i \hat{\bm{r}}^i \hat{\bm{r}}^j \lim_{T \to \infty}
\int_0^T dt\, t^2
\left( \langle\!\langle g_s \bm{E}_1^i (t) g_s \bm{E}_1^j (0) \rangle\!\rangle
- \langle\!\langle g_s \bm{E}_1^i (t)  \rangle\!\rangle
\langle\!\langle g_s \bm{E}_1^j (0) \rangle\!\rangle
\right)
\nonumber \\
&&
+ i \hat{\bm{r}}^i \hat{\bm{r}}^j \lim_{T \to \infty}
\int_0^T dt\, t^2
\left( \langle\!\langle g_s \bm{E}_1^i (t) g_s \bm{E}_2^j (0) \rangle\!\rangle
- \langle\!\langle g_s \bm{E}_1^i (t)  \rangle\!\rangle
\langle\!\langle g_s \bm{E}_2^j (0) \rangle\!\rangle
\right), \quad\quad
\end{eqnarray}
\end{subequations}
where $\langle\!\langle \cdots \rangle\!\rangle
\equiv \langle \cdots W_{r \times T} \rangle/ \langle W_{r \times T} \rangle$,
$\hat{\bm{r}} = \bm{r}/|\bm{r}|$,
the angular brackets $\langle \cdots \rangle$ stand
for the average over the Yang-Mills action, and $g_s \bm{E}_1(t)$ ($g_s
\bm{E}_2(t)$) are insertions of the chromoelectric field $\bm{E}^i = G^{i0}$ at
time $t$ on the quark (antiquark) line of the Wilson loop, with
$G^{\mu \nu}$ being the gluon field-strength tensor.
The superscript WL denotes that the potential is obtained in Wilson loop
matching.
The complicated expressions for $V^{(2)}_{r} (r)$ and $V_{S^2}^{(2)} (r)$ 
can be found in ref.~\cite{Pineda:2000sz}. 
The short-distance behavior of the potentials in Wilson loop matching can be
obtained by computing the nonperturbative definitions in perturbative 
QCD~\cite{Peset:2015vvi}. We list the results at leading nonvanishing orders in
$\alpha_s$: 
\begin{subequations}
\label{eq:potentials_wilson_short}
\begin{eqnarray}
V^{(1)} (r) \big|_{\rm pert}^{\rm WL} &=& 
- \frac{\alpha_s^2 C_F C_A}{2 r^2} + O(\alpha_s^3), 
\\
V^{(2)}_{r} (r) \big|_{\rm pert}^{\rm WL} &=& 
\pi \alpha_s C_F \delta^{(3)} (\bm{r}) + O(\alpha_s^2), 
\\
V_{p^2}^{(2)} (r) \big|_{\rm pert}^{\rm WL} &=& 
- \frac{\alpha_s C_F}{r} + O(\alpha_s^2), 
\\
V_{S^2}^{(2)} (r) \big|_{\rm pert}^{\rm WL} &=& 
\frac{4 \pi \alpha_s C_F}{3} \delta^{(3)} (\bm{r}) 
+ O(\alpha_s^2). 
\end{eqnarray}
\end{subequations}

The potentials from on-shell matching in eq.~(\ref{eq:potentials_pert}) and
the potentials from Wilson loop matching in
eq.~(\ref{eq:potentials_wilson_short}) are related by unitary transformations,
as described in section~\ref{sec:unitary_transformations}.

\section{Short-distance coefficients}
\label{appendix:sdcs}

In this appendix, we list the NRQCD factorization formulas and 
SDCs for the decay constants and decay rates that we
consider in sec.~\ref{sec:results}. 
The NRQCD factorization formula for the decay constant $f_V$ of a vector
quarkonium $V$ reads
\begin{equation}
\label{eq:fac_vectordecayconstant}
f_V = \frac{\sqrt{2 m_V}}{m_V} 
\left( c_v 
\langle 0 | \chi^\dag \bm{\epsilon} \cdot \bm{\sigma} \psi | V \rangle
+ \frac{d_v}{m^2} 
\langle 0 | \chi^\dag \bm{\epsilon} \cdot \bm{\sigma}
(-\tfrac{i}{2} \overleftrightarrow{\bm{D}} )^2
\psi | V \rangle
+ O(v^3) 
\right),  
\end{equation}
where $m_V$ is the mass of the quarkonium $V$, and 
the SDCs $c_v$ and $d_v$ are given in the $\overline{\rm MS}$ scheme 
by~\cite{Barbieri:1975ki,Celmaster:1978yz,Czarnecki:1997vz,Beneke:1997jm,  
Keung:1982jb,Luke:1997ys,Bodwin:2008vp} 
\begin{subequations}
\label{eq:sdcs_veccurrent}
\begin{eqnarray}
\label{eq:sdcs_veccurrent1}
c_v &=& 1 - \frac{2 \alpha_s(m) C_F}{\pi} + 
\left( \frac{\alpha_s(m)}{\pi} \right)^2 
\bigg[ 
C_F^2 c_{v,A} + C_F C_A c_{v,NA}
\nonumber \\ && \hspace{32ex}
+ C_F T_F n_f c_{v,L} 
+ C_F T_F c_{v,H}
\bigg]  +O(\alpha_s^3), 
\\
d_v &=& - \frac{1}{6} + 
\frac{2 \alpha_s C_F}{9 \pi} \left(1 - 3 \log \frac{m^2}{\Lambda^2} \right)
+ O(\alpha_s^2), 
\end{eqnarray}
\end{subequations}
and 
\begin{subequations}
\begin{eqnarray}
c_{v,A} &=& 
\frac{23}{8} - \frac{\zeta(3)}{2} 
+ \pi^2 \log 2
- \frac{76 \pi^2 }{36}
+\frac{\pi^2}{6} \log \frac{m^2}{\Lambda^2},
\\
c_{v,NA} &=& 
- \frac{151}{72} - \frac{13}{4} \zeta(3)
- \frac{5 \pi^2}{6} \log 2
+ \frac{89 \pi^2}{144}
+ \frac{\pi^2}{4} \log \frac{m^2}{\Lambda^2}, 
\\
c_{v,L} &=& \frac{11}{18}, 
\\
c_{v,H} &=& - \frac{2 \pi^2}{9} + \frac{22}{9}. 
\end{eqnarray}
\end{subequations}
Here, $\Lambda$ is the scale at which the NRQCD LDMEs are renormalized. 
The expression for $c_v$ in eq.~(\ref{eq:sdcs_veccurrent1}) is valid 
when $\alpha_s$ is evaluated in the $\overline{\rm MS}$ scheme at the scale
$m$. Since the QCD renormalization scale dependence cancels order by order in 
$c_v$, eq.~(\ref{eq:sdcs_veccurrent1}) is still valid if we replace 
$\alpha_s(m)$ by $\alpha_s(\mu_R)$ and add 
$- \frac{2 \alpha_s C_F}{\pi} \times \frac{\alpha_s \beta_0}{4 \pi}
\log (\mu_R^2/m^2)$, which compensates for the running of $\alpha_s$. 
We note that the order-$\alpha_s^3$ correction to $c_v$ have been obtained in
ref.~\cite{Marquard:2014pea}. 

The NRQCD factorization formula for the decay constant $f_P$ of a
pseudoscalar quarkonium $P$ reads
\begin{equation}
f_P = \frac{\sqrt{2 m_P}}{m_P}
\left( c_p
\langle 0 | \chi^\dag \psi | P \rangle
+ \frac{d_p}{m^2}
\langle 0 | \chi^\dag 
(-\tfrac{i}{2} \overleftrightarrow{\bm{D}} )^2
\psi | P \rangle
+ O(v^3)
\right),
\end{equation}
where $m_P$ is the mass of the quarkonium $P$, and 
$c_p$ and $d_p$ read, in the $\overline{\rm MS}$ scheme~\cite{Braaten:1995ej,Kniehl:2006qw,Wang:2017bgv},
\begin{subequations}
\label{eq:sdcs_pscurrent}
\begin{eqnarray}
\label{eq:sdcs_pscurrent1}
c_p &=& 1 - \frac{3 \alpha_s(m) C_F}{2 \pi} +
\left( \frac{\alpha_s(m)}{\pi} \right)^2
\bigg[
C_F^2 c_{p,A} + C_F C_A c_{p,NA}
\nonumber \\ && \hspace{20ex}
+ C_F T_F n_f c_{p,L}
+ C_F T_F c_{p,H}
+ C_F T_F X_{\rm sing}^{(p)} 
\bigg]  +O(\alpha_s^3),
\\
d_p &=& - \frac{1}{2} + O(\alpha_s),
\end{eqnarray}
\end{subequations}
and
\begin{subequations}
\begin{eqnarray}
c_{p,A} &=&
\frac{29}{16}- \frac{79}{8} \zeta(2) + 6 \zeta(2) \log 2
+ \frac{9}{2} \zeta(3) +3 \zeta(2) 
\log \frac{m^2}{\Lambda^2},
\\
c_{p,NA} &=& - \frac{17}{48} +\frac{17}{8} \zeta(2) - 6 \zeta(2) \log 2
-3 \zeta(3) + \frac{3}{2} \zeta(2) \log \frac{m^2}{\Lambda^2},
\\
c_{p,L} &=& \frac{1}{12},
\\
c_{p,H} &=& \frac{43}{12} - 2 \zeta(2), 
\\
X_{\rm sing}^{(p)} 
&=& \frac{5}{4} \zeta(2)+3 \zeta(2) \log 2 - \frac{21}{8} \zeta(3) 
+ \frac{3}{4} i \pi \zeta(2). 
\end{eqnarray}
\end{subequations}
Again, $\Lambda$ is the scale at which the NRQCD LDMEs are renormalized. 
The imaginary part in $X_{\rm sing}^{(p)}$ arises from the process 
$Q \bar Q \to gg \to Q \bar Q$, where the gluons are on 
shell~\cite{Kniehl:2006qw}. 

To the best of the author's knowledge, 
the order-$\alpha_s$ correction to $d_p$ has not been computed yet.
Since the QCD renormalization scale dependence cancels order by order in
$c_p$, eq.~(\ref{eq:sdcs_pscurrent1}) is still valid if we replace
$\alpha_s(m)$ by $\alpha_s(\mu_R)$ and add
$- \frac{3 \alpha_s C_F}{2 \pi} \times \frac{\alpha_s \beta_0}{4 \pi}
\log (\mu_R^2/m^2)$, which compensates for the running of $\alpha_s$.

Finally, the two-photon decay rate of a pseudoscalar quarkonium $P$ is given by 
\begin{equation}
\Gamma(P \to \gamma \gamma) 
= 
\frac{8 \pi \alpha^2 e_Q^4}{m_P^2} 
\left| c_{\gamma \gamma} 
\langle 0 | \chi^\dag \psi | P \rangle
+ \frac{d_{\gamma \gamma} }{m^2}
\langle 0 | \chi^\dag
(-\tfrac{i}{2} \overleftrightarrow{\bm{D}} )^2
\psi | P \rangle
+ O(v^3)
\right|^2,
\end{equation}
where the SDCs $c_{\gamma \gamma}$ and $d_{\gamma \gamma}$ are given 
in the $\overline{\rm MS}$ scheme by~\cite{Harris:1957zza, Barbieri:1979be,
Hagiwara:1980nv,Keung:1982jb,Bodwin:1994jh,Czarnecki:2001zc,Feng:2015uha,
Jia:2011ah,Guo:2011tz}
\begin{subequations}
\label{eq:sdcs_psdecay}
\begin{eqnarray}
\label{eq:sdcs_psdecay1}
c_{\gamma \gamma} &=& 1 - \frac{\alpha_s(m) C_F}{\pi} 
\left( \frac{\pi^2}{8}-\frac{5}{2} \right) 
\nonumber \\ && 
+
\left( \frac{\alpha_s(m)}{\pi} \right)^2
\bigg[
\frac{\pi^2}{2} C_F \left( C_F + \frac{C_A}{2} \right) 
\log \left( \frac{m^2}{\Lambda^2} \right) 
+ f^{(2)}_{\rm reg}  
+ f^{(2)}_{\rm lbl}  
\bigg]  +O(\alpha_s^3),
\\
\label{eq:sdcs_psdecay2}
d_{\gamma \gamma} &=& - \frac{1}{6} 
+ \frac{\alpha_s C_F}{\pi} 
\left( - \frac{7}{36} - \frac{4}{3} \log 2 - \frac{\pi^2}{16} - \frac{2}{3} 
\log \frac{m^2}{\Lambda^2} \right) 
+ O(\alpha_s^2).
\end{eqnarray}
\end{subequations}
Here, $\Lambda$ is the scale at which the NRQCD LDMEs are renormalized. 
The constants $f^{(2)}_{\rm reg}$ and $f^{(2)}_{\rm lbl}$ have been determined
numerically in ref.~\cite{Feng:2015uha}, which read  
\begin{subequations}
\begin{eqnarray}
f^{(2)}_{\rm reg} &=& 
-21.10789797(4) \, C_F^2 - 4.79298000(3) \,C_F C_A 
+ 0.223672013(2) \, C_F T_F n_H 
\nonumber \\ && 
- \left(\frac{13}{144} \pi^2 + \frac{2}{3} \log(2) 
+ \frac{7}{24} \zeta(3) - \frac{41}{36} \right) C_F T_F n_f , 
\\
f^{(2)}_{\rm lbl} &=& \left[ 0.73128459 
+ i \pi \left( \frac{\pi^2}{9} - \frac{5}{3} \right) \right]
C_F T_F
\sum_{q} \left( \frac{e_q}{e_Q} \right)^2 
\nonumber \\ && 
+ \big( 0.64696557+2.07357556 \, i \big) \, C_F T_F n_H,
\end{eqnarray}
\end{subequations}
where $n_H = 1$, 
$e_q$ is the fractional charge of the light quark with flavor $q$,
and the sum runs over $n_f$ light quark flavors. 
The term $f^{(2)}_{\rm lbl}$ in $c_{\gamma \gamma}$ originates from the process 
where the $Q \bar Q$ decays into $gg$, which then decays into 
$\gamma \gamma$ via a quark loop~\cite{Feng:2015uha}. The imaginary parts
in $f^{(2)}_{\rm lbl}$ arise from the region
of loop momenta where the intermediate particles are on shell. 
Since the QCD renormalization scale dependence cancels order by order in
$c_{\gamma \gamma}$, eq.~(\ref{eq:sdcs_psdecay1}) is still valid if we replace
$\alpha_s(m)$ by $\alpha_s(\mu_R)$ and add
$- \frac{\alpha_s C_F}{\pi} \left( \frac{\pi^2}{8} - \frac{5}{2} \right) 
\times \frac{\alpha_s \beta_0}{4 \pi}
\log (\mu_R^2/m^2)$.

\section{Wavefunctions at the origin in perturbative QCD}
\label{appendix:perttest}

If we ignore the nonperturbative long-distance behavior of the static
potential, so that $V_{\rm LO} (r) = -\alpha_s C_F/r$, the Schr\"odinger
equation can be solved exactly, and the $S$-wave contribution to the Green's 
function in position space is known analytically:
\begin{eqnarray}
\label{eq:cgreen_analytic}
G^S (\bm{r}',\bm{r};E) 
&=& - \frac{\alpha_s C_F m^2}{4 \pi} 
\Gamma(-\lambda) 
\exp \left({- \frac{1}{2 \lambda} \alpha_s C_F m (r_<+r_>)} \right)
\nonumber \\ && \times 
{}_1F_1 (1-\lambda;2;\alpha_s C_F m r_</\lambda) 
\,U(1-\lambda;2;\alpha_s C_F m r_>/\lambda), 
\end{eqnarray}
where $r_< = {\rm min}(|\bm{r}|,|\bm{r}'|)$,
$r_> = {\rm max}(|\bm{r}|,|\bm{r}'|)$, 
$\lambda = \alpha_s C_F/\sqrt{-4 E/m}$, and 
\begin{subequations}
\begin{eqnarray}
{}_1F_1(a;b;z) &=& \sum_{k=0}^\infty \frac{(a)_k}{(b)_k} \frac{z^k}{k!},
\\
U(a;b;z) &=& \frac{1}{\Gamma(a)}
\int_0^\infty dt\, e^{-zt} t^{a-1} (1+t)^{b-a-1}.
\end{eqnarray}
\end{subequations}
This result can be obtained by solving the differential equation in
eq.~(\ref{eq:greenequations}) analytically. 
The bound states can be identified from the poles of $\Gamma(-\lambda)$, 
which are located at $\lambda = n$ with principal quantum numbers  
$n=1,2,3,\ldots$. 
The reduced Green's functions can also be obtained analytically 
from eq.~(\ref{eq:cgreen_analytic}). 
This makes possible analytical calculations of the corrections to the
$S$-wave wavefunctions at the origin. 
Such a calculation has been done in the context of heavy quark 
pair production near threshold in perturbative QCD in
refs.~\cite{Hoang:1998xf,Melnikov:1998ug,Penin:1998kx,Hoang:1999zc,
Melnikov:1998pr,Yakovlev:1998ke,Beneke:1999qg,Nagano:1999nw,Penin:1998mx,
Penin:2004ay}. 
We use the known results in perturbative QCD to check the numerical
procedure used in this paper for computing the divergent 
corrections to the wavefunctions at the origin that originate from the 
$1/m$ and $1/m^2$ potentials.

The explicit analytical expressions for the 
two-loop non-Coulombic corrections to the wavefunctions at the origin from
the $1/m$ and $1/m^2$ potentials can be found in ref.~\cite{Penin:1998kx} 
for the spin-triplet state. 
Rather than comparing the wavefunctions at the origin, which depends on 
the renormalization scheme, it is simpler to compare the corrections to the 
decay constant $f_V$, which is scale and scheme independent. 
At leading order in $\alpha_s$ and $v$, $f_V$ for the spin-triplet $nS$
state is given in perturbative QCD by 
\begin{equation}
f_V^{\rm LO} \big|_{\textrm{pert}} = 
\sqrt{\frac{2 N_c}{m}} 
| \Psi_n^{\rm LO} (0)| , 
\end{equation}
where $| \Psi_n^{\rm LO} (0)|^2 = (\alpha_s C_F m)^3/(8 \pi n^3)$.
The
corrections to the wavefunctions at the origin coming from the $1/m$ and $1/m^2$
potentials are given in ref.~\cite{Penin:1998kx} by 
\begin{eqnarray}
\label{eq:wforigin_pert}
| \Psi_n (0)|
&=& | \Psi_n^{\rm LO} (0)|
\times \bigg\{ 
1 - \frac{1}{2} 
\alpha_s^2 C_F
\bigg[
\frac{15 C_F}{8 n^2} + 
\left( \frac{2}{3} C_F + C_A \right)
\nonumber \\ && \hspace{25ex} \times 
\left( H_{n-1} - \frac{1}{n} 
- \log \left( \frac{2 \mu_f n}{\alpha_s C_F m} \right)
\right) 
\bigg] + \cdots
\bigg\}, 
\end{eqnarray}
where $\mu_f$ is a factorization scale, and the ellipsis represent the 
Coulombic corrections that we neglect. The wavefunctions at the origin in 
ref.~\cite{Penin:1998kx} are renormalized in a scheme that is different from 
the $\overline{\rm MS}$ scheme that we use in this paper, so it is not
possible to compare eq.~(\ref{eq:wforigin_pert}) directly with the results
in this paper.
This scheme dependence cancels in the decay constant 
against the scheme dependence in the hard matching coefficient 
given by eq.~(2) in ref.~\cite{Penin:1998kx}, which is obtained from the 
direct matching procedure~\cite{Hoang:1997ui}. 
Since the non-Coulombic corrections are proportional to $\alpha_s^2 C_F^2$
and $\alpha_s^2 C_F C_A$, we only need to keep the contributions that are
proportional to $C_F^2$ and $C_F C_A$ in the loop corrections to the 
hard matching coefficient. 
By combining the corrections to the wavefunctions at the origin 
and the hard matching coefficient, 
we obtain the two-loop non-Coulombic correction given by 
\begin{eqnarray}
\label{eq:fVanalytical_nonCoulombic}
\delta_f^{\rm NC} &=& 
- \frac{1}{2}
\alpha_s^2 C_F \bigg[
\frac{15 C_F}{8 n^2}
+ \left( \frac{2}{3} C_F + C_A \right)
\left( H_{n-1} - \frac{1}{n} 
- \log \left( \frac{2 \mu_f n }{\alpha_s C_F m} \right)
\right) \bigg]
\nonumber \\ && 
+ \left( \frac{\alpha_s}{\pi} \right)^2 
\bigg[ \bigg(
\frac{23}{8} - \frac{\zeta(3)}{2} + \frac{2 \pi^2}{3} \log 2- 
\frac{35 \pi^2}{36}
+ \frac{\pi^2 }{6} \log \frac{m^2}{\mu_f^2} 
\bigg) C_F^2 
\nonumber \\ && \hspace{10ex} 
+ \bigg(
- \frac{151}{72} - \frac{13}{4} \zeta(3) - \frac{4 \pi^2}{3} \log 2
+ \frac{179 \pi^2}{144} 
+ \frac{\pi^2}{4} \log \frac{m^2}{\mu_f^2} 
\bigg)
C_F C_A \bigg], \quad 
\end{eqnarray}
where the last two lines correspond to the $C_F^2$ and 
the $C_F C_A$ terms of the two-loop corrections to the hard
matching coefficients in ref.~\cite{Penin:1998kx}. 
The $\mu_f$ dependence cancels exactly between the 
non-Coulombic corrections to the wavefunctions at the origin and the 
two-loop corrections to the hard matching coefficient. 
Note that the last two lines of eq.~(\ref{eq:fVanalytical_nonCoulombic})
differ from the $C_F^2$ and the $C_F C_A$ terms of the two-loop corrections to
the SDC $c_v$ in eq.~(\ref{eq:sdcs_veccurrent1}). 
This reflects the difference between the renormalization scheme 
used in ref.~\cite{Penin:1998kx} and the $\overline{\rm MS}$ scheme used 
in this work. 
This analytical result can be compared with the numerical calculation 
in this paper, which is given by 
\begin{eqnarray}
\label{eq:fVnumerical_nonCoulombic}
\delta_f^{\rm NC} &=& 
\delta_\Psi^{\rm NC} \big|_{\rm pert} 
+ \left( \frac{\alpha_s}{\pi} \right)^2 
\left( C_F^2 c_{v,A} + C_F C_A c_{v,NA} \right) , 
\end{eqnarray}
where $\delta_\Psi^{\rm NC} \big|_{\rm pert}$ is equal to 
eq.~(\ref{eq:NCcorr}), 
except that we set $V_{\rm LO} (r) = -\alpha_s C_F/r$,
$V_{p^2}^{(2)} (0) = 0$, and we take the 
perturbative $1/m$ and $1/m^2$ potentials in eq.~(\ref{eq:potentials_pert}). 
The dependence on $\Lambda$ cancels exactly in 
eq.~(\ref{eq:fVnumerical_nonCoulombic})
between $\delta_\Psi^{\rm NC} \big|_{\rm pert}$ and 
the SDCs $c_{v,A}$ and $c_{v,NA}$. 

\begin{figure}[tbp]
\centering
\includegraphics[width=.55\textwidth]{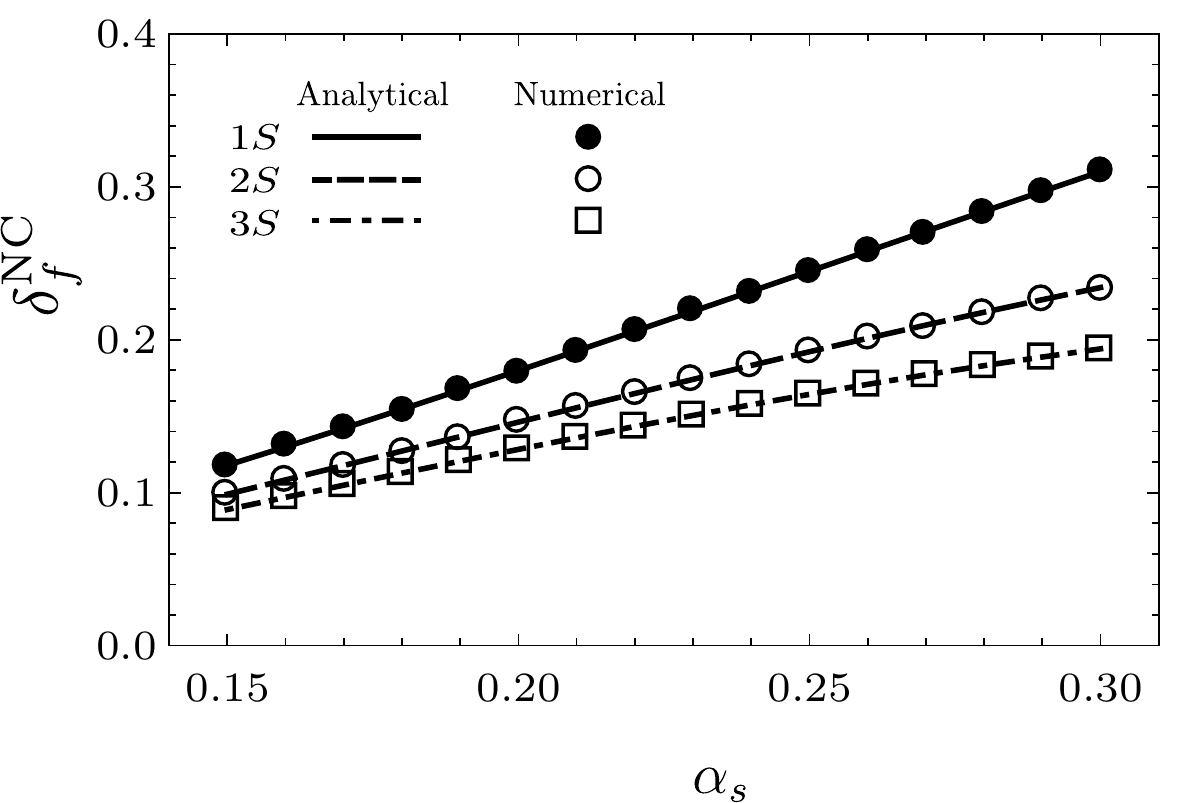}
\caption{\label{fig:perttest}
Comparison of numerical calculations and analytical results for 
the non-Coulombic corrections $\delta_f^{\rm NC}$
to the decay constant of vector quarkonium 
in perturbative QCD
for various values of $\alpha_s$. 
Numerical results are shown as 
filled circles ($1S$), open circles ($2S$), and open squares ($3S$).
The analytical results are shown as solid line ($1S$), 
dashed line ($2S$), and dot-dashed line ($3S$). 
}
\end{figure}

We compare the numerical calculation in 
eq.~(\ref{eq:fVnumerical_nonCoulombic})
with the analytical result in eq.~(\ref{eq:fVanalytical_nonCoulombic}) 
for $n=1$, 2, and 3 in fig.~\ref{fig:perttest}. 
We set $m = 4.743$~GeV, $r_0 = 10^{-4}$~GeV$^{-1}$, 
and vary $\alpha_s$ between $0.15$ and $0.3$.  
The agreement between the numerical calculations and the analytical results 
is better than 1\%. 
This agreement demonstrates the validity of the numerical calculation 
in this work.

\bibliography{drpnrqcd_v3.bib}
\bibliographystyle{JHEP}

\end{document}